\documentclass[conference]{IEEEtran} 
\pdfoutput=1
\IEEEoverridecommandlockouts
\usepackage[noadjust]{cite}
\usepackage{amsmath,amssymb,amsfonts}
\usepackage{algorithmic}
\usepackage{graphicx}
\usepackage{textcomp}
\usepackage{xcolor}
\usepackage{multirow}
\usepackage{subcaption}
\usepackage{cleveref}
\usepackage{listings}
\usepackage{float}
\usepackage{makecell}
\usepackage{hhline}
\usepackage{listings}
\usepackage{adjustbox}
\usepackage{balance}
\usepackage[T1]{fontenc}

\def\BibTeX{{\rm B\kern-.05em{\sc i\kern-.025em b}\kern-.08em
    T\kern-.1667em\lower.7ex\hbox{E}\kern-.125emX}}

\begin{document}
\renewcommand\citepunct{, }
\title{PolyFrame: A Retargetable Query-based Approach\\ to Scaling DataFrames (Extended Version)
% \thanks{The  work  reported  in  this  paper  was  supported  in  part  by [ICS grant], Thai government scholarship, and [BREN].}
}

\author{\IEEEauthorblockN{Phanwadee Sinthong}
\IEEEauthorblockA{\textit{Dept. of Computer Science}\\
\textit{University of California, Irvine}\\
psinthon@uci.edu}
\and
\IEEEauthorblockN{Michael J. Carey}
\IEEEauthorblockA{\textit{Dept. of Computer Science}\\
\textit{University of California, Irvine}\\
mjcarey@ics.uci.edu}
}

\maketitle

\begin{abstract}
In the last few years, the field of data science has been growing rapidly as various businesses have adopted statistical and machine learning techniques to empower their decision making and applications. Scaling data analysis, possibly including the application of custom machine learning models, to large volumes of data requires the utilization of distributed frameworks. This can lead to serious technical challenges for data analysts and reduce their productivity. AFrame, a Python data analytics library, is implemented as a layer on top of Apache AsterixDB, addressing these issues by incorporating the data scientists' development environment and transparently scaling out the evaluation of analytical operations through a Big Data management system. While AFrame is able to leverage data management facilities (e.g., indexes and query optimization) and allows users to interact with a very large volume of data, the initial version only generated SQL++ queries and only operated against Apache AsterixDB. In this work, we describe a new design that retargets AFrame’s incremental query formation to other query-based database systems as well, making it more flexible for deployment against other data management systems with composable query languages.
\end{abstract}

% \begin{IEEEkeywords}
% DataFrames, distributed data management, large-scale data analysis, data science, benchmark
% \end{IEEEkeywords}
\section{Introduction}
In this era of big data, extracting useful patterns and intelligence for improved decision-making is becoming a standard requirement for many businesses. The growing interest in interpreting large volumes of user-generated content on social media sites for purposes ranging from business advantages to societal insights motivates the development of data analytic tools. The requirements that large-scale modern data analysis imposes on these tools are not met by a single system. Data scientists are thus required to integrate and maintain multiple separate platforms, such as HDFS~\cite{hdfs}, Spark~\cite{spark}, and TensorFlow~\cite{tensorflow}, which demands systems expertise from analysts who should instead be focused on data modeling, selection of machine learning techniques, and data exploration.

AFrame~\cite{aframe} is a data exploration library that provides a Pandas-like DataFrame experience on top of Apache AsterixDB~\cite{asterixweb}. AFrame leverages distributed data storage and management in order to accommodate the rapid rate and volume at which modern data arrives. Storing such massive data in a traditional file system is no longer an ideal solution because analysis then requires complete file scans to retrieve even a modest subset of the data. Database management systems are able to store, manage, and utilize indexes and query optimization to efficiently retrieve subsets of the data, enabling much more interactive data manipulation.

This project is an effort to make AFrame more flexible to enable a wider audience in the data science community to leverage its scale-independent data analysis and data management capabilities. This is achievable by abstracting AFrame's existing language translation layer and retargeting its incremental query formation mechanism to operate against other database systems. We establish a set of rewrite rules to provide an easily extensible template for supporting other composable query languages, thus allowing AFrame to operate against other query-based database systems. As a proof-of-concept, we have applied our language rewrite rules to four different query languages SQL++~\cite{sql_pp}, SQL~\cite{sql},  MongoDB's Query Language~\cite{mongo_aggregation}, and Cypher~\cite{cypher} to retarget AFrame to work against AsterixDB~\cite{asterixweb}, PostgreSQL~\cite{postgres}, MongoDB~\cite{mongodb}, and Neo4j~\cite{neo4j,neo4j_book} respectively.

In this work we have re-architected AFrame to make it retargetable to other query-based database systems to allow their users to accomplish large-scale data analysis. The contributions of the resulting PolyFrame system are the following:

\begin{enumerate}
    
 \item We enable large-scale data analysis using a Pandas-like syntax on a variety of query-based database systems of choice. 

\item We identify common mapping rules between dataframe operations and database queries. This allows the system to reuse any combinations of the rules to construct queries that represent the supported dataframe operations.

\item We extract and separate generic and language-specific rules to make it easy to introduce a new language, as the query composition mechanism is separated from the query syntax.

\item We decompose complex Pandas dataframe operations (e.g., get\_dummies, describe) into a sequence of simple operations via generic rewrite rules allows PolyFrame to utilize subqueries, which provides a simple localized model for  language-specific mappings.

 \item We support user-defined rewrites to allow users to specify their own custom rewrite rules to leverage a system's language-specific optimizations. 
\end{enumerate}

The rest of this paper is organized as follows: Section 2 discusses background and related work. Section 3 provides an overview of our retargetable query-based design along with examples. Section 4 describes a set of performance experiments and results from running PolyFrame against different backend database systems. We conclude and describe open research problems in Section 5.

% Due to the eager-evaluation nature of Pandas Dataframes, AFrame converts Pandas Dataframes' operations into nested queries in order to maintain the operation sequence. However, it is able to efficiently operate on data at scale by leveraging AsterixDB’s query optimizer. As a result, our retargetable database systems are assumed to be those that can operate against composable query languages and are equipped with an efficient and powerful query optimizer. The main contribution of this work is providing a `scale-independent' user experience when moving from a local exploratory data analysis environment to a large-scale distributed workflow.

\section{Background and Related Work}
We are extending AFrame (to create PolyFrame) by making it platform-independent, via flexible language rewrite rules, in order to create a framework that can help support end-to-end large-scale data analysis with data scientists' familiar syntax against a variety of backend databases. Here we review the basics of AFrame's architecture, its current backend (AsterixDB), and related work.
% \vspace{-1em}
\subsection{Apache AsterixDB}
Apache AsterixDB is an open source Big Data Management System (BDMS)~\cite{asterixdb}. It provides distributed data management for large-scale, semi-structured data using the AsterixDB Data Model (ADM), which is a superset of JSON. In AsterixDB, data records are stored in datasets. Each dataset has a datatype that describes the stored data. ADM datatypes are open so that users can provide a minimal description even when the stored data can have additional contents. 

\subsection{AFrame}
AFrame~\cite{aframe,arxiv} is a library that provides a Pandas DataFrame~\cite{pandasdataframe} based syntax to interact with data in Apache AsterixDB. AFrame targets data scientists who are already familiar with Pandas DataFrames. It works on distributed data by connecting to AsterixDB using its RESTful API.  Inspired in part by Spark, AFrame leverages lazy evaluation to take advantage of AsterixDB's query optimizer. AFrame operations are incrementally translated into SQL++ queries that are sent to AsterixDB only when final results are actually called for.

\subsection{Related Work}
There are several scalable Pandas-like dataframe implementations. These libraries either provide similar Pandas interface on distributed compute engine or run multiple Pandas dataframes in parallel to speed up computation and provide a mean to explore and manipulate data larger than the available memory. However, to our knowledge, there has not been an effort to develop a dataframe interface directly on top of database systems where large volumes of data are stored. Below we compare and contrast PolyFrame with some of the existing scalable dataframe libraries and polystore systems. 

\textbf{Pandas}: Pandas~\cite{pandas} is a Python data analytics framework that reads data from various file formats and creates a Python object, a DataFrame, with rows and columns similar to Excel. Pandas works with Python machine learning libraries such as Scikit-Learn~\cite{scikit} and it can also be integrated with scientific visualization tools such as Jupyter  notebooks~\cite{jupyter}. The rich set of features that are available in Pandas makes it one of the most preferred and widely used tools in data exploration. However, its limitation lies in its lack of scalability, as its strength has typically been for in-memory computation on a single machine. In addition, Pandas' internal data representation is inefficient as Wes McKinney (Pandas' creator) stated in~\cite{mckinney} that a "rule of thumb for pandas is that you should have 5 to 10 times as much RAM as the size of your dataset". 
% Pandas does not provide either data storage or support for interacting with a large amount of (local or distributed) data

% \textbf{Apache Hive}~\cite{hive} is a data warehouse infrastructure tool which is used for processing large amounts of data stored in a distributed file system. It allows query-based user interactions with archived distributed data residing in HDFS. In order to leverage Hive's processing power, extensive knowledge of SQL is essential. A limitation of Hive is in the absence of its own data storage, making Hadoop installation and scan-based query processing its core requirements.

\textbf{Apache Spark}: Apache Spark~\cite{apachespark} is a general-purpose cluster-based computing platform that provides in-memory parallel data processing on clusters with scalability and fault tolerance. Spark also provides DataFrames~\cite{sparkdf}, an API roughly similar to Pandas Dataframes, built for distributed structured data manipulation. However, Spark's DataFrame syntax is different from Pandas' Dataframe syntax in several respects due to its lazy evaluation approach. As a result, Koalas~\cite{koalas}, a new open source project, was established to allow for easier transitioning from Pandas to Spark. Koalas provides a Pandas-like Dataframe API and uses Spark for evaluation under the hood. Although Spark supports fast computation, querying, and accessing of structured data, it does not provide its own data storage or data management.
% It typically works together with HDFS.

\textbf{Modin}:
Modin~\cite{modin}, previously named Pandas on Ray, is another recent attempt to make Pandas DataFrames work on big data by supporting the Pandas syntax and distributing the data and operations using a shared memory framework called Ray~\cite{ray}. Ray employs a distributed scheduler to manage a cluster's resources. By running on Ray, Modin automatically utilizes all the available cores on a machine to execute Pandas operations in parallel. However, Modin does not provide data storage, and it uses Pandas internally (which does not address Pandas' high memory consumption issue).

\textbf{Polystores}: Polystore systems (e.g., BigDAWG, BigIntegrator, and Polybase) provide integrated and transparent access to multiple data stores with heterogeneous storage engines through a common language. In~\cite{principles}, polystores are categorized into three different groups based on the level of coupling with the underlying data stores: loosely coupled, tightly coupled, and hybrid systems. These systems typically share a common mediator-wrapper architecture in which a mediator process accepts input queries, interacts with data stores to obtain and merge the results, and delivers the results to the user. The differences between polystore systems and PolyFrame are their interactions and intended usages. PolyFrame provides a common language of DataFrame operations for users to interact with a query-based database system of choice. PolyFrame does not aim to communicate or orchestrate queries between multiple data stores.

% 
% \vspace{-1em}

\section{PolyFrame System Overview}
% In order to extend AFrame to accommodate ta broader subset of the data science community, we propose an architecture to retarget AFrame's operations to other database systems as well. Here we first describe the AFrame query formation process and the discuss the proposed language abstraction.
% \vspace{-1em}
In this section, we briefly describe PolyFrame's architecture, its incremental query formation process, and the new language rewrite component which is an architectural extension to make AFrame extensible for deployment against different widely used query-based database systems.

\subsection{PolyFrame Overview}

AFrame provides a Pandas-like DataFrame experience while scaling out the evaluation of its operations over a large cluster. The current implementation of AFrame does this by operating against Apache AsterixDB~\cite{asterixdb} and incrementally constructing nested SQL++ queries in order to mimic Pandas’ (eager) evaluation characteristics while enabling the lazy execution of the resulting queries. In order to make AFrame language-independent, PolyFrame separates its query language rewriting component from the original incremental query formation process. 
\begin{figure}[h!]
\centering
  \includegraphics[width=0.48 \textwidth]{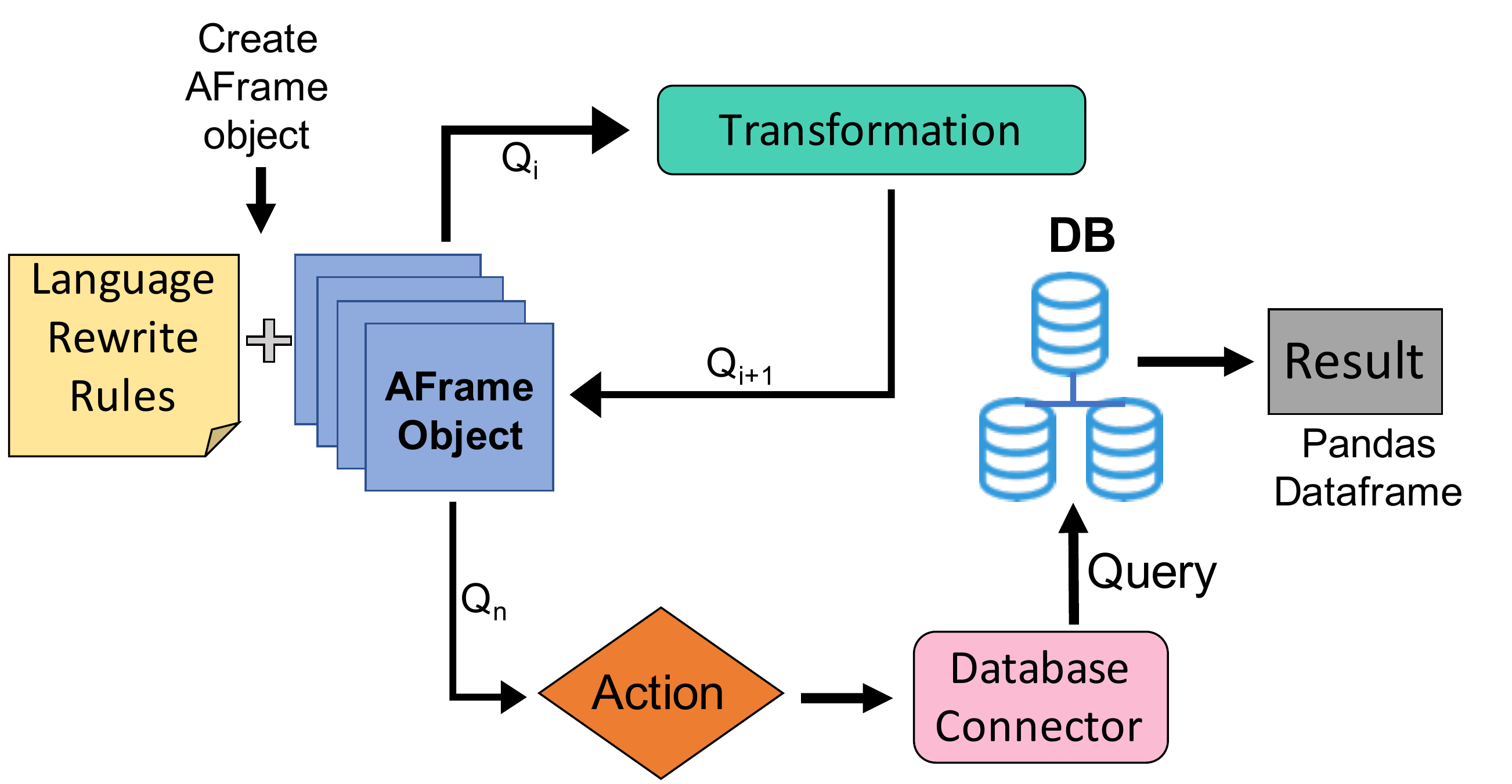}
  \caption{AFrame's New Architecture (PolyFrame)}
  \label{fig:workflow}
\end{figure}

Figure~\ref{fig:workflow} outlines the new AFrame architecture (PolyFrame). An AFrame runtime object is created using a set of language-specific rewrite rules by selecting from those that we provide (e.g., SQL++, SQL, Cypher, MongoDB) or providing a set of custom rules. Inspired by Spark, each operation in AFrame can be categorized as either a transformation or an action. Transformations are operations that transform data. These operations are functions that take an underlying query (Qi) from an AFrame object and produce a new AFrame object with a new underlying query (Qi+1). Transformation operations will not trigger query evaluation, hence AFrame does not produce any intermediate results. Actions are operations that trigger query evaluation. This is done through a database connector which sends the underlying query (Qn) of an AFrame object to a database. The database connector is an abstract class in AFrame that makes connections to database engines. It also performs AFrame initialization, pre-processing of queries before sending them to the database, and post processing of queries' results from the database. A new database connector can be included by providing an implementation of these three required methods. AFrame's query results are then returned in the form of a Pandas Dataframe which is useful when further visualization is desired.  

% The current implementation of AFrame does this by operating against Apache AsterixDB~\cite{asterixdb} and incrementally constructing nested SQL++ queries in order to mimic Pandas’ (eager) evaluation characteristics and record the order of operations. However, it utilizes lazy evaluation to take advantage of AsterixDB’s capability for query optimization. 

% Figure~\ref{fig:op_chain} shows an example of three SQL++ queries generated as a result of Pandas DataFrame operations. The DataFrame operations are listed on top of each AFrame object (the numbered rectangles). The corresponding SQL++ queries are listed below the objects. The first AFrame object is created by passing in the dataverse and dataset names of an existing dataset in AsterixDB. Notice how each subsequent SQL++ query is composed using the query resulting from the previous operation~\cite{aframe}.

\subsection{Incremental Query Formation}
\begin{figure}[h!]
\centering
  \includegraphics[width=0.48 \textwidth]{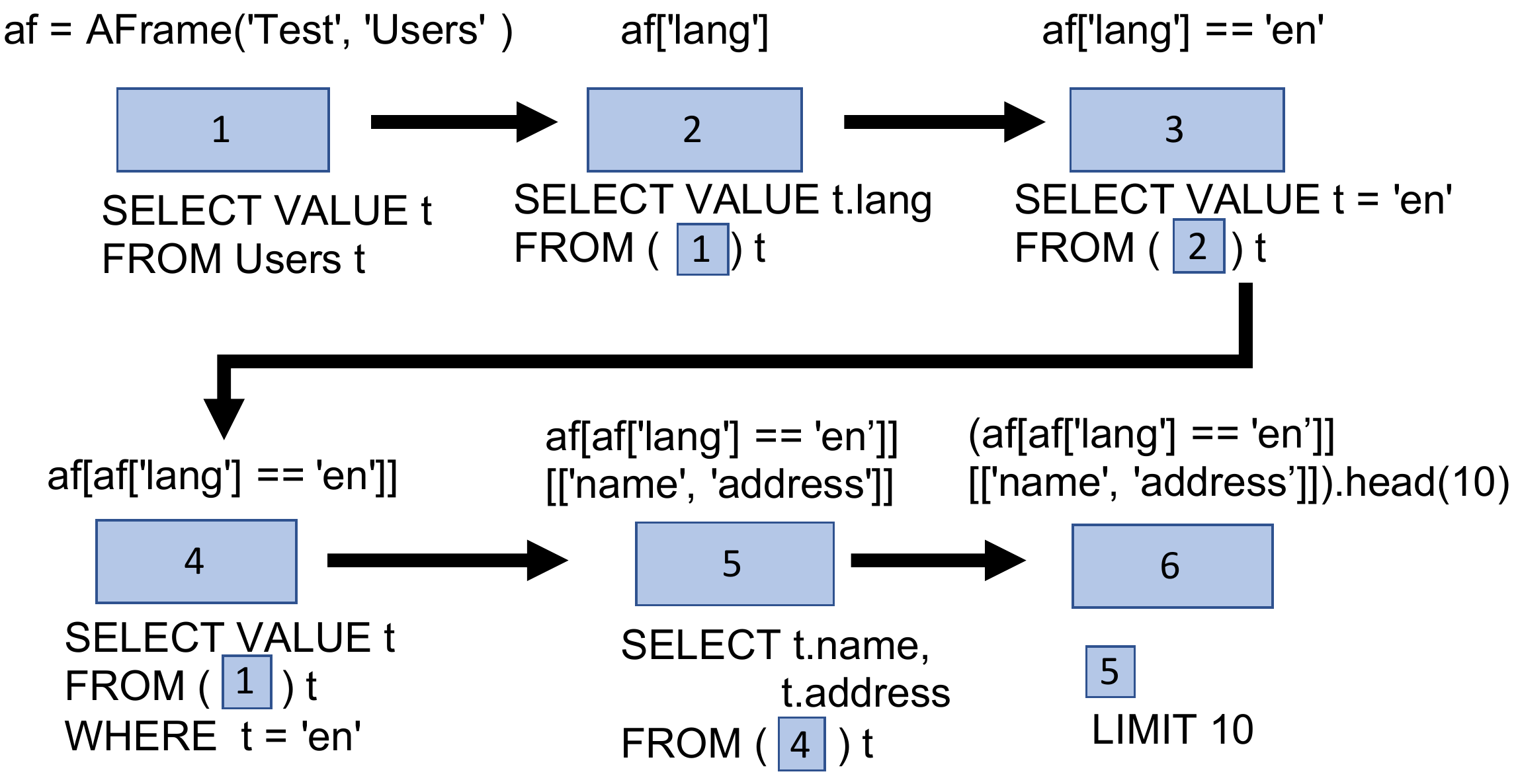}
  \caption[]{Incremental Query Formation\footnotemark}
  \label{fig:op_chain}
\end{figure}

PolyFrame incrementally constructs queries in order to mimic Pandas’ eager evaluation characteristics and record the order of operations. However, it utilizes lazy evaluation to take advantage of databases’ query optimization. Figure~\ref{fig:op_chain} shows an AsterixDB example of six SQL++ queries generated as a result of Pandas DataFrame operations. The Dataframe operations are listed on top of each PolyFrame object (the numbered rectangles). The corresponding SQL++ queries are listed below the objects. The first PoloFrame object (marked 1) is created by passing in the dataverse and dataset name of an existing dataset in AsterixDB. Notice how each subsequent SQL++ query is composed from the query resulting from the previous operation. Operations 1 to 5 are transformations. For these types of operations, PolyFrame does not load any data into memory nor execute any query. Operation 6 (which is asking for a sample of 10 records) is an action that triggers the actual query evaluation. For this operation, PolyFrame appends a `LIMIT 10' clause to the underlying query and makes a connection to a database (AsterixDB in this case) to send the underlying query and retrieve its results.

% AFrame supports DataFrame interactions with machine learning models through AsterixDB's user-defined functions~\cite{deem}.

\subsection{Query Rewrite}

% A unique characteristic of pandas dataframe is the transformation operations return a new dataframe object unless specified otherwise. Therefore, in PolyFrame, query rewriting happens to every dataframe operation regardless of whether the query will be sent over to an underlying database or not. 
\footnotetext{Dataframe number 4 is actually derived from dataframe number 1 but it has the same condition as dataframe number 3.}

\begin{figure}[h!]
\centering
  \includegraphics[width=0.45\textwidth]{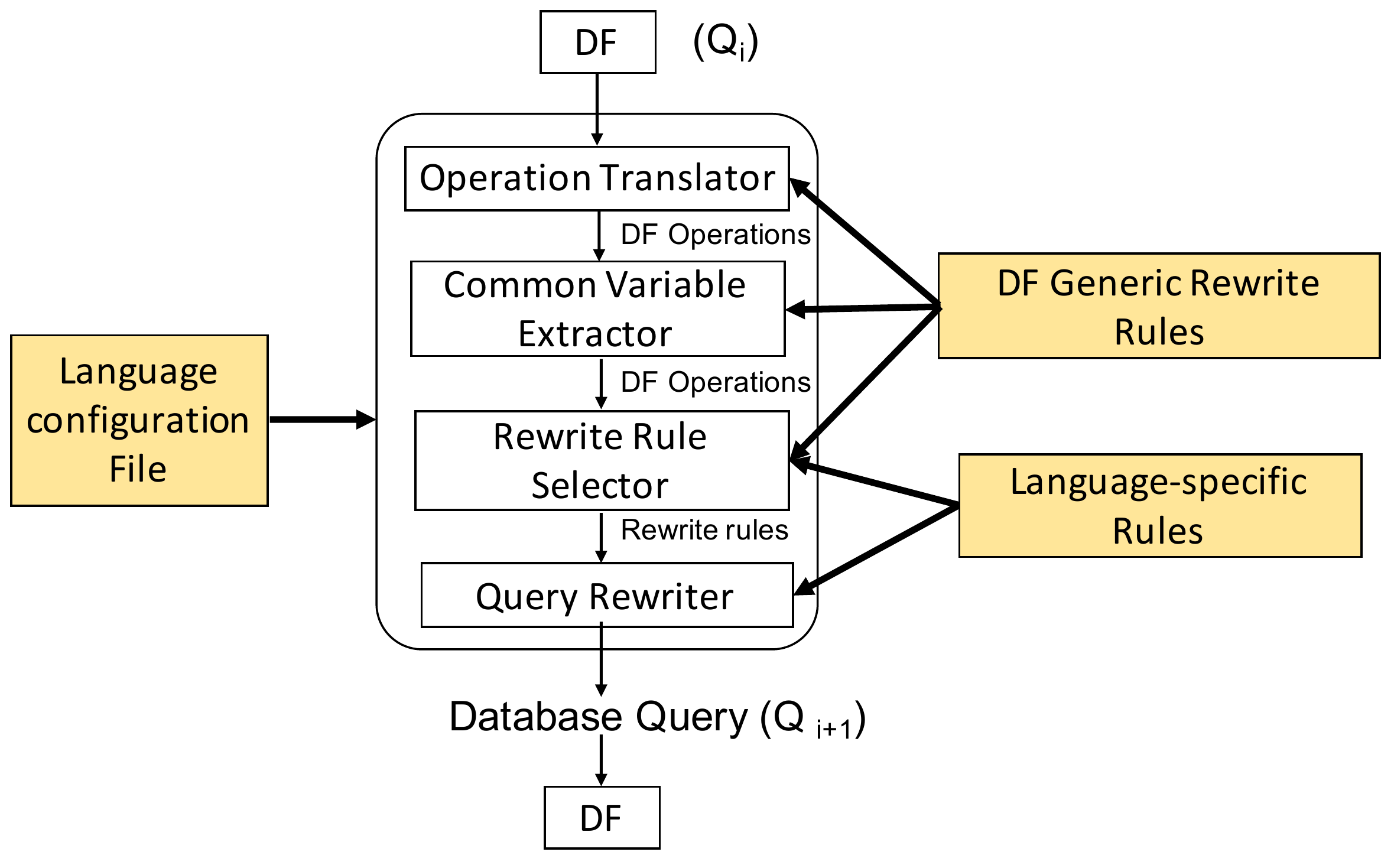}
  \vspace{-1em}
  \caption{Flowchart of a query rewrite}
  \label{fig:flowchart}
  \vspace{-1em}
\end{figure}

Originally, in AFrame, each supported dataframe operation has within its implementation an equivalent SQL++ query that we embedded into the operation. However, this design made AFrame language-dependent. It relied only on features available in AsterixDB. In order to separate the query formation process from the language syntax, we re-architected AFrame and established two sets of rewrite rules that govern how each query is constructed for a particular dataframe operation. Figure~\ref{fig:flowchart} shows the sequence of steps in PolyFrame’s query rewriting process. In PolyFrame, a language configuration file contains query mappings that the system uses during the query formation process. Each PolyFrame object has an underlying query (Qi). When an operation is called on a PolyFrame object, the underlying query is passed into the rewriting process. A dataframe operation is inspected, and if possible, decomposed into multiple simple dataframe operations. Variables from each dataframe operation will also be extracted. For example, df[`name’] is a projection on an attribute `name’, so `name’ is a common variable extracted from the project operation. The system uses generic rewrite rules (described in Section~\ref{generic}) to map each operation to a set of language-specific rules (described in Section~\ref{language_specific}). Query rewriting is then performed on each identified language-specific rule using string pattern matching to replace each token with the extracted common variables. The result is a new database query (Qi+1) encapsulated in a new PolyFrame object.\footnotetext{Dataframe number 4 is actually derived from dataframe number 1 but it has the same condition as dataframe number 3.}
\vspace{-1em}

\subsection{Language Rewrite Rules} 

% Today’s DataFrames and their implementations target highly regular, tabular data. We want to support analyses of social media and IoT data, which are much less regular (e.g., nested and varying from one item to another). We are developing DataFrame extensions for semi-structured (e.g., JSON) data. 

Our work in creating PolyFrame has been to separate AFrame from AsterixDB so that AFrame's benefits can be applied to other database platforms as well. In order to preserve AFrame’s incremental query formation and subquery characteristics, we target query languages that are composable. Another important requirement that any of AFrame’s target database systems must satisfy is an efficient query optimizer. Executing subqueries without any optimization could result in unnecessary data scans that would significantly affect performance. Fortunately, this latter requirement is already an important property of many database systems and is an ongoing area of research.

PolyFrame, our language-independent version of AFrame, provides an extensible database connector and incrementally generates queries based on a set of rewrite rules. Support for each new query language is based on a language mapping file that PolyFrame utilizes at runtime. Currently, we support rewrites of general Pandas Dataframe operations such as selection, projection, join, group by, aggregation, and sorting. A challenge in generating common rewrite rules for AFrame is distinguishing between configurable and general components across various languages. We have established two main types of rewrite rules. One type is language-specific rules and the other type is generic rules. 

\subsubsection{\textbf{Language-specific rules}}\label{language_specific} These are rewrite rules for translating Dataframe operations into (sub) queries that have to be defined in a language configuration file due to the potential syntax differences across various languages. These rules are defined in such a way that they can be combined to create complex queries. In addition to the general Dataframe operations that we support, we require rules for translating arithmetic operations (addition, subtraction, multiplication, division, etc.), aggregation (e.g., sum, average, count, min, and max), comparison statements (equal, not equal, greater than, less than, etc.), logical operations (and, or, and not), and attribute aliases. We include a sample subset of these rewrite rules from the MongoDB and Cypher configuration files in sections~\ref{sec:cypher_rewrites} and \ref{sec:mongo_rewrites} of the Appendix. A challenge in establishing a set of language-specific rewrite rules is identifying the granularity of the rules while maintaining efficiency. Our goal is to identify common components that are shared across languages.

\begin{figure}[h!]
\centering
  \includegraphics[width=0.49 \textwidth]{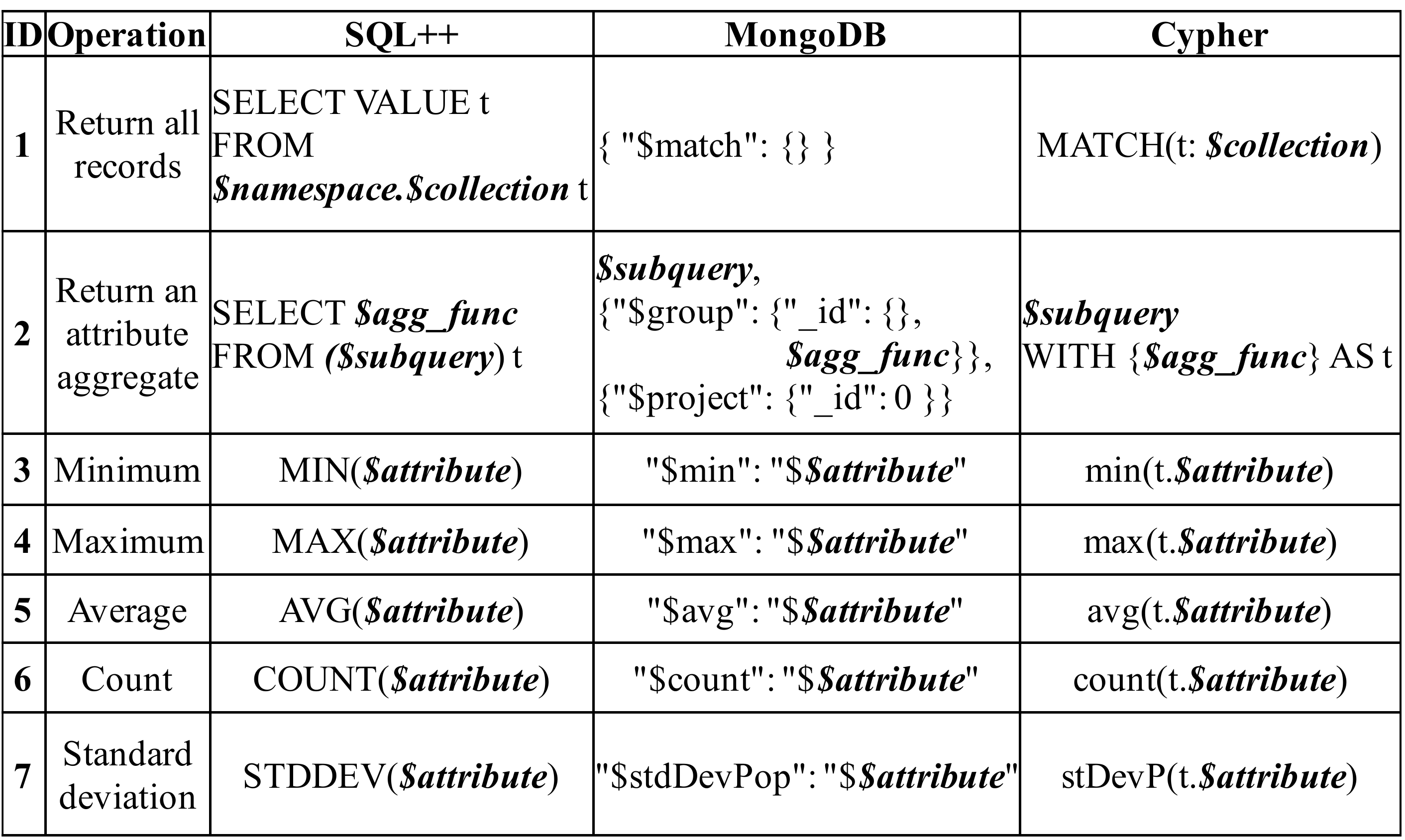}
  \caption{Sample Rewrite Rules (rewrite variables are italicized)}
  \label{fig:rewrite}
\end{figure}

\begin{table*}[h!]
\huge
\resizebox{\textwidth}{!}{%

\begin{tabular}{|l|l|l|l|l|l|}
\hline
\multicolumn{1}{|c|}{ID} & \multicolumn{1}{|c|}{\begin{tabular}[c]{@{}c@{}}AFrame\\ Operation\end{tabular}} & \multicolumn{1}{c|}{SQL++} & \multicolumn{1}{c|}{SQL} & \multicolumn{1}{c|}{MongoDB} & \multicolumn{1}{c|}{Cypher} \\ \hline
1 & af = AFrame(`Test',`Users') & \begin{tabular}[c]{@{}l@{}}SELECT VALUE t\\ FROM \colorbox{pink}{Test.Users} t\end{tabular} & 
\begin{tabular}[c]{@{}l@{}}SELECT *\\ FROM \colorbox{pink}{Test.Users} \end{tabular} & \begin{tabular}[c]{@{}l@{}} \{``\$match": \{\} \} \end{tabular} & \begin{tabular}[c]{@{}l@{}}MATCH(t:  \colorbox{pink}{Users})\end{tabular}\\ \hline

2 & af[`lang'] & \begin{tabular}[c]{@{}l@{}}SELECT t.\colorbox{pink}{lang} \\ FROM (\emph{\textbf{1}}) t\end{tabular} & \begin{tabular}[c]{@{}l@{}}SELECT t.\colorbox{pink}{lang} \\ FROM (\emph{\textbf{1}}) t\end{tabular} & \begin{tabular}[c]{@{}l@{}} 
 \emph{\textbf{1}}, \\
\{``\$project": \{ ``\colorbox{pink}{lang}": 1 \} \}  \end{tabular} 
& \begin{tabular}[c]{@{}l@{}} \emph{\textbf{1}}\\ WITH t\{\`{}\colorbox{pink}{lang}\`{}: t.\colorbox{pink}{lang}\} \end{tabular}\\ \hline

3 & af[`lang'] == `en' & \begin{tabular}[c]{@{}l@{}}SELECT VALUE\\ t.\colorbox{pink}{lang} = \colorbox{pink}{``en"}\\ FROM (\emph{\textbf{2}}) t\end{tabular} & \begin{tabular}[c]{@{}l@{}}SELECT\\ t.\colorbox{pink}{lang} = \colorbox{pink}{``en"} \\ FROM (\emph{\textbf{2}}) t\end{tabular} & 
\begin{tabular}[c]{@{}l@{}} \emph{\textbf{2}}, \\
\{``\$project": \{``is\_eq": \{``\$eq": {[}``\colorbox{pink}{lang}", \colorbox{pink}{``en"}{]}\}\}\}  \end{tabular}
& \begin{tabular}[c]{@{}l@{}} \emph{\textbf{2}} \\WITH t\{\`{}is\_eq\`{}: t.\colorbox{pink}{lang} = \colorbox{pink}{"en"}\}\end{tabular}\\ \hline

4 & af[af[`lang'] == `en'] & \begin{tabular}[c]{@{}l@{}}SELECT VALUE t \\ FROM (\emph{\textbf{1}}) t \\ WHERE t.\colorbox{pink}{lang} = \colorbox{pink}{"en"}\end{tabular} & \begin{tabular}[c]{@{}l@{}}SELECT t.* \\ FROM (\emph{\textbf{1}}) t \\ WHERE t.\colorbox{pink}{lang} = \colorbox{pink}{"en"}\end{tabular} & 
\begin{tabular}[c]{@{}l@{}} \emph{\textbf{1}}, \\
\{"\$match": \{ "\$expr": \{ "\$eq": {[}"\colorbox{pink}{lang}", \colorbox{pink}{"en"}{]} \} \} \} \end{tabular}
& \begin{tabular}[c]{@{}l@{}} \emph{\textbf{1}} \\WITH t WHERE t.\colorbox{pink}{lang} = \colorbox{pink}{"en"}\end{tabular}
\\ \hline

5 & \begin{tabular}[c]{@{}l@{}} af[af[`lang'] == `en']\\  $[[$`name', `address'$]]$\end{tabular} & \begin{tabular}[c]{@{}l@{}}SELECT t.\colorbox{pink}{name},\\     t.\colorbox{pink}{address}\\ FROM (\emph{\textbf{4}}) t\end{tabular} & \begin{tabular}[c]{@{}l@{}}SELECT t.\colorbox{pink}{name}, \\     t.\colorbox{pink}{address} \\ FROM (\emph{\textbf{4}}) t\end{tabular} & \begin{tabular}[c]{@{}l@{}} 
\emph{\textbf{4}}, \\
\{"\$project": \{ "\colorbox{pink}{name}": 1,  "\colorbox{pink}{address}": 1\} \} \end{tabular} 
& \begin{tabular}[c]{@{}l@{}}\emph{\textbf{4}}\\ WITH t\{\`{}\colorbox{pink}{name}\`{}:t.\colorbox{pink}{name}, \`{}\colorbox{pink}{address}\`{}:t.\colorbox{pink}{address}\} \end{tabular}\\\hline

6 & \begin{tabular}[c]{@{}l@{}} af[af[`lang'] == `en']\\  $[[$`name', `address'$]]$.head(10)\end{tabular} & \begin{tabular}[c]{@{}l@{}}\emph{\textbf{5}}\\ LIMIT \colorbox{pink}{10};\end{tabular} & \begin{tabular}[c]{@{}l@{}}\emph{\textbf{5}}\\ LIMIT \colorbox{pink}{10}; \end{tabular} & \begin{tabular}[c]{@{}l@{}} \emph{\textbf{5}}, \\
\{ ``\$project": \{ ``\_id": 0 \} \},\\
\{ ``\$limit" : \colorbox{pink}{10} \} \end{tabular} 
& \begin{tabular}[c]{@{}l@{}} \emph{\textbf{5}}\\ RETURN t \\LIMIT \colorbox{pink}{10} \end{tabular}\\\hline

\end{tabular}
}

\caption{PolyFrame's Incremental Query Formation}
\label{tab:rewrite-rules}
\end{table*}

Figure~\ref{fig:rewrite} displays examples of our language-specific rewrite rules. For these particular examples, SQL happens to share the same syntax as SQL++ for all operations except operation 1, so due to space limitations we only show SQL++, MongoDB, and Cypher examples. Operation number 2, for example, requires each language to return a value of an aggregate function of a particular attribute. There are two variables (italicized) that will be rewritten for this operation, `\$subquery' and `\$agg\_func'. A previous operation's underlying query will replace the value of the variable `\$subquery' and one of the aggregate functions (e.g., operations 3-7) will replace the variable `\$agg\_func'. As indicated in the same figure, aggregate functions also require a rewrite for a variable labeled `\$attribute'. This variable will be rewritten to the name of an attribute. For example, to get the minimum value of `age' from a dataset named `Users' in a database named `Test', PolyFrame will combine the rewrite results of operations 1, 2, and 3. First it will rewrite the variable `\$namespace' of operation 1 with `Test' and the variable `\$collection' with `Users'. The rewritten result of operation 1 will replace the variable `\$subquery of operation 2'. It will then rewrite the variable `\$attribute' in operation 3 with the value `age' and use operation 3 to replace the value of the variable `\$agg\_func' in operation 2.

\subsubsection{\textbf{Generic rules}} \label{generic} Generic rules are rewrite rules that are not explicitly defined in a system-specific PolyFrame language configuration file. These are for complex queries that are equivalent to Pandas-specific functions such as describe() and get\_dummies(). Generic rules are composed of several language-specific rules. We construct generic rules by decomposing Pandas' complex functions into a chain of basic Pandas operations which are then translated via the existing language-specific rewrite rules. For example, the function `describe()' in Pandas displays statistics for each attribute in a Dataframe. In PolyFrame, we construct this function by combining operations 1-7 in Figure~\ref{fig:rewrite} together to form a query that asks for aggregate values (min, max, average, count, and standard deviation) of specified attributes. We chain together operations 3-7 according to a pre-defined language-specific attribute rewrite rule and use them to rewrite operation 1 and 2. Thus, instead of creating a rule for each function of Pandas Dataframe, these generic rules allow PolyFrame to efficiently utilize common components to form more complex queries that perform the desired function.

% To demonstrate the generality of our approach, we eventually plan to also target other query languages and database platforms such as PartiQL (used by Amazon) and Cypher (used by Neo4j).

\subsection{PolyFrame Examples}

To demonstrate the generality of our approach, we have implemented a first prototype of the PolyFrame architecture operates against AsterixDB, PostgreSQL~\cite{postgres}, MongoDB~\cite{mongodb} and Neo4j~\cite{neo4j} by translating Dataframe operations into SQL++ for AsterixDB, nested SQL queries for PostgreSQL, MongoDB aggregation pipeline stages for MongoDB, and Cypher queries using `WITH' statements for Neo4j. Table~\ref{tab:rewrite-rules} displays query rewrites for SQL++, regular SQL, MongoDB, and Cypher that correspond to the PolyFrame operations in Figure~\ref{fig:op_chain}. The highlighted parts of each query are generated by PolyFrame's query translation process, while the non-highlighted parts come directly from the provided language-specific rewrite rules. The bold italicized numbers are operation IDs. These IDs refer to query results from the indicated operation. We can see that SQL++ has much in common with SQL, but some differences exist due to the different data models of the two languages. The MongoDB and Cypher rewrites are very different, but the passed-in operation parameters are the same across all four languages. The full finished products of operation 6 rewritten in each of the languages can be found in section~\ref{sec:language_translate} of the Appendix.

For MongoDB, PolyFrame uses its aggregation pipeline in order to obtain the incremental query formation leveraged for AFrame. As a result, certain optimizations might be limited for particular operations in a pipeline (as described in MongoDB's documentation). Operation 1 in Table~\ref{tab:rewrite-rules} for MongoDB does not have any variable rewritten because our MongoDB rewrite rules are pipeline stages and pipeline constructions are handled through its database connector. Figure~\ref{mongo_rewrite} displays a complete MongoDB aggregation pipeline for operation 6 from Table~\ref{tab:rewrite-rules}. Notice here that we include a `\{"\$project":\{"\_id":0\}\}' statement as part of the MongoDB's rewrite rule to exclude the MongoDB object identifier attribute `\_id' from the final results. We include this statement as part of the Pandas `head()' method rewrite rule because this attribute is an internal attribute of MongoDB. This attribute is the last attribute to be excluded in the pipeline because its presence in the pipeline enables index usage (if any).

\begin{figure}[h!]

\begin{lstlisting}[basicstyle={\small}]
Test.Users.aggregate([
{"$match":{}},
{"$match":{"$expr":{"$eq":["$lang","en"]}}},
{"$project":{"name": 1, "address": 1}},
{"$project":{"_id": 0}},
{"$limit" :10}
])
\end{lstlisting}
\caption{MongoDB Aggregation Pipeline Example}
\label{mongo_rewrite}
\end{figure}

\section{Experiments}
% In order to demonstrate our PolyFrame implementation and empirically validate the generality of our language-rewrite approach working against different database systems, we conducted a set of experiments using the DataFrame benchmark detailed in~\cite{arxiv}. That DataFrame benchmark was originally developed to evaluate AFrame and to compare AFrame's performance with that of other DataFrame libraries. However, we conducted this experiment here as a demonstration of our new architecture rather than to compare the performance of the different database systems. The experiment also illustrates that our implementation is able to take advantage of the different databases' optimizations (such as indexes, joins, index-only queries, etc.) to efficiently compute results in comparison to Pandas DataFrames. As such, we ran the benchmark on Pandas and on PolyFrame operating on top of AsterixDB, MongoDB, PostgreSQL, and Neo4j.

In order to demonstrate the value of database-backed dataframes and to empirically validate the generality of our language-rewrite approach working against different database systems, we have conducted two sets of experiments. One set illustrates the performance differences between a distributed data processing framework (Spark) that can consume the data from database systems and a framework (PolyFrame) that uses a database system to process the data. The other set of experiments illustrates our new architecture operating against different database systems to compute results and compare that to Pandas Dataframes. We conducted our experiments using the Dataframe benchmark detailed in~\cite{aframe}. That Dataframe benchmark was originally developed to evaluate AFrame and to compare its performance with that of other Dataframe libraries. Note that we use the benchmark here as a demonstration of our new architecture (not to compare the performance of the different database systems).Performance results for AFrame versus distributed dataframe alternatives are available in~\cite{aframe}.

\subsection{DataFrame Benchmark}
To our knowledge, there is no standard benchmark for evaluating dataframe libraries. Therefore, when we first created AFrame we also implemented our own Dataframe benchmark to evaluate AFrame's performance. We use the same benchmark here to evaluate PolyFrame. The DataFrame Benchmark used for these experiments is designed to evaluate the performance of dataframe libraries using Wisconsin benchmark data~\cite{wisconsin} of various sizes in both local and distributed environments. Our original DataFrame Benchmark issues a set of 12 analytical Pandas DataFrame expressions to evaluate whether or not a dataframe library is able to handle exploratory data manipulation operations on large volumes of data efficiently. Here we have added one more expression to the set, which involves missing data, in order to more closely mimic real-world data characteristics. An important feature of the benchmark is that it presents two separate timing comparisons. One is the total runtime, which includes both the DataFrame creation time and the expression runtime, and the other is the expression-only runtime. This is done to reflect the impact of the schema inferencing process which can be time-consuming for some DataFrame libraries~\cite{arxiv,aframe}. Also, depending on the nature of a given analysis, the DataFrame creation time can dominate the actual expression evaluation time. The benchmark timing points for Pandas and PolyFrame are listed in section~\ref{sec:timingAppendix} of the Appendix.

\subsection{Benchmark Datasets}
The DataFrame benchmark issues its expressions against a synthetically generated Wisconsin benchmark dataset. This dataset allows us to precisely control the selectivity percentages, to generate data with uniform value distributions, and to broadly represent data for general analysis use cases. A specification of the attributes in the Wisconsin benchmark's dataset is displayed in Table~\ref{fig:wisconsin_benchmark}. For our use case, we modified the Wisconsin dataset to include missing values in some of its attributes. The unique2 attribute is a declared key and is ordered sequentially, while the unique1 attribute has a dense sent of unique values that are randomly distributed.  The other numerical attributes are used to provide access to a known percentage of values in the dataset. The dataset also contains three string attributes: stringu1, stringu2, and string4. We used a JSON data generator to generate Wisconsin datasets of various sizes ranging from 1 GB (0.5 million records) to 40 GB (20 million records). 

\begin{table}[h]
\resizebox{0.45\textwidth}{0.11\textheight}{%
\begin{tabular}{|l|l|l|}
\hline
Attribute name & Attribute domain & Attribute value \\ \hline
uniquel & O..(MAX-1) & unique, random \\
unique2 & O..(MAX-1) & unique, sequential \\
two & 0..1 & uniquel mod 2 \\
four & 0..3 & uniquel mod 4 \\
ten & 0..9 & uniquel mod 10 \\
twenty & 0..19 & uniquel mod 20 \\
onePercent & 0..99 & uniquel mod 100 \\
tenPercent & 0..9 & uniquel mod 10 \\
twentyPercent & 0..4 & uniquel mod 5 \\
fiftyPercent & 0..1 & uniquel mod 2 \\
unique3 & O..(MAX-1) & uniquel \\
evenOnePercent & 0,2,4, ...,198 & onePercent*2 \\
oddOnePercent & 1,3,5, ...,199 & (onePercent *2)+ 1 \\
stringul & per template & derived from uniquel \\
stringu2 & per template & derived from unique2 \\
string4 & per template & cyclic: A, H, O, V \\ \hline
\end{tabular}%
}
\caption{Scalable Wisconsin benchmark: attributes~\cite{wisconsin}}
\label{fig:wisconsin_benchmark}
\vspace{-2em}
\end{table}

\subsection{Benchmark Expressions}
The DataFrame benchmark introduced in ~\cite{aframe} has a set of 12 analytical expressions that are aimed at evaluating the scalability of DataFrame libraries. As mentioned before, we modified the benchmark by adding another expression to cover a case with missing data which is common in modern data like social media or IoT. Table~\ref{tab:operations} displays the complete set of the updated DataFrame benchmark’s expressions (expression 13 is the added expression). PolyFrame's actual generated benchmark queries for each of the four languages (SQL++, SQL, Cypher, and MongoDB Query Language) can be found in sections~\ref{sec:sqlpp_queries},~\ref{sec:sql_queries},~\ref{sec:cypher_queries}, and~\ref{sec:mongo_queries} of the Appendix.

\lstdefinestyle{pythonNoFramestyle}
      {  language = python,
        showstringspaces=false,
        basicstyle=\ttfamily,
        keywordstyle=\color{blue},
        otherkeywords = {min,max,groupby,head,sort_values,map,agg,len,merge,read_json,read.json},
        morekeywords = [2]{min},
        morekeywords = [3]{groupby},
        morekeywords = [4]{head}
        % frame=single
    }
\begin{table}[h]

\begin{adjustbox}{width=0.49\textwidth}
\begin{tabular}{|l|l|l|}
\hline
\textbf{ID} & \textbf{Operation}  & \textbf{DataFrame Expression} \\ \hline
1 & Total Count & \begin{lstlisting}[style=pythonNoFramestyle]
len(df)
\end{lstlisting} \\ \hline
2 & Project   &  \begin{lstlisting}[style=pythonNoFramestyle]
df[[`two',`four']].head()\end{lstlisting} \\ \hline
3 & Filter \& Count  &  \begin{lstlisting}[style=pythonNoFramestyle]
len(df[(df[`ten'] == x) 
    & (df[`twentyPercent'] == y) 
    & (df[`two'] == z)])
\end{lstlisting}\\ \hline
4 & Group By   &  \begin{lstlisting}[style=pythonNoFramestyle]
df.groupby(`oddOnePercent').agg(`count')
\end{lstlisting} \\ \hline
5 & Map Function  &  \begin{lstlisting}[style=pythonNoFramestyle]
df[`stringu1'].map(str.upper).head()
\end{lstlisting} \\ \hline
6 & Max     &  \begin{lstlisting}[style=pythonNoFramestyle]
df[`unique1'].max()
\end{lstlisting} \\ \hline
7 & Min &  \begin{lstlisting}[style=pythonNoFramestyle]
df[`unique1'].min()
\end{lstlisting} \\ \hline
8 & Group By \& Max       &  \begin{lstlisting}[style=pythonNoFramestyle]
df.groupby(`twenty')[`four'].agg(`max')
\end{lstlisting} \\ \hline
9 & Sort    &  \begin{lstlisting}[style=pythonNoFramestyle]
df.sort_values(`unique1',ascending=False).head()
\end{lstlisting} \\ \hline
10& Selection     &  \begin{lstlisting}[style=pythonNoFramestyle]
df[(df[`ten'] == x)].head()
\end{lstlisting} \\ \hline
11& Range Selection      &  \begin{lstlisting}[style=pythonNoFramestyle]
len(df[(df[`onePercent'] >= x) 
    & (df[`onePercent'] <= y)])
\end{lstlisting} \\ \hline
12& Join \& Count    &  \begin{lstlisting}[style=pythonNoFramestyle]
len(pd.merge(df, df2, 
            left_on=`unique1',
            right_on=`unique1',
            how=`inner',hint=`index'))
\end{lstlisting} \\ \hline
13 & Count Missing Value & \begin{lstlisting}[style=pythonNoFramestyle]
len(df[df[`tenPercent'].isna()])
\end{lstlisting} \\ \hline

\end{tabular}
\end{adjustbox}
\caption{Dataframe Benchmark Operations (df, df2 = DataFrame objects, x,y,z = variables representing random values within an attribute's range)}
\label{tab:operations}
\vspace{-1em}
\end{table}

% \begin{figure*}[h!]
%      \centering
%   \begin{subfigure}[t]{0.75\textwidth}
%         \includegraphics[trim=0.5cm 1.5 0 2,width=\textwidth,height=0.5cm]{figures/legend.pdf}%
%     \end{subfigure}

%     \begin{subfigure}[t]{0.45\textwidth}
%         \includegraphics[trim=1.5 1.5 0cm 1.5,width=\textwidth,height=4cm]{figures/EXP1-6.pdf}
%         \caption{Expression 1-6 total times}
%         \label{fig:1-6total}
%     \end{subfigure}
%   \hspace{0.2cm}
%     \begin{subfigure}[t]{0.47\textwidth}
%         \includegraphics[trim=0.5cm 1.5 0.3cm 1.5,width=\textwidth,height=4cm]{figures/EXP7-13.pdf}
%         \caption{Expression 7-13 total times}
%         \label{fig:7-13total}
%     \end{subfigure}
%     \hfill

%     \begin{subfigure}[t]{0.43\textwidth}
%         \includegraphics[trim=0.9cm 1.5 0.5cm 1.5,width=\textwidth,height=4cm]{figures/EXP1-6_WO.pdf}
%         \caption{Expression 1-6 expression-only times}
%         \label{fig:1-6task}
%     \end{subfigure}
%     \hspace{0.2cm}
%     \begin{subfigure}[t]{0.47\textwidth}
%         \includegraphics[trim=0cm 1.5 0.5cm 1.5,width=\textwidth,height=4cm]{figures/EXP7-13_WO.pdf}
%         \caption{Expression 7-13 expression-only times}
%         \label{fig:7-13task}
%     \end{subfigure}
%     \caption{XS Results of Single Node Evaluation (*=value where the bar ends)}
%     \vspace{-1em}
%     \label{fig:single_node_xs}
% \end{figure*}

\subsection{Experimental Setup}
In order to present a reproducible evaluation environment, we set up a benchmark cluster using Amazon EC2 m4.large machines. Each machine has 8 GB of memory, 100 GB of SSD, and the Ubuntu Linux operating system.

We used the Wisconsin benchmark data generator to generate the data in JSON file format in 5 different sizes ranging from 1GB (0.5 million records) to 10 GB (5 million records). We labeled these as the XS through XL datasets as shown in Table~\ref{tab:onenode}.

\begin{table}[h]
\resizebox{0.48\textwidth}{!}{%
\begin{tabular}{l|l|l|l|l|l|}
\cline{2-6}
 & \multicolumn{5}{c|}{\textbf{Dataset Name}} \\ \cline{2-6} 
 & \multicolumn{1}{c|}{\textbf{XS}} & \multicolumn{1}{c|}{\textbf{S}} & \multicolumn{1}{c|}{\textbf{M}} & \multicolumn{1}{c|}{\textbf{L}} & \multicolumn{1}{c|}{\textbf{XL}} \\ \hline
\multicolumn{1}{|l|}{Number of Records} & 0.5 mil  & 1.25 mil  & 2.5 mil  & 3.75 mil & 5 mil \\ \hline
\multicolumn{1}{|l|}{JSON File Size} & 1 GB  & 2.5 GB  & 5 GB  & 7.5 GB & 10 GB \\ \hline
\end{tabular}%
}
\caption{Single Node's Dataset Summary (mil = million)}
\label{tab:onenode}
\vspace{-1em}
\end{table}

\subsubsection{\textbf{Comparison with Spark (Single Node)}} These experiments are included for the benefit of readers who may wonder why Spark plus its database connectivity are not the ultimate scaling answer. We conducted these Spark experiments on a single node using two different data access methods for PySpark dataframes reading from a MongoDB instance. We used the MongoDB Spark connector provided by MongoDB to read the data and create a Spark dataframe object. For the first data access method (labeled `Spark’), we used the connector to connect to the database and directly read the data from it. For the second data access method (labeled `Spark+MongoDB pipeline’), we provided the connector with a MongoDB aggregation pipeline for each query as a part of each operation's dataframe creation process. This is an optimization that Spark supports to push down a database query and thus utilize database optimizations in order to lower the amount of data transferred from a database. The pipelines that we issued to Spark are the same ones that PolyFrame automatically generated, and both Spark and PolyFrame were connected to the same MongoDB instance with its storage on the same machine.

\subsubsection{\textbf{PolyFrame Heterogeneity (Single and Multi-Node)}}
To assess the benefits and feasibility of PolyFrame, we ran the DataFrame benchmark on Pandas and on PolyFrame operating on the four different database systems detailed below:
\begin{itemize}
    \item AsterixDB: v.0.9.5 with data compression enabled
    \item PostgreSQL: v.12
    \item MongoDB: v.4.2 Community edition
    \item Neo4j: v.3.5.14 Community edition
\end{itemize}

For the multi-node benchmark, we ran the benchmark only on PolyFrame operating on top of AsterixDB, MongoDB, and Greenplum (esentially parallel PostgreSQL). The community version of Neo4j does not run on sharded clusters. Since Greenplum uses PostgreSQL v.9.5 (which is different from the PostgreSQL version that we used for the single node evaluation), some of PostgreSQL's latest optimizations were not available to it. As a result, we also ran Greenplum on a single node before conducting its multi-node experiment. The evaluated cluster sizes ranged from 2-4 nodes. 

Speedup and scaleup are the two preferred and widely used metrics to evaluate the processing performance of distributed systems, so we evaluated PolyFrame on each multi-node system using these two metrics. Table~\ref{tab:cluster} displays the dataset sizes that we used for each of the cluster experiment evaluations. The aggregate memory listed is the sum of all of the available memory in the cluster. We conducted speedup experiments using a fixed-size dataset while increasing the number of processing machines from one up to four. For the scaleup experiments, we increased both the number of processing machines and the amount of data proportionally to measure each system's performance.

\begin{table}[h]
\resizebox{0.48\textwidth}{!}{%
\begin{tabular}{l|l|l|l|l|}
\cline{2-5}
 & \textbf{1 node} & \textbf{2 nodes} & \textbf{3 nodes} & \textbf{4 nodes} \\ \hline
\multicolumn{1}{|l|}{\textbf{Aggregate Memory}} & 8 GB & 16 GB & 24 GB & 32 GB \\ \hline
\multicolumn{1}{|l|}{\textbf{Speedup: JSON File Size}} & 10 GB & 10 GB & 10GB & 10 GB \\ \hline
\multicolumn{1}{|l|}{\textbf{Scaleup: JSON File Size}} & 10 GB & 20 GB & 30GB & 40 GB \\ \hline
\end{tabular}%
}
\caption{Multi-Node Experiment Setup}
\label{tab:cluster}
\end{table}

\subsection{Experimental Results}
Here we present benchmark results for both the PolyFrame comparison with Spark and PolyFrame on different database systems.

\subsubsection{\textbf{Comparison with Spark (Single Node)}}

In Big Data analytics, Apache Spark is currently one of the most popular frameworks. Although Spark can connect to various database systems to consume data, it cannot fully utilize the optimizations available from these database systems. Spark is optimized for in-memory data processing, and it requires its input data to be loaded from the database system into its environment in order to then efficiently operate on it. However, modern database systems already provide advanced query optimization, efficient storage layout and indexes, and effective data retrieval techniques that are important when querying large volumes of data. 

% In order to demonstrate values of our database-backed dataframe implementation, we conducted a set of experiments to compare PolyFrame and Spark operating on the same MongoDB instance.

\begin{figure}[h!]
\centering
  \includegraphics[width=0.48 \textwidth]{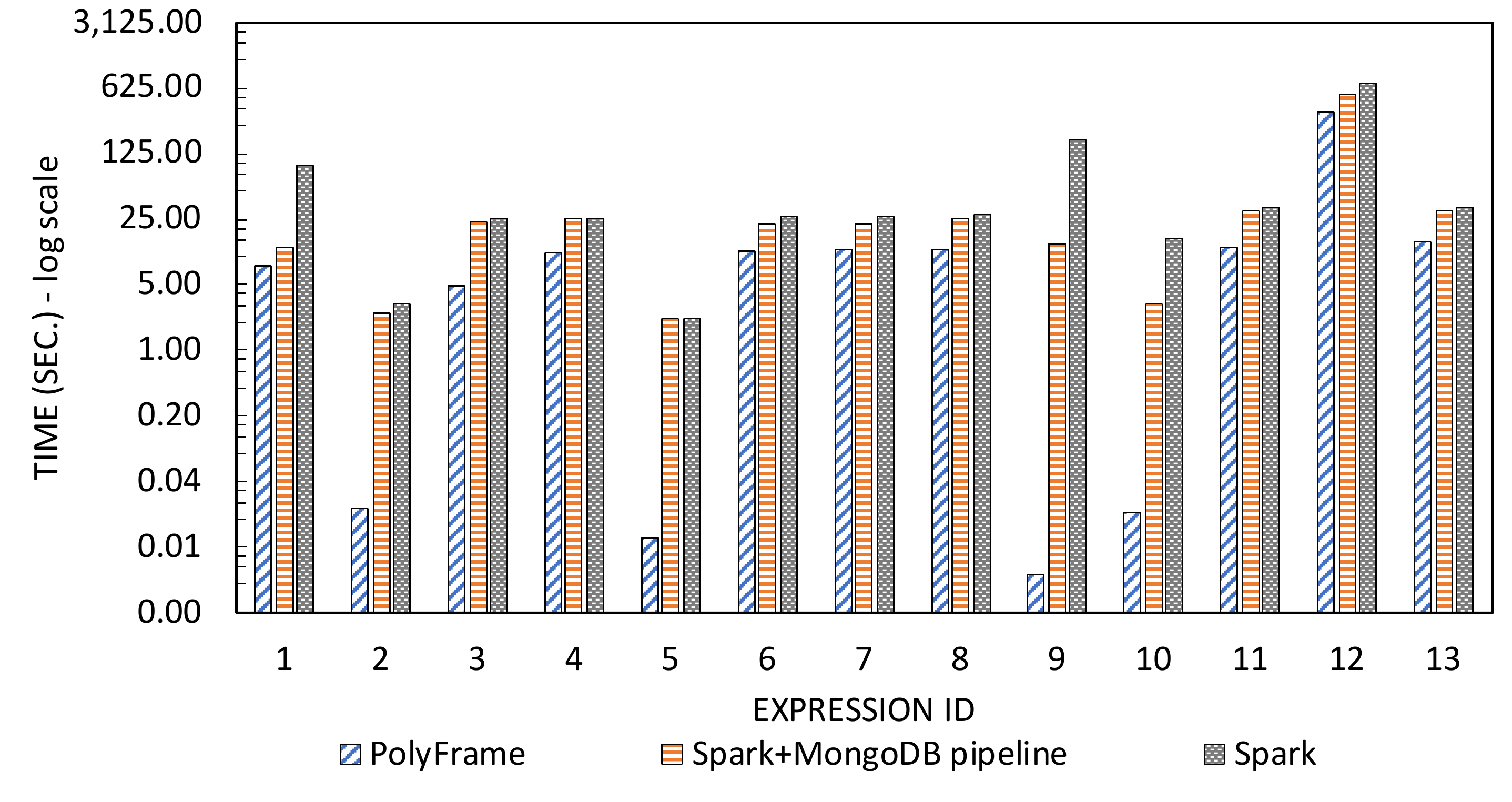}
%   \vspace{-1.5em}
  \caption{Performance comparison with Spark}
%   \vspace{-1em}
  \label{fig:spark}
\end{figure}

For this experiment, we ran the benchmark on all the dataset sizes, we will first present the results for the XL dataset (which exceeds a single node memory capacity) and then describe a few of the operations' results in detail. Figure~\ref{fig:spark} shows the results for PolyFrame and Spark for all of the dataframe benchmark's expressions. It is important to note that the plot is in log scale, which understates the significant differences in runtimes. PolyFrame performed the best across all of the expressions (lower is better). It was faster than both variants of Spark. 

\begin{figure*}[h!]
\centering
    \begin{subfigure}[t]{0.48\textwidth}
        \includegraphics[trim=0.3cm 1.5 0cm 1.5,width=\textwidth,height=4.5cm]{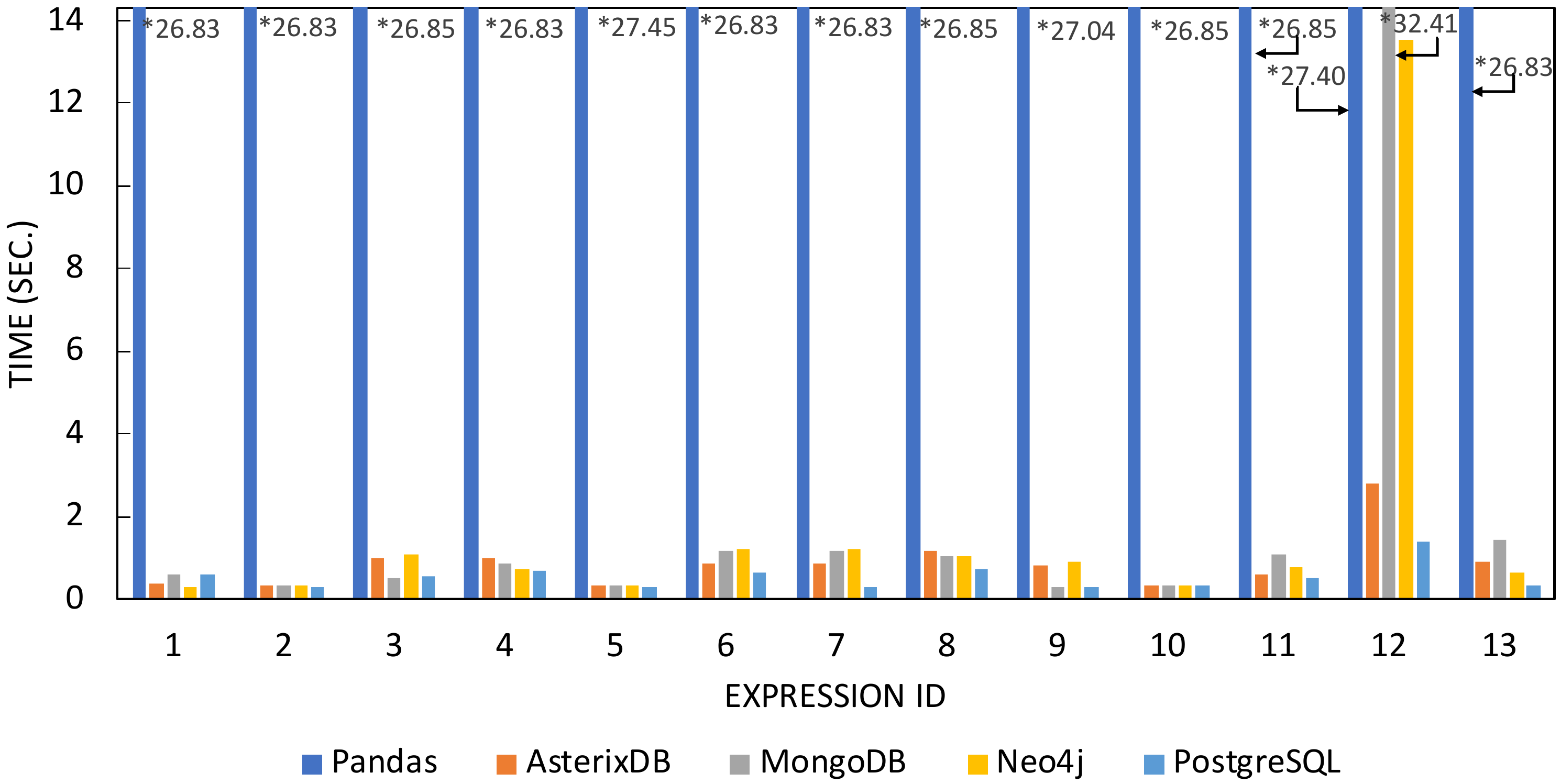}
        \caption{Expression 1-13 total times}
        \vspace{-1em}
        \label{fig:1-13total}
    \end{subfigure}
    \begin{subfigure}[t]{0.5\textwidth}
        \includegraphics[trim=0.3cm 1.5 0cm 1.5,width=\textwidth,height=4.5cm]{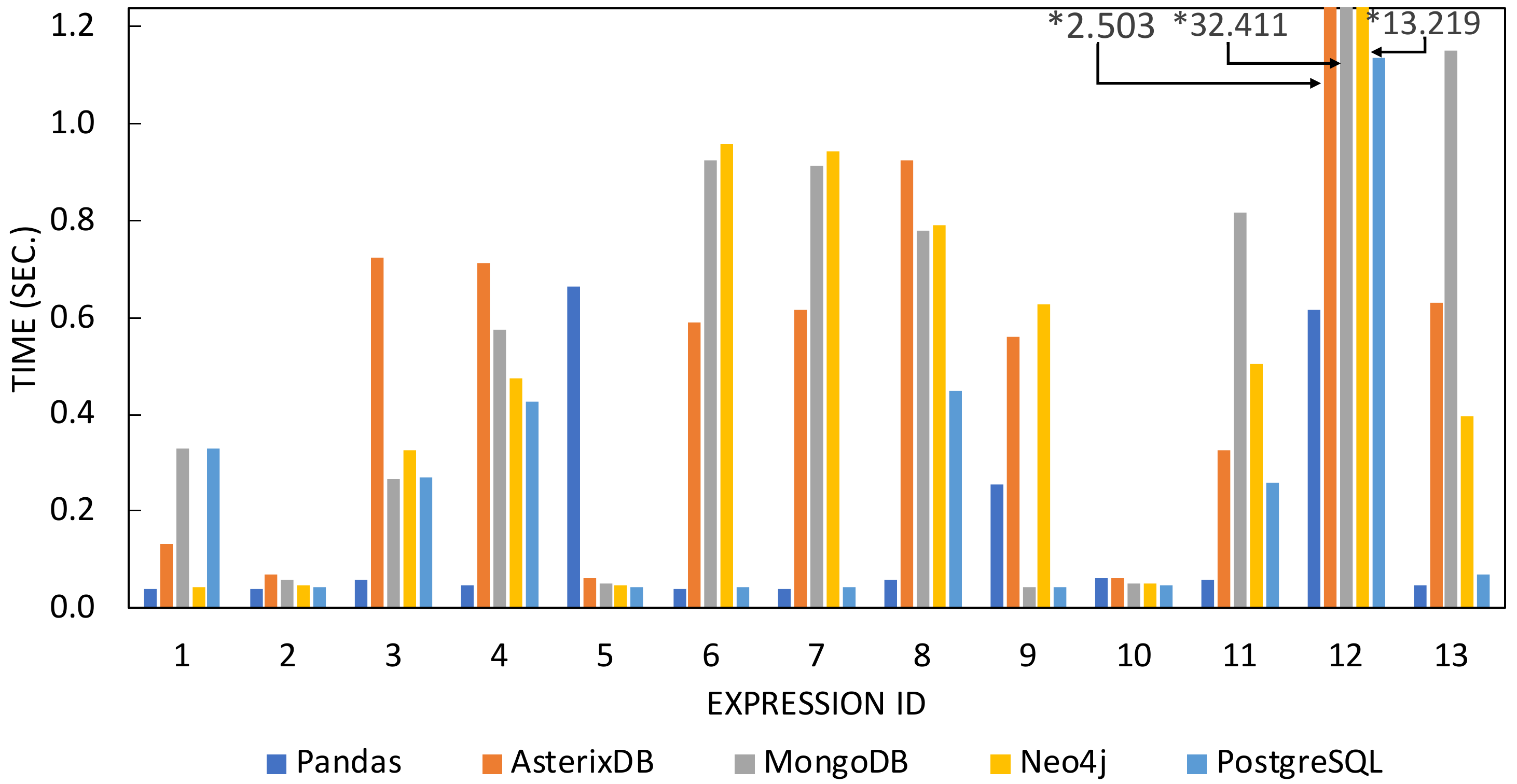}
        \caption{Expression 1-13 expression-only times}
        % \vspace{-1em}
        \label{fig:1-13task}
    \end{subfigure}
    \caption{XS Results of Single Node Evaluation (*=value where the bar ends)}
    \label{fig:single_node_xs}
    % \vspace{-1em}
\end{figure*}

\begin{figure}[h!]
\centering
    \begin{subfigure}[t]{0.45\textwidth}
        \includegraphics[trim=0.3cm 1.5 0cm 1.5,width=\textwidth,height=0.4cm]{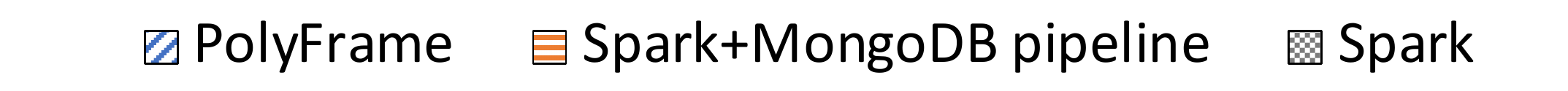}
    \end{subfigure}
    
    \begin{subfigure}[t]{0.235\textwidth}
        \includegraphics[trim=0.5cm 1.5 0.2cm 1.5,width=\textwidth,height=3.5cm]{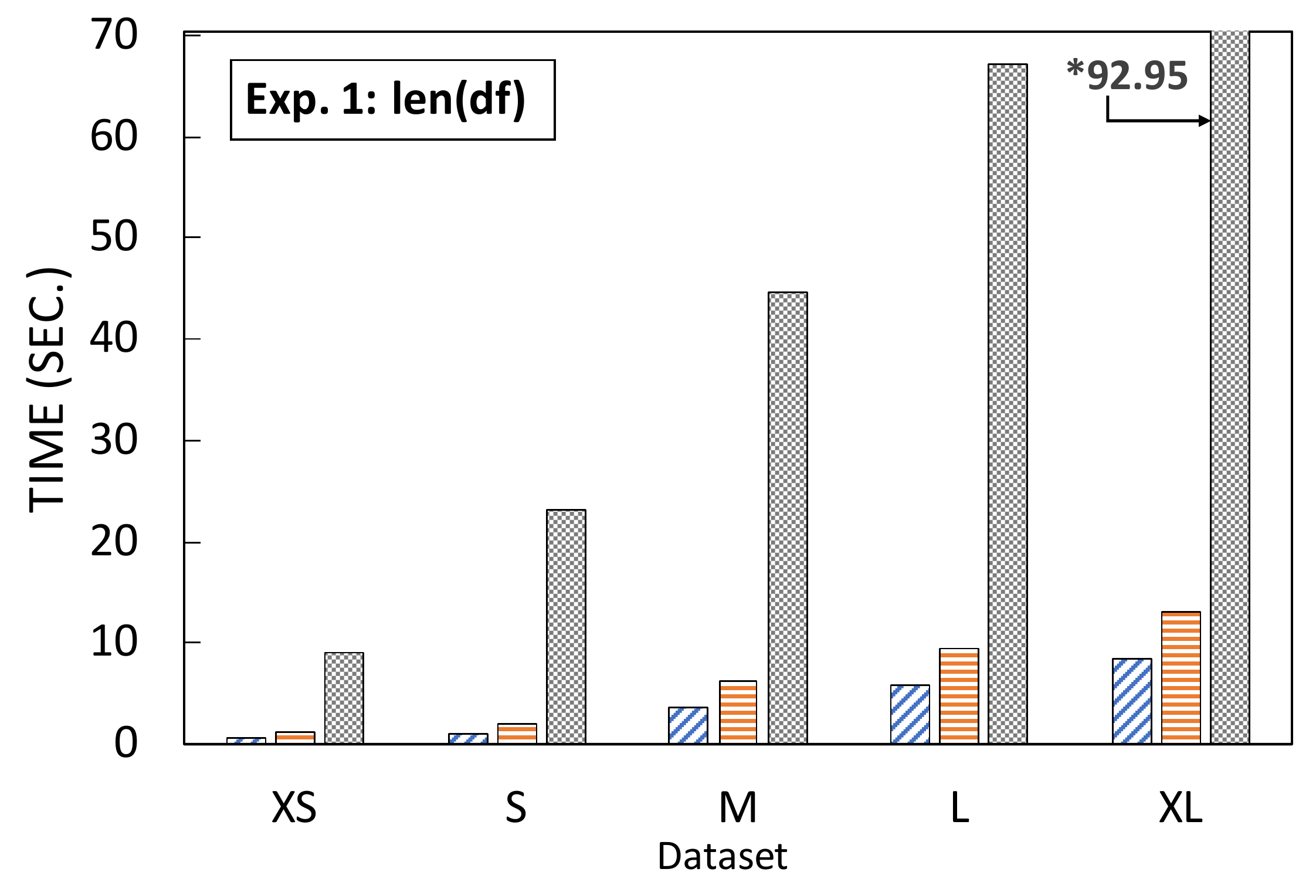}
        \caption{Expression 1}
        \label{fig:spark1}
    \end{subfigure}
    \begin{subfigure}[t]{0.23\textwidth}
        \includegraphics[trim=0.3cm 1.5 1cm 1.5,width=\textwidth,height=3.5cm]{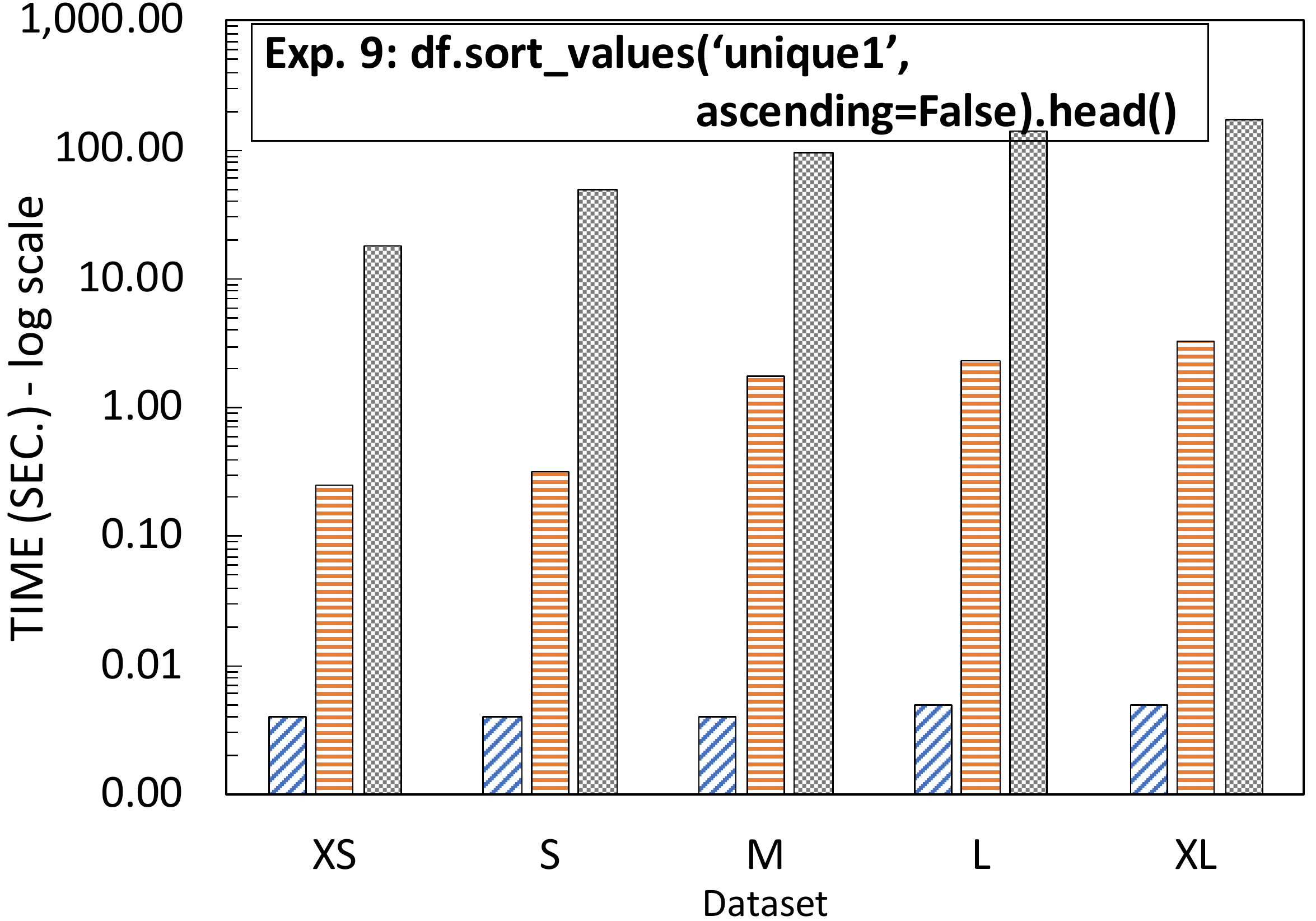}
        \caption{Expression 9}
        \label{fig:spark9}
    \end{subfigure}
    % \vspace{-1em}
    \caption{Selected single node Spark and PolyFrame comparisons (*=value where the bar ends)}
    % \vspace{-1em}
  \label{fig:spark1-9}
\end{figure}

Spark was significantly slower than PolyFrame in all of the cases, even when operating on the same MongoDB instance, partly due to the data transfer time between MongoDB and Spark. PolyFrame sends queries to MongoDB directly, without first loading any data into memory for processing. This lets MongoDB process the operations and only return the queries' results. This design allows PolyFrame to take advantage of MongoDB's database optimizations and to avoid loading large amounts of data into memory.

Spark with MongoDB pipelines had better performance than regular Spark because it limits the amount of data needed to be transferred from the database into the Spark environment for processing. However, one can see that even with  passed-down pipelines, Spark was still slower than PolyFrame. This is because the MongoDB pipelines that are passed through the connector are applied to each data partition, and not to the whole dataset. The number of data partitions is determined by MongoDB's partitioner in order to optimize data transfers to multiple Spark workers\footnote{Spark spawns a worker per core on a single machine, so MongoDB's connector defaults to partitioning the data in order to achieve maximum parallelism with Spark.}. Post-processing is then done at the Spark level. We will describe some of the test cases and their results in more detail next.

%  \vspace{-1em}

Figure~\ref{fig:spark1} displays the performance of expression 1 on all dataset sizes. This expression asks for the count of records from the dataset. Spark was significantly slower than PolyFrame here because it has to first load the data input into memory to perform the operation, and the data size was larger than the available memory. When a pipeline is passed from Spark to MongoDB, its performance was an order of magnitude faster. This is because MongoDB then performed count operations and only sent the numbers of records from each of its data partitions back to Spark to be aggregated. PolyFrame performed the best because it avoids transferring any data and having to perform any aggregation outside the database.

Figure~\ref{fig:spark9} displays the performance result for expression 9, which asks for the five records with the highest values in a unique field. Because of this operation, we created an index on the sort attribute at the database level. MongoDB was able to then use this index to perform a backward index scan to retrieve the requested records. PolyFrame’s generated MongoDB pipeline allows for such an optimization, which resulted in the best performance across all data sizes. However, we can see that Spark even with the MongoDB pipeline was still significantly slower than PolyFrame issuing the equivalent query. This is because, even when using the index, the sorted records from all partitions have to be globally sorted again in the Spark environment. However, the Spark+pipeline results were orders of magnitude faster than when Spark read the entire data from MongoDB and performed the sort operation itself. 

%  \vspace{-1em}

\subsubsection{\textbf{PolyFrame Heterogeneity (single node results)}}

An important point to note here is that we began with a single node evaluation primarily to compare PolyFrame's database system-based lazy evaluation with Pandas' eager in-memory evaluation approach. First, we executed the benchmark on the XS dataset as a preliminary test before running it on the other bigger datasets. As mentioned, the DataFrame benchmark separately presents the DataFrame creation time and the expression-only runtime. Figure~\ref{fig:single_node_xs} presents the XS results for the single node evaluation.~\ref{fig:1-13total} displays the total runtimes for expressions 1-13, and~\ref{fig:1-13task} displays the expression-only runtimes for the expressions. The total evaluation times of Pandas were significantly higher than all variants of PolyFrame because Pandas has to load the entire dataset into memory to create its DataFrames. For the expression-only times, Pandas was then the fastest to complete most of the operations due to having the data already available in memory, except for expressions 5 and 10 where Pandas suffered due to eagerly evaluating sub-components of the expressions. In contrast, PolyFrame operating on top of the four database systems did not incur any DataFrame creation times. The four PolyFrame variants were all able to execute all benchmark expressions. The runtime results among these four database systems vary due to their each having different optimizations (more on that shortly).

\begin{figure*}[h!]
     \centering
    \begin{subfigure}[t]{0.75\textwidth}
        \includegraphics[trim=1.5 1.5 0cm 1.5,width=\textwidth,height=0.5cm]{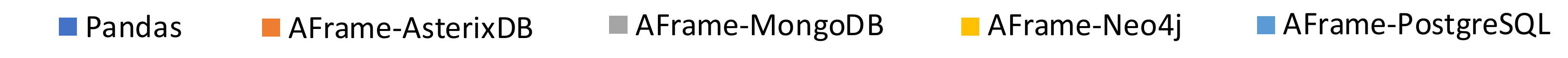}
    \end{subfigure}
    % \hspace{15cm}
    \begin{subfigure}[t]{0.41\textwidth}
        \includegraphics[trim=1.5 1.5 0cm 1.5,width=\textwidth,height=3.9cm]{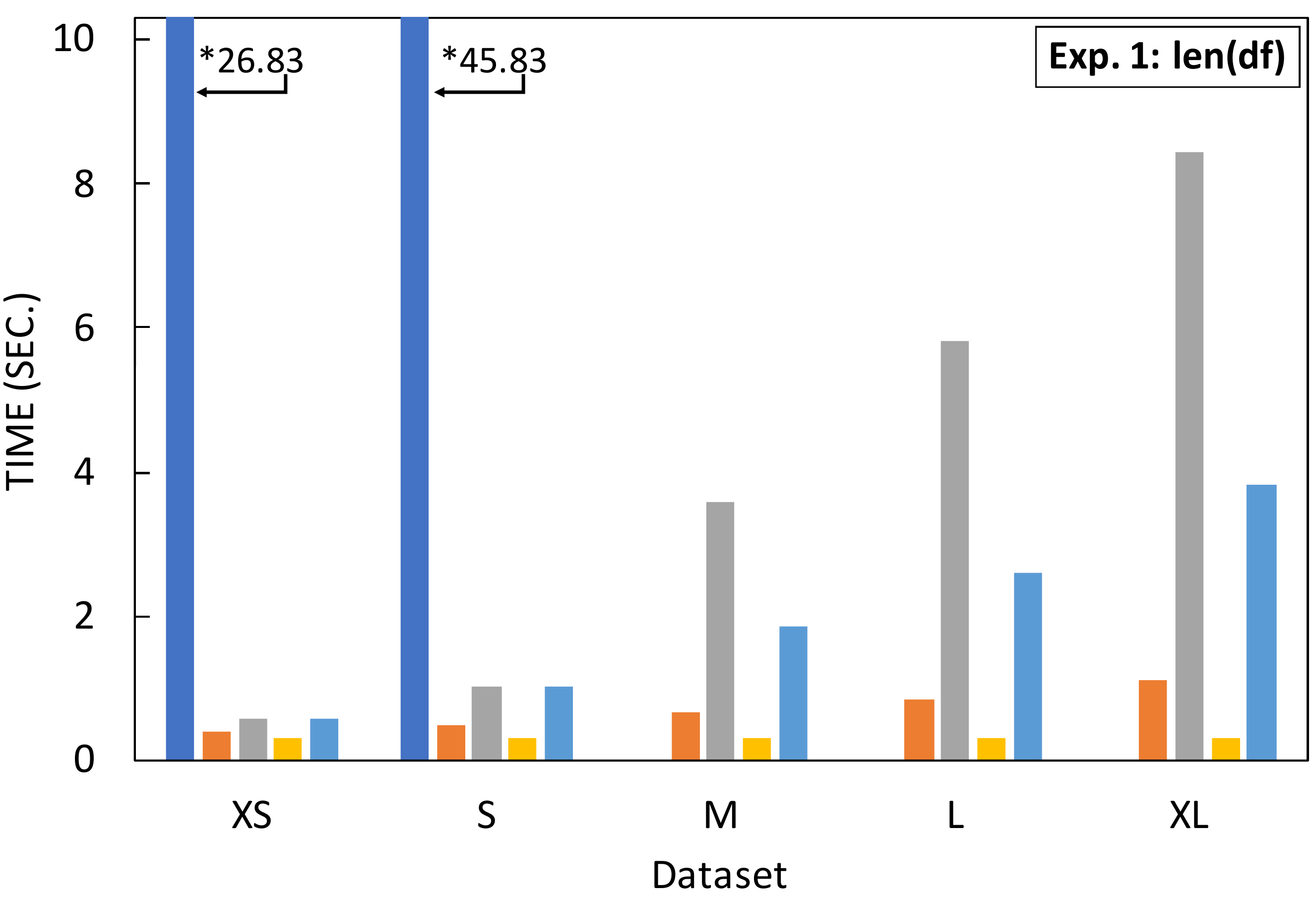}
        \caption{Expression 1: total times}
        \label{fig:q1_single}
    \end{subfigure}
    \begin{subfigure}[t]{0.41\textwidth}
        \includegraphics[trim=1.5 1.5 0cm 1.5,width=\textwidth,height=3.9cm]{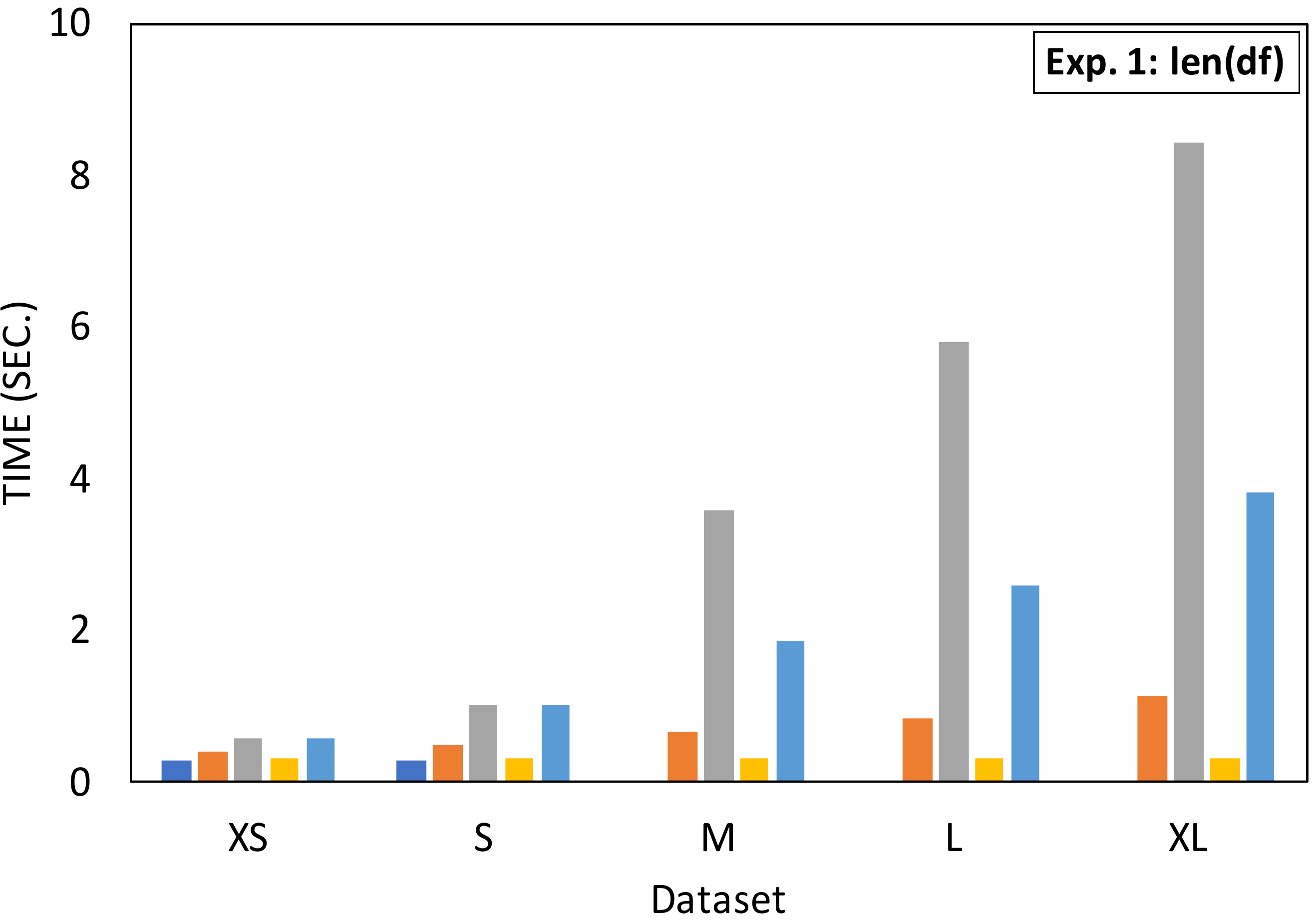}
        \caption{Expression 1: expression-only times}
        \label{fig:q1_single_wo}
    \end{subfigure}
    
    \begin{subfigure}[t]{0.41\textwidth}
        \includegraphics[trim=1.5 1.5 0cm 1.5,width=\textwidth,height=3.9cm]{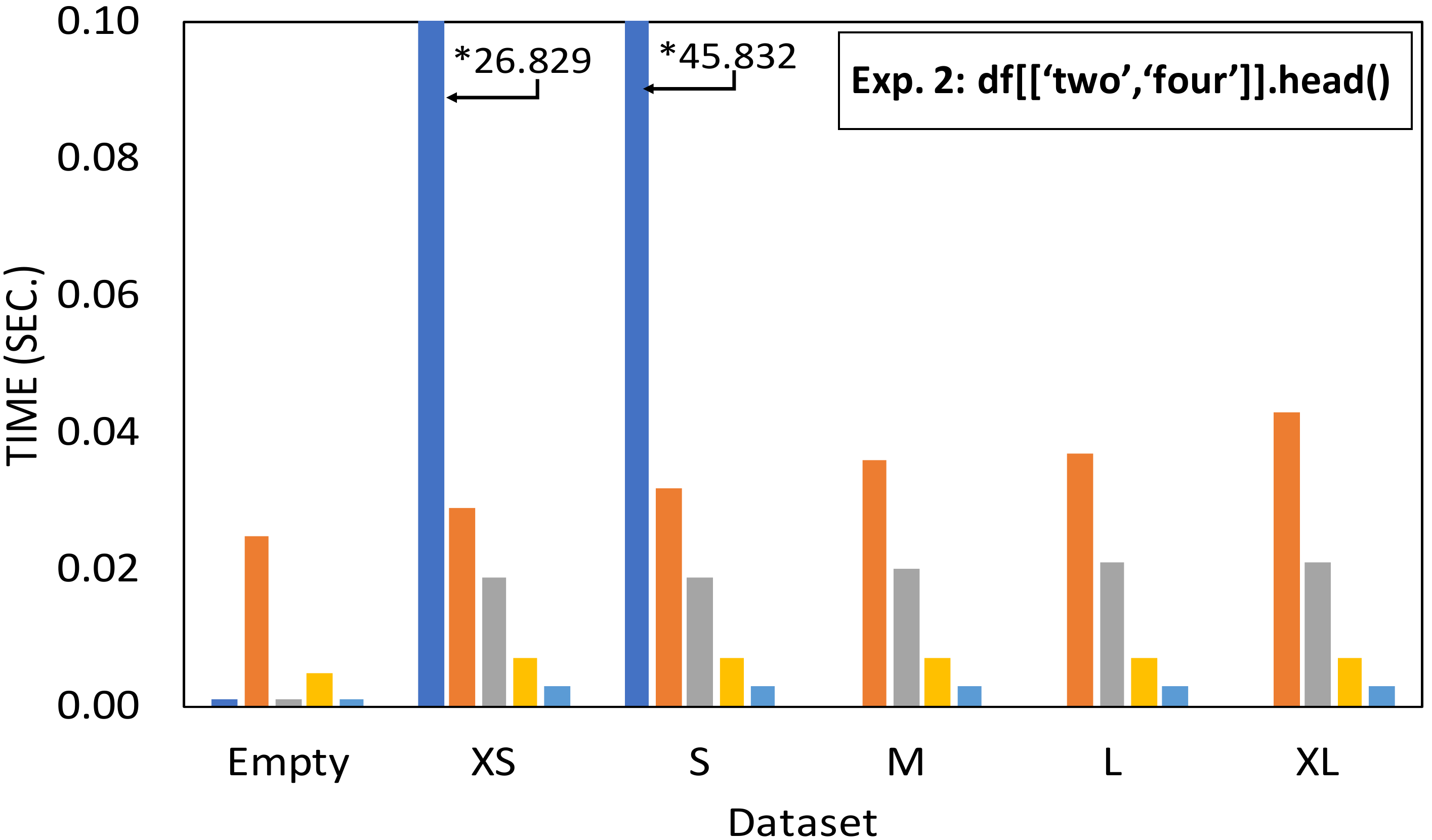}
        \caption{Expression 2: total times}
        \label{fig:q2_single}
    \end{subfigure}
    \begin{subfigure}[t]{0.41\textwidth}
        \includegraphics[trim=1.5 1.5 0cm 1.5,width=\textwidth,height=3.9cm]{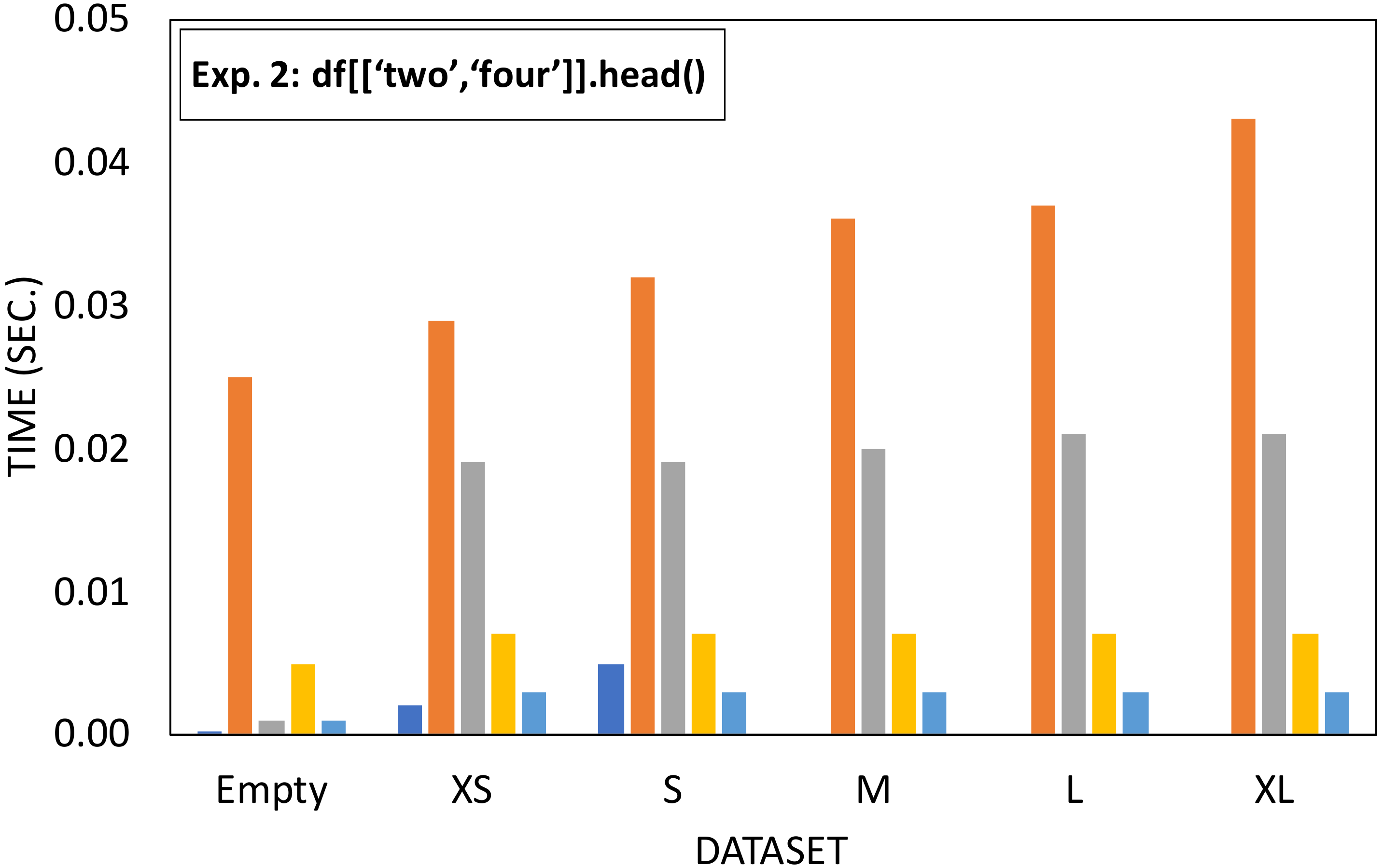}
        \caption{Expression 2: expression-only times}
        \label{fig:q2_single_wo}
    \end{subfigure}
    
    \begin{subfigure}[t]{0.41\textwidth}
        \includegraphics[trim=1.5 1.5 0cm 1.5,width=\textwidth,height=3.9cm]{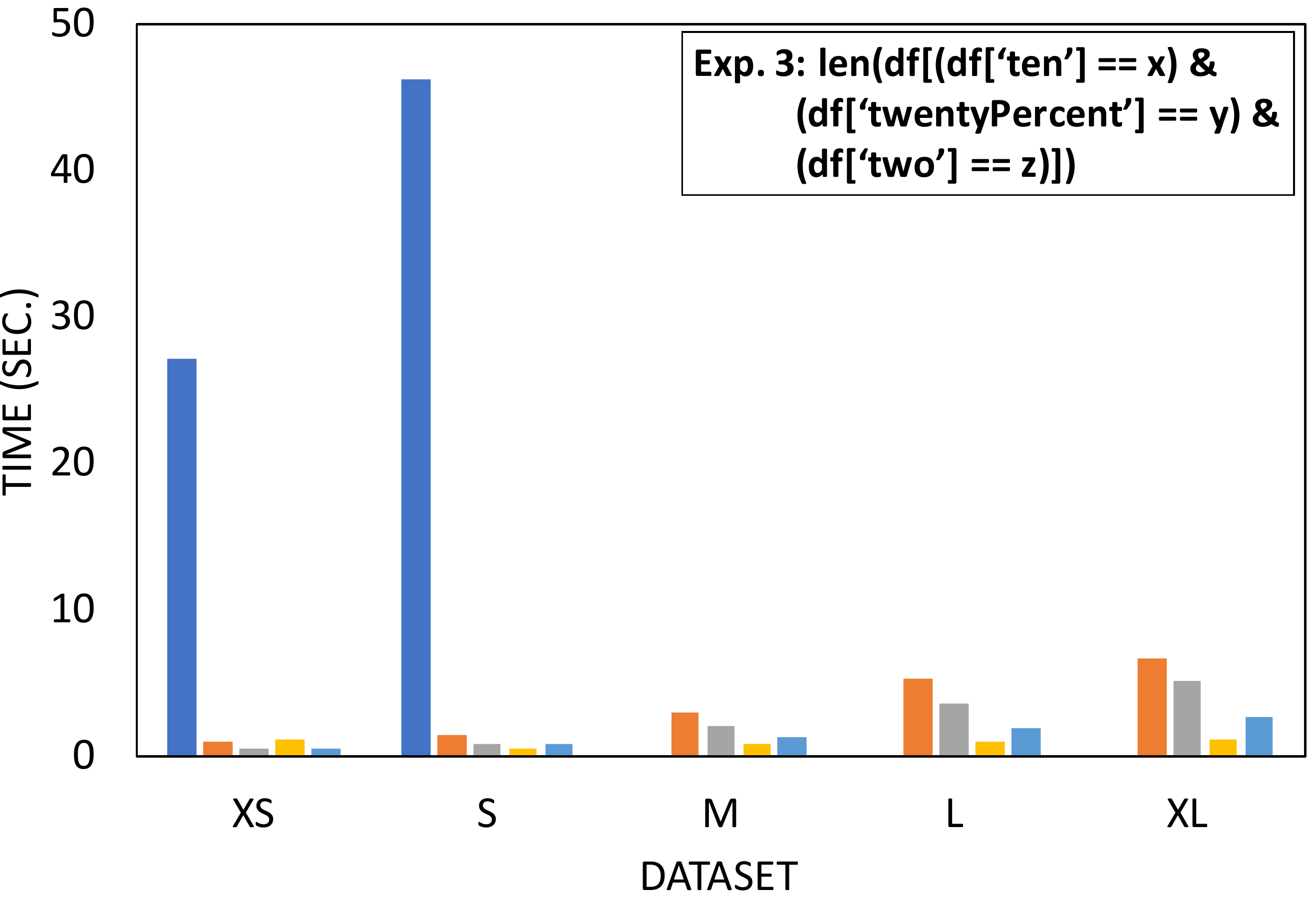}
        \caption{Expression 3: total times}
        \label{fig:q3_single}
    \end{subfigure}
    \begin{subfigure}[t]{0.41\textwidth}
        \includegraphics[trim=1.5 1.5 0cm 1.5,width=\textwidth,height=3.9cm]{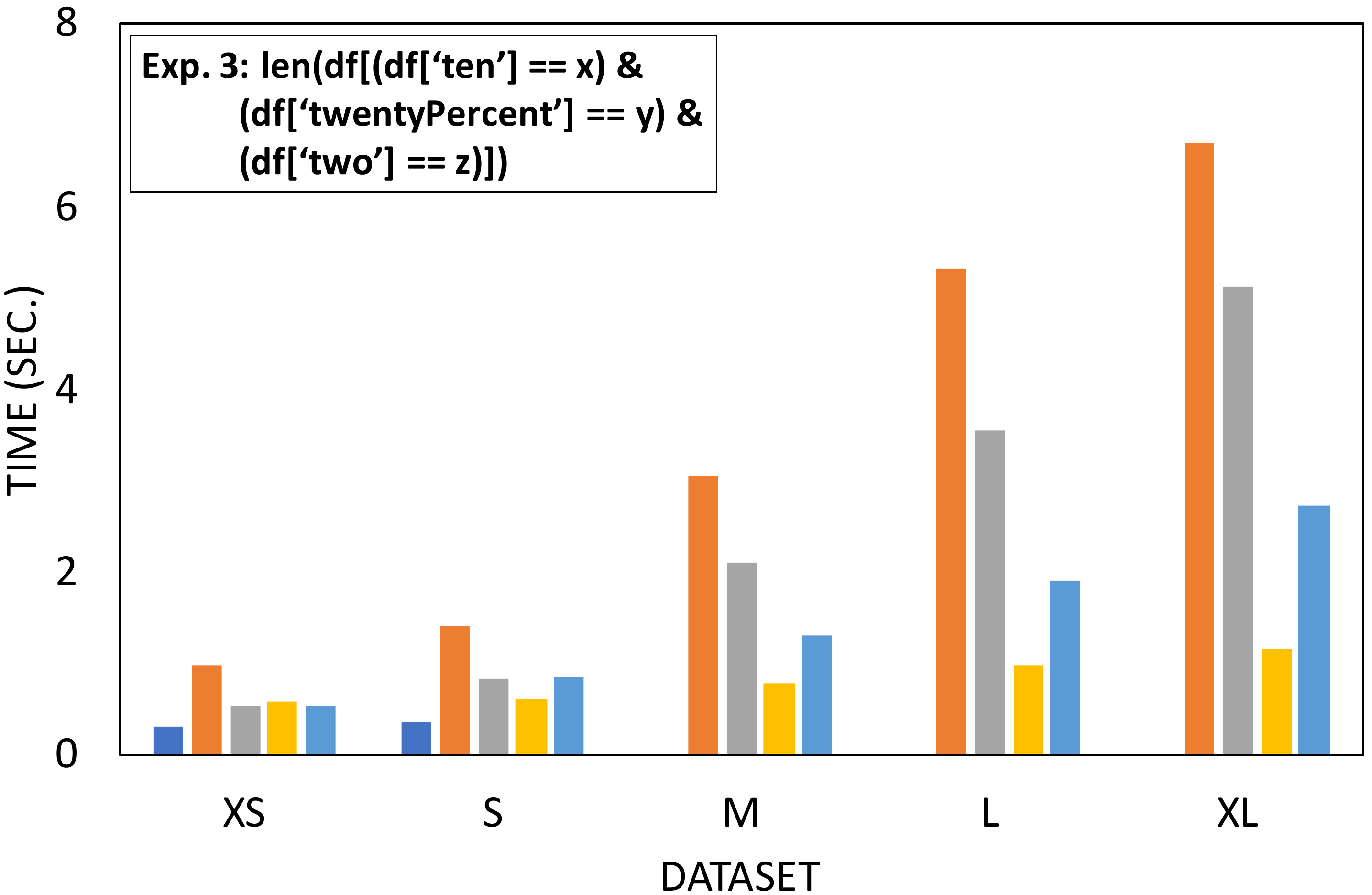}
        \caption{Expression 3: expression-only times}
        \label{fig:q3_single_wo}
    \end{subfigure}

    \begin{subfigure}[t]{0.41\textwidth}
        \includegraphics[trim=1.5 1.5 0cm 1.5,width=\textwidth,height=3.9cm]{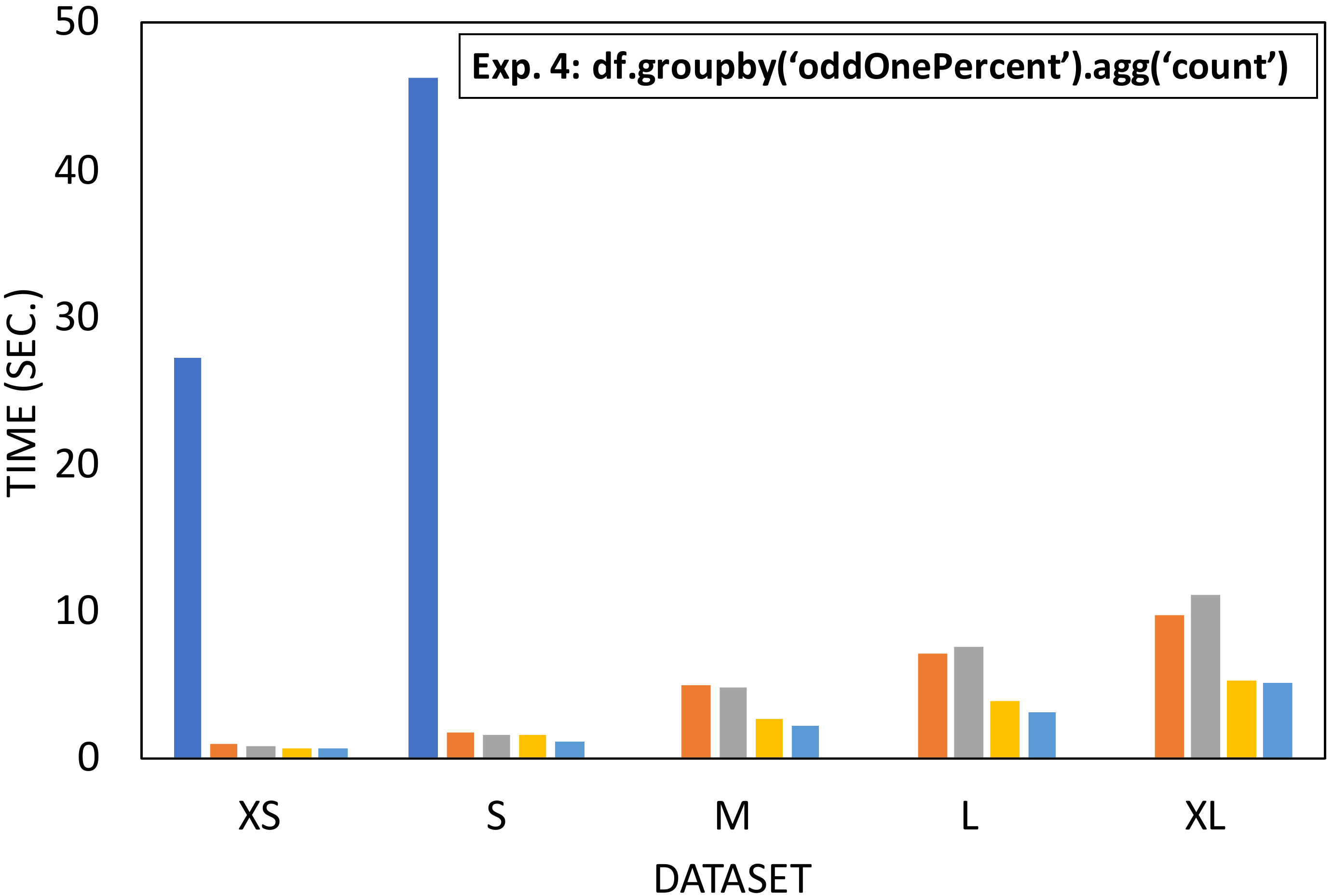}%
        \caption{Expression 4: total times}
        \label{fig:q4_single}
    \end{subfigure}
    \begin{subfigure}[t]{0.41\textwidth}
        \includegraphics[trim=1.5 1.5 0cm 1.5,width=\textwidth,height=3.9cm]{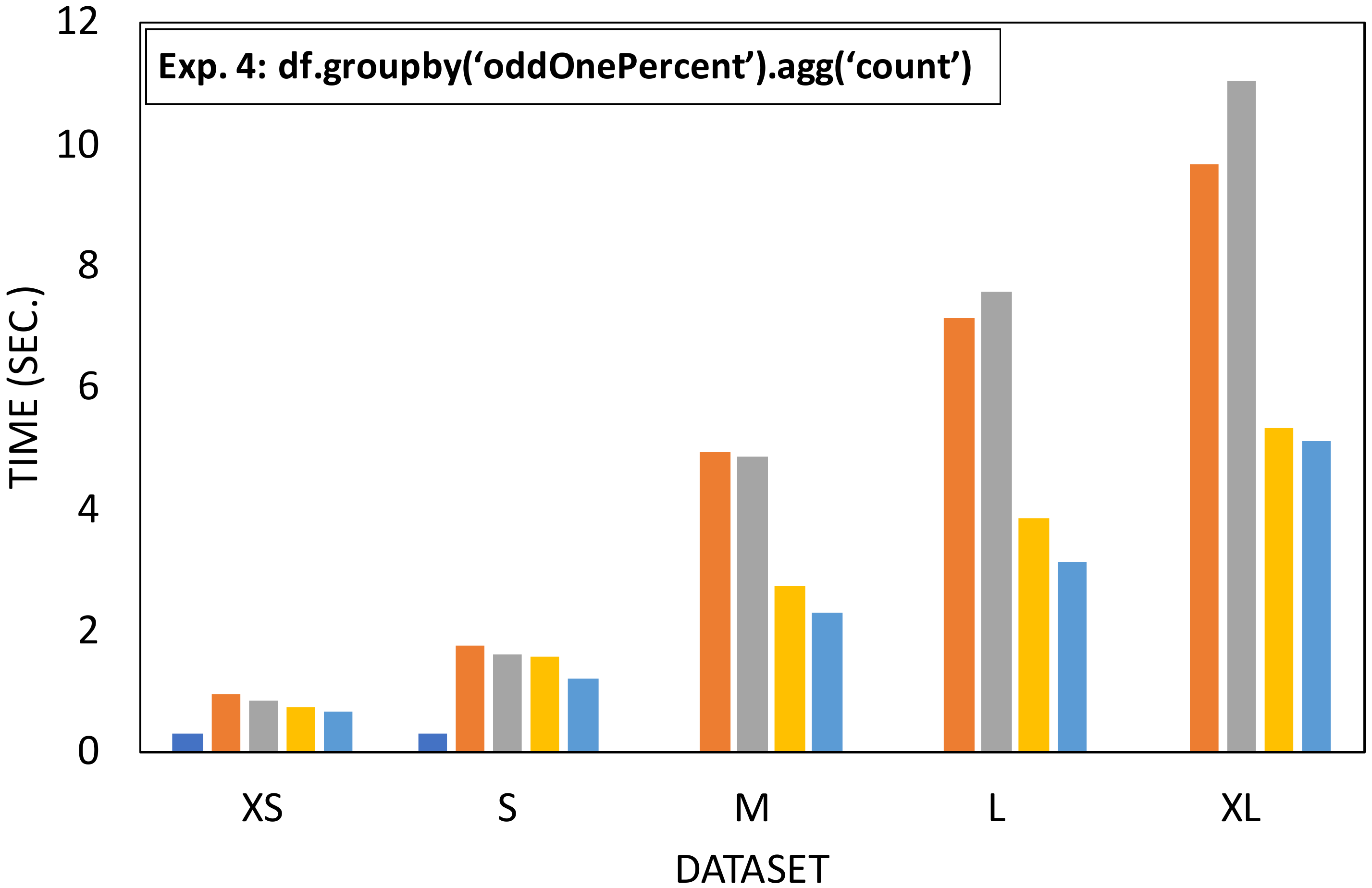}%
        \caption{Expression 4: expression-only times}
        \label{fig:q4_single_wo}
    \end{subfigure}

    \begin{subfigure}[t]{0.41\textwidth}
        \includegraphics[trim=1.5 1.5 0cm 1.5,width=\textwidth,height=3.9cm]{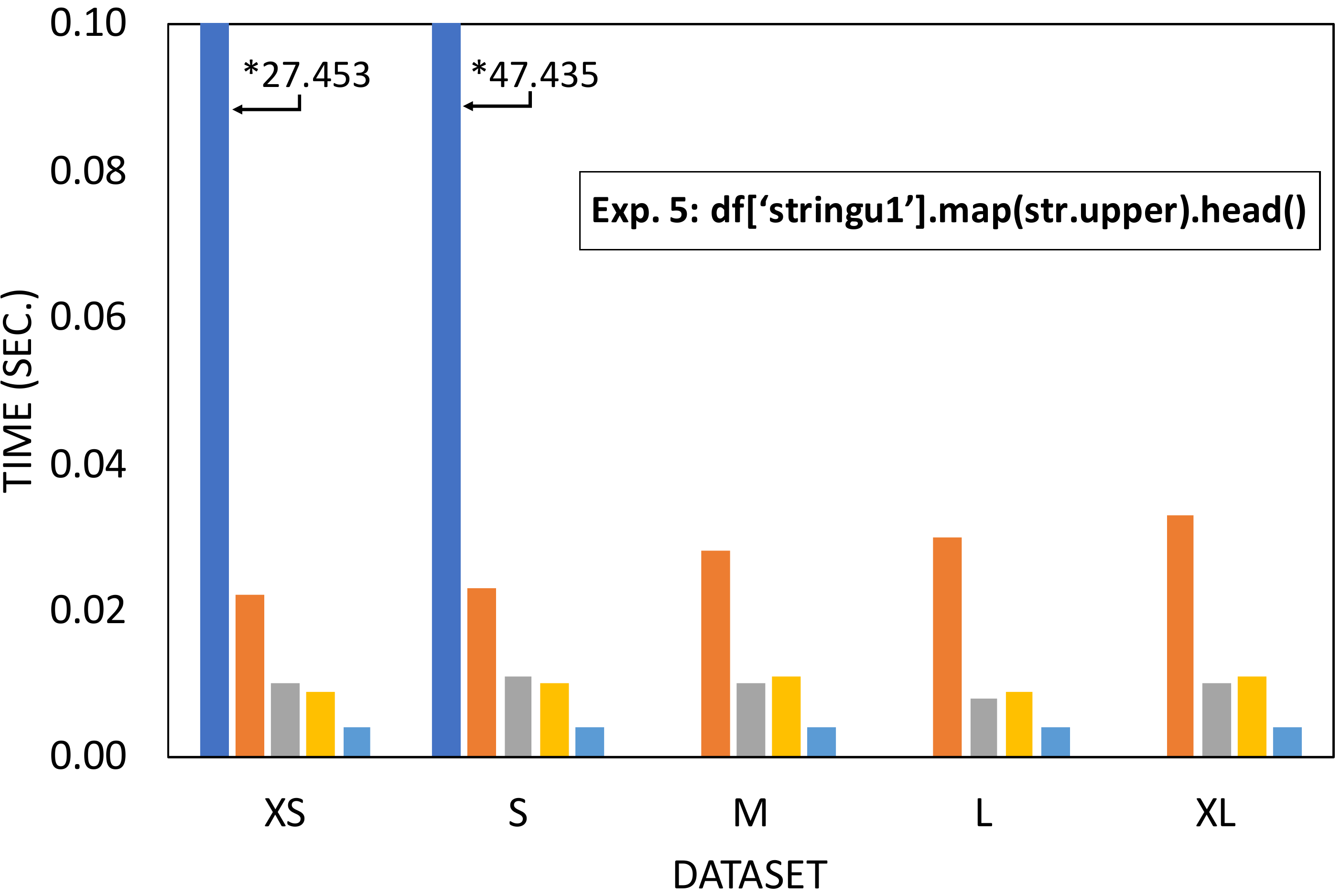}
        \caption{Expression 5: total times}
        \label{fig:q5_single}
    \end{subfigure}
    \begin{subfigure}[t]{0.41\textwidth}
        \includegraphics[trim=1.5 1.5 0cm 1.5,width=\textwidth,height=3.9cm]{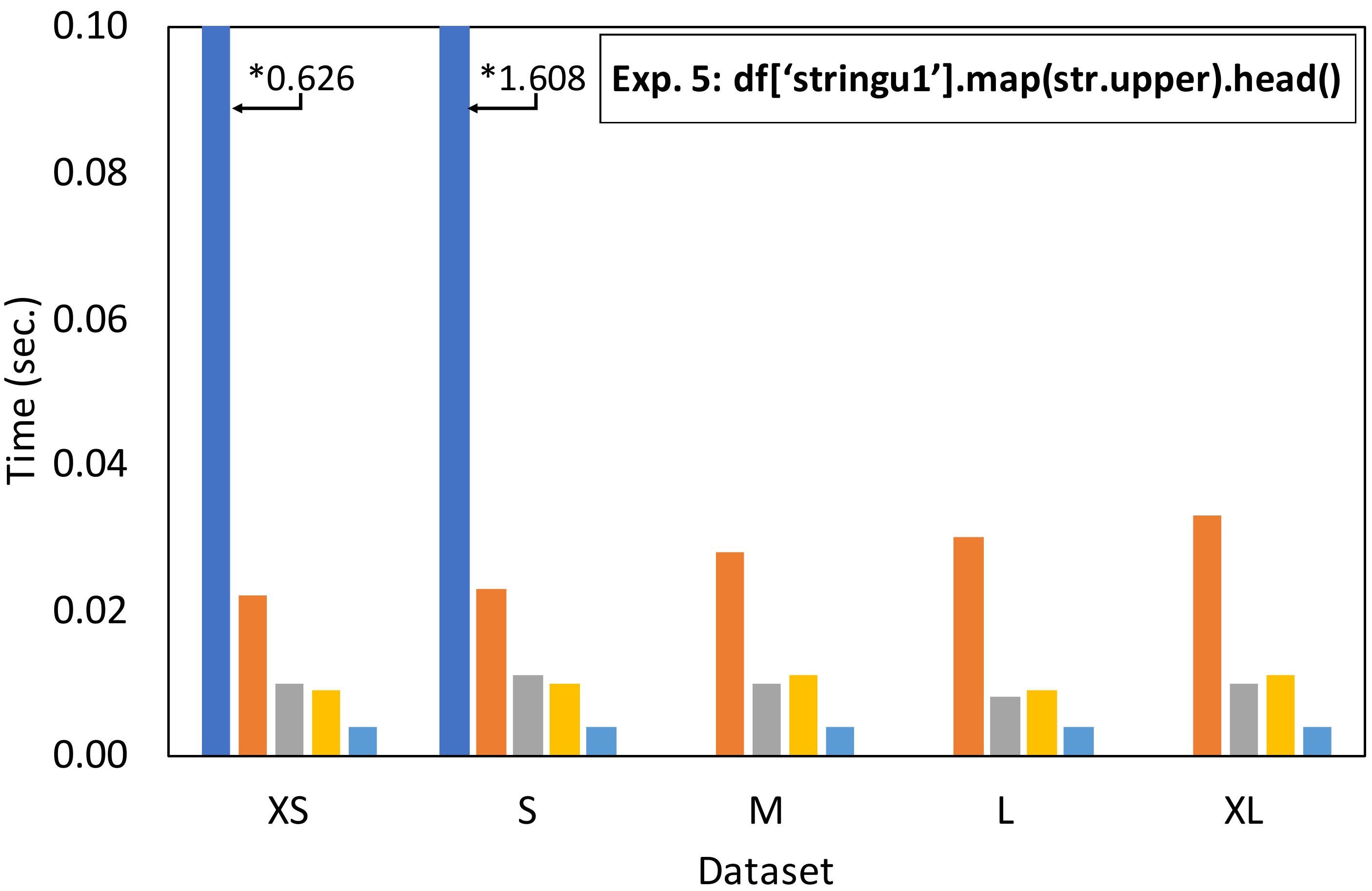}
        \caption{Expression 5: expression-only times}
        \label{fig:q5_single_wo}
    \end{subfigure}

    \caption{Exp.1-5 Single Node Evaluation Results (*=value where the bar ends)}

    \label{fig:1-5_single_results}
    % \vspace{-1.5em}
\end{figure*}

\begin{figure*}[h!]
     \centering
    \begin{subfigure}[t]{0.75\textwidth}
        \includegraphics[trim=1.5 1.5 0cm 1.5,width=\textwidth,height=0.5cm]{figures/legend.pdf}
    \end{subfigure}
    % \hspace{15cm}
    \begin{subfigure}[t]{0.41\textwidth}
        \includegraphics[trim=1.5 1.5 0cm 1.5,width=\textwidth,height=3.9cm]{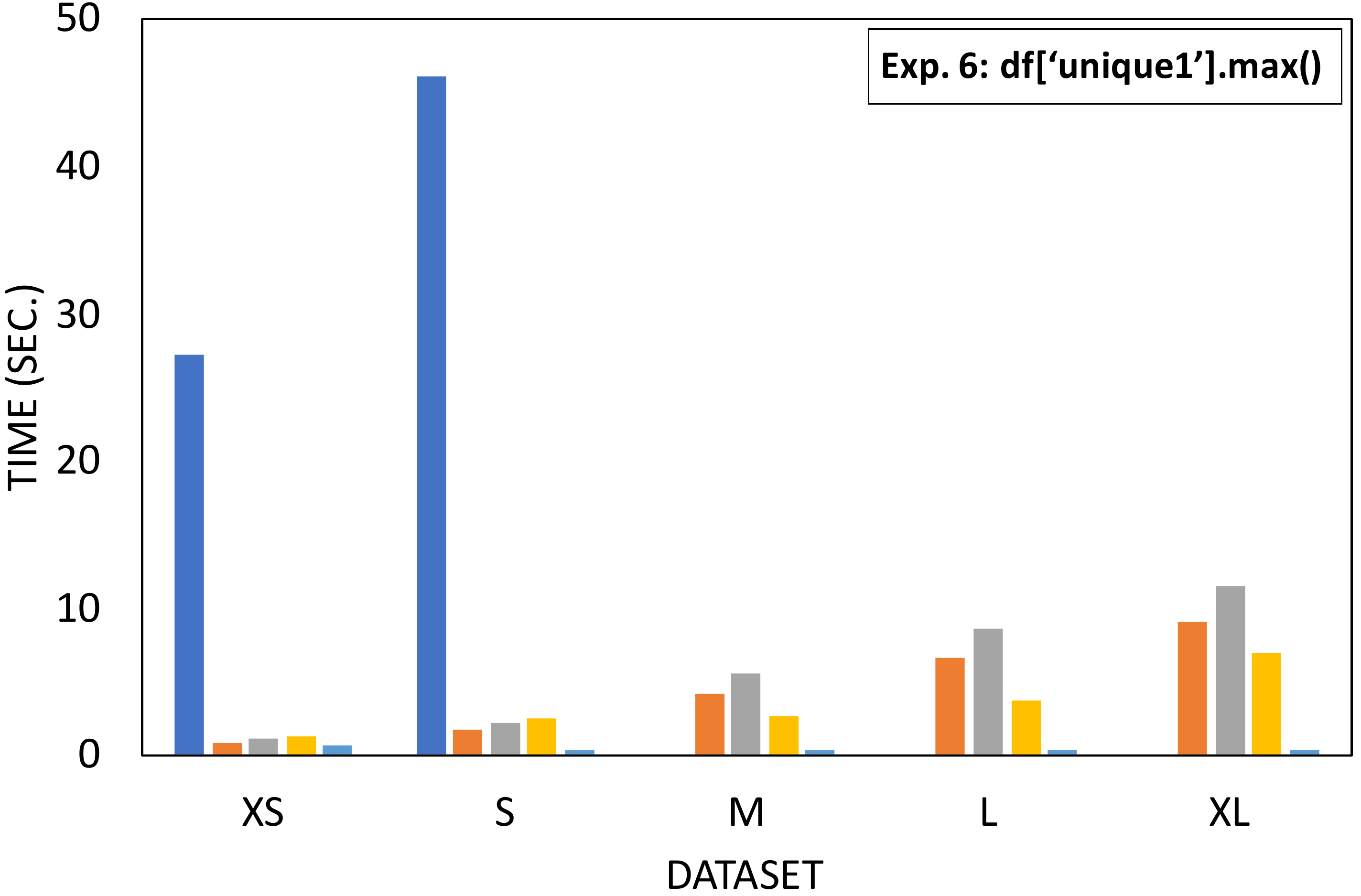}
        \caption{Expression 6: total times}
        \label{fig:q6_single}
    \end{subfigure}
    \begin{subfigure}[t]{0.41\textwidth}
        \includegraphics[trim=1.5 1.5 0cm 1.5,width=\textwidth,height=3.9cm]{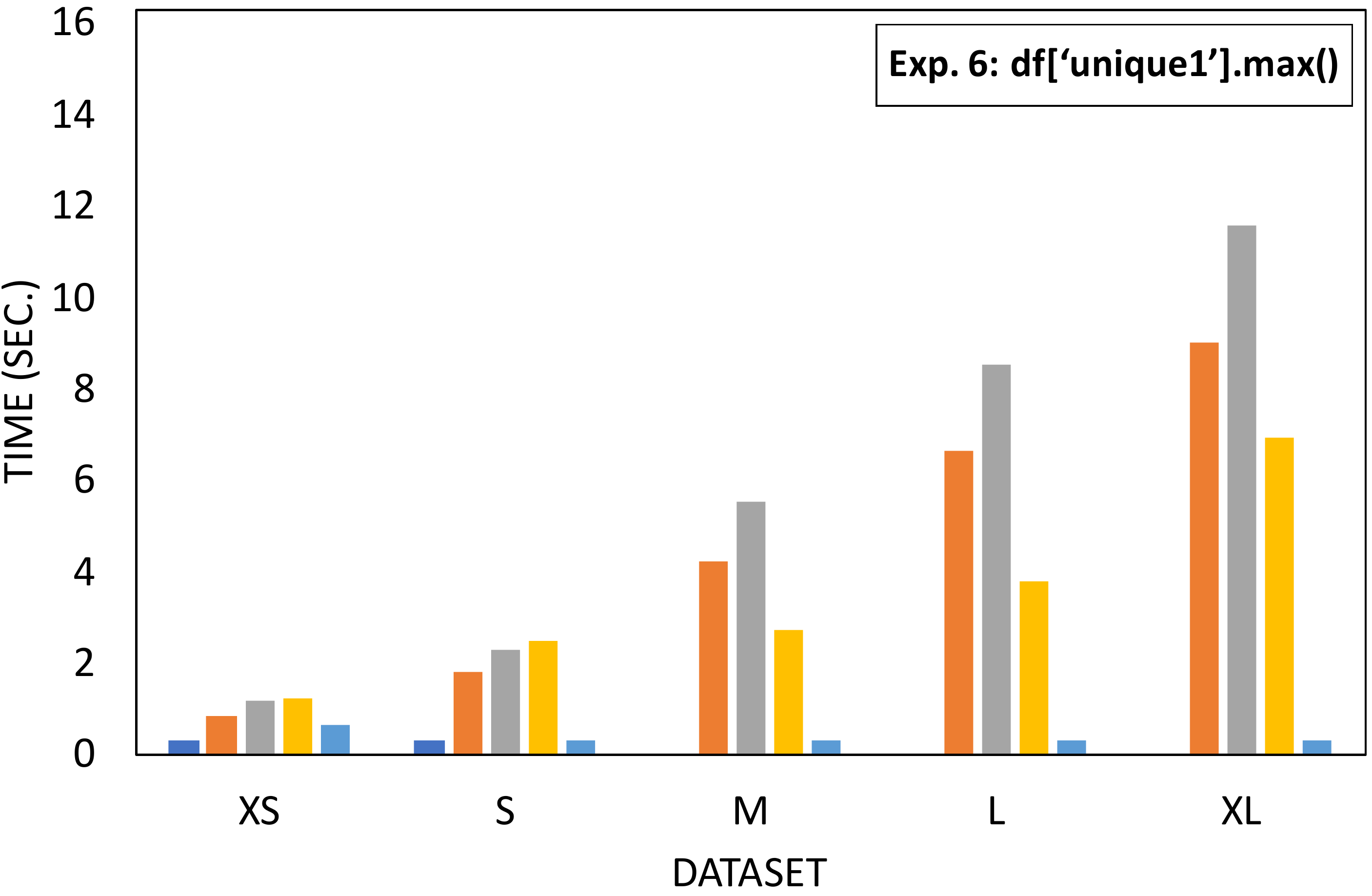}
        \caption{Expression 6: expression-only times}
        \label{fig:q6_single_wo}
    \end{subfigure}
    
    \begin{subfigure}[t]{0.41\textwidth}
        \includegraphics[trim=1.5 1.5 0cm 1.5,width=\textwidth,height=3.9cm]{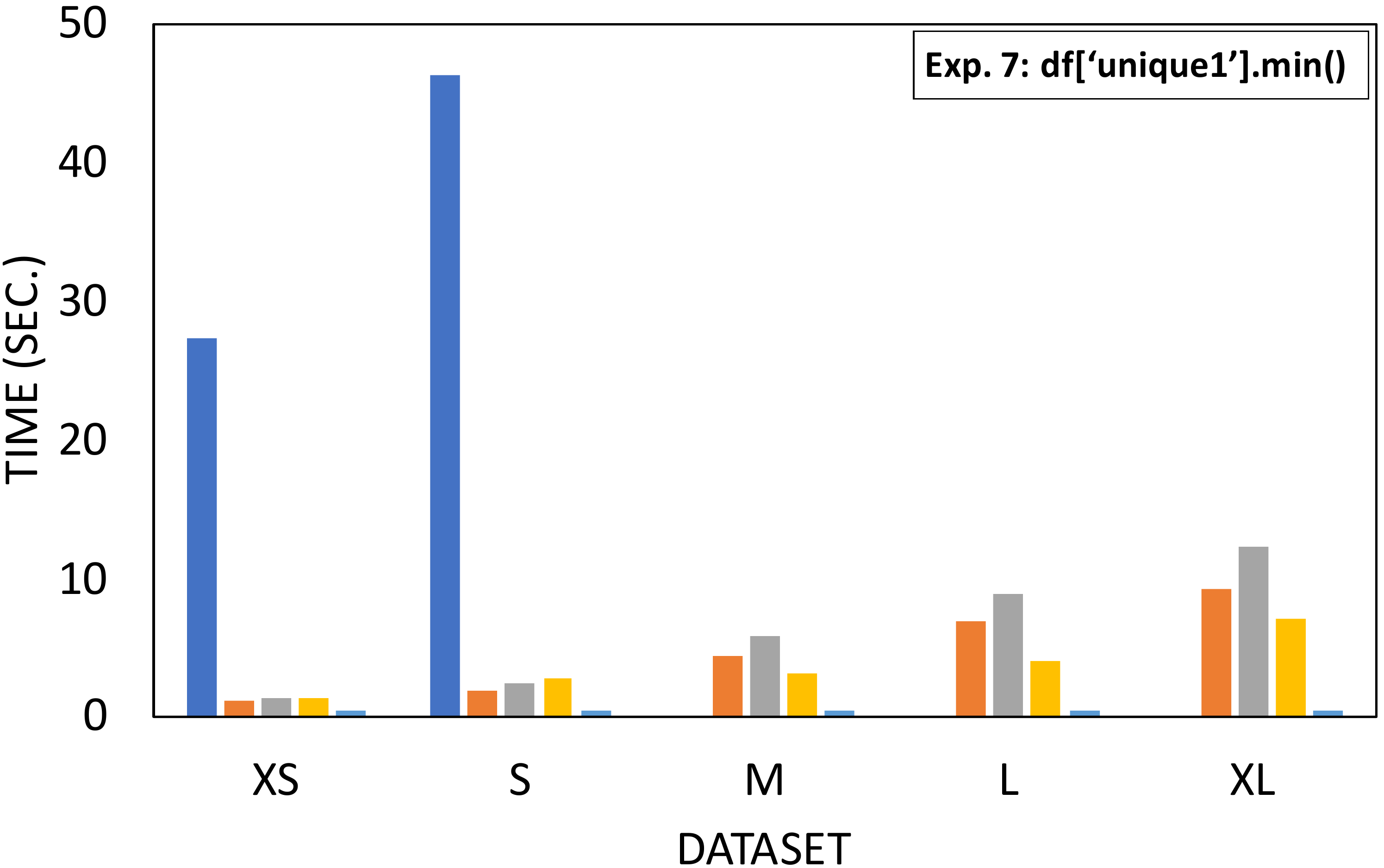}
        \caption{Expression 7: total times}
        \label{fig:q7_single}
    \end{subfigure}
    \begin{subfigure}[t]{0.41\textwidth}
        \includegraphics[trim=1.5 1.5 0cm 1.5,width=\textwidth,height=3.9cm]{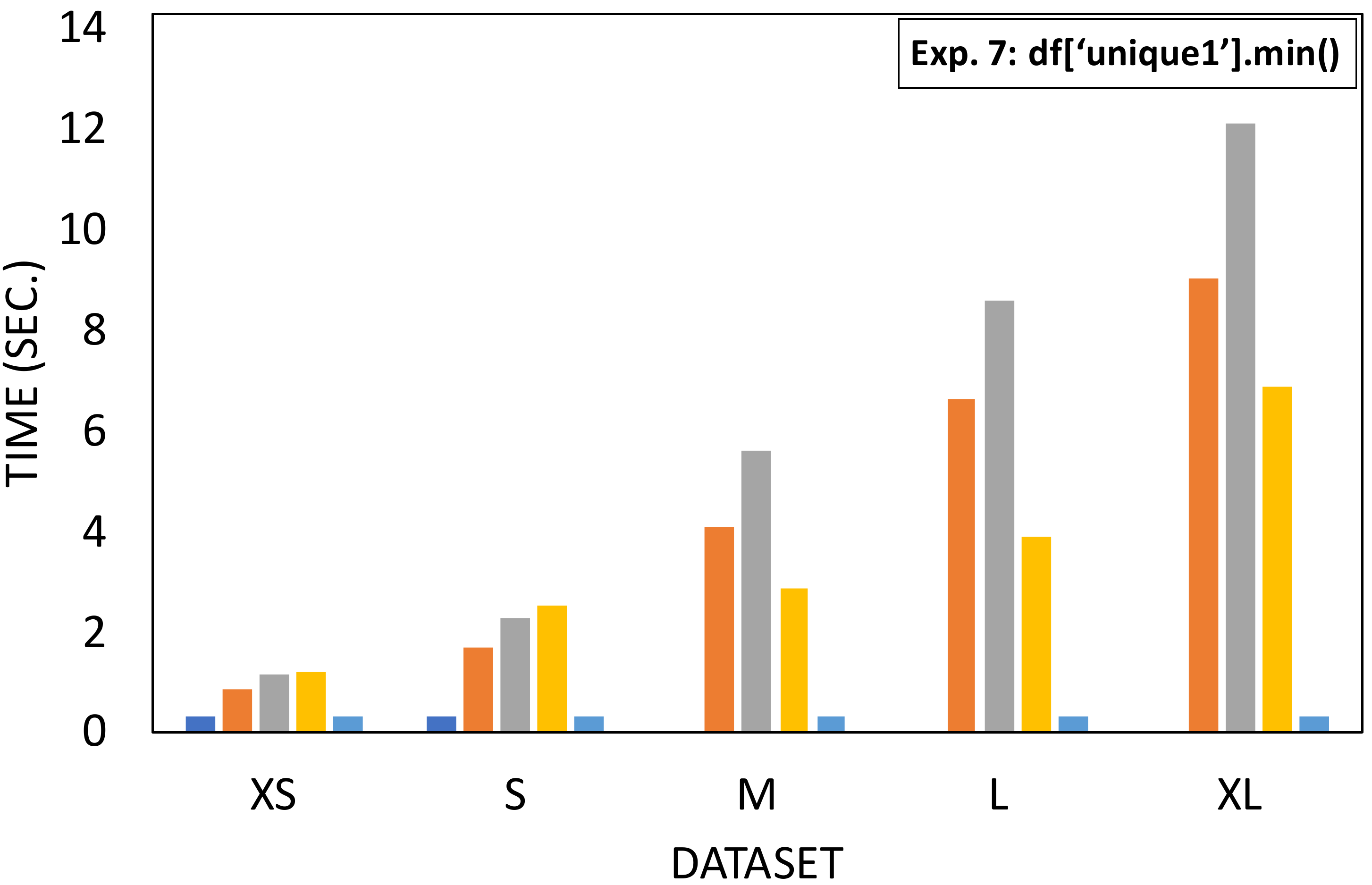}
        \caption{Expression 7: expression-only times}
        \label{fig:q7_single_wo}
    \end{subfigure}
    
    \begin{subfigure}[t]{0.41\textwidth}
        \includegraphics[trim=1.5 1.5 0cm 1.5,width=\textwidth,height=3.9cm]{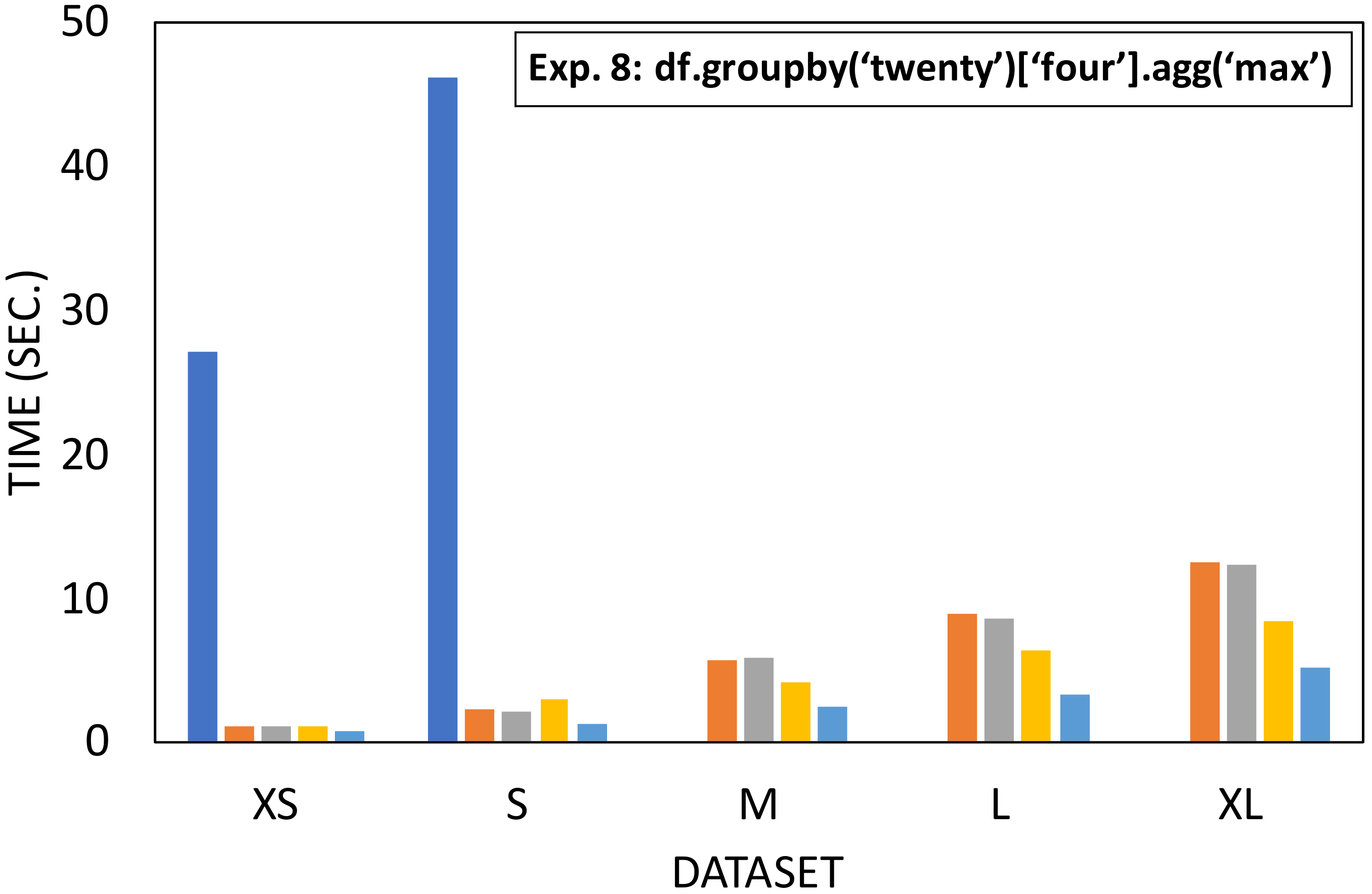}
        \caption{Expression 8: total times}
        \label{fig:q8_single}
    \end{subfigure}
    \begin{subfigure}[t]{0.41\textwidth}
        \includegraphics[trim=1.5 1.5 0cm 1.5,width=\textwidth,height=3.9cm]{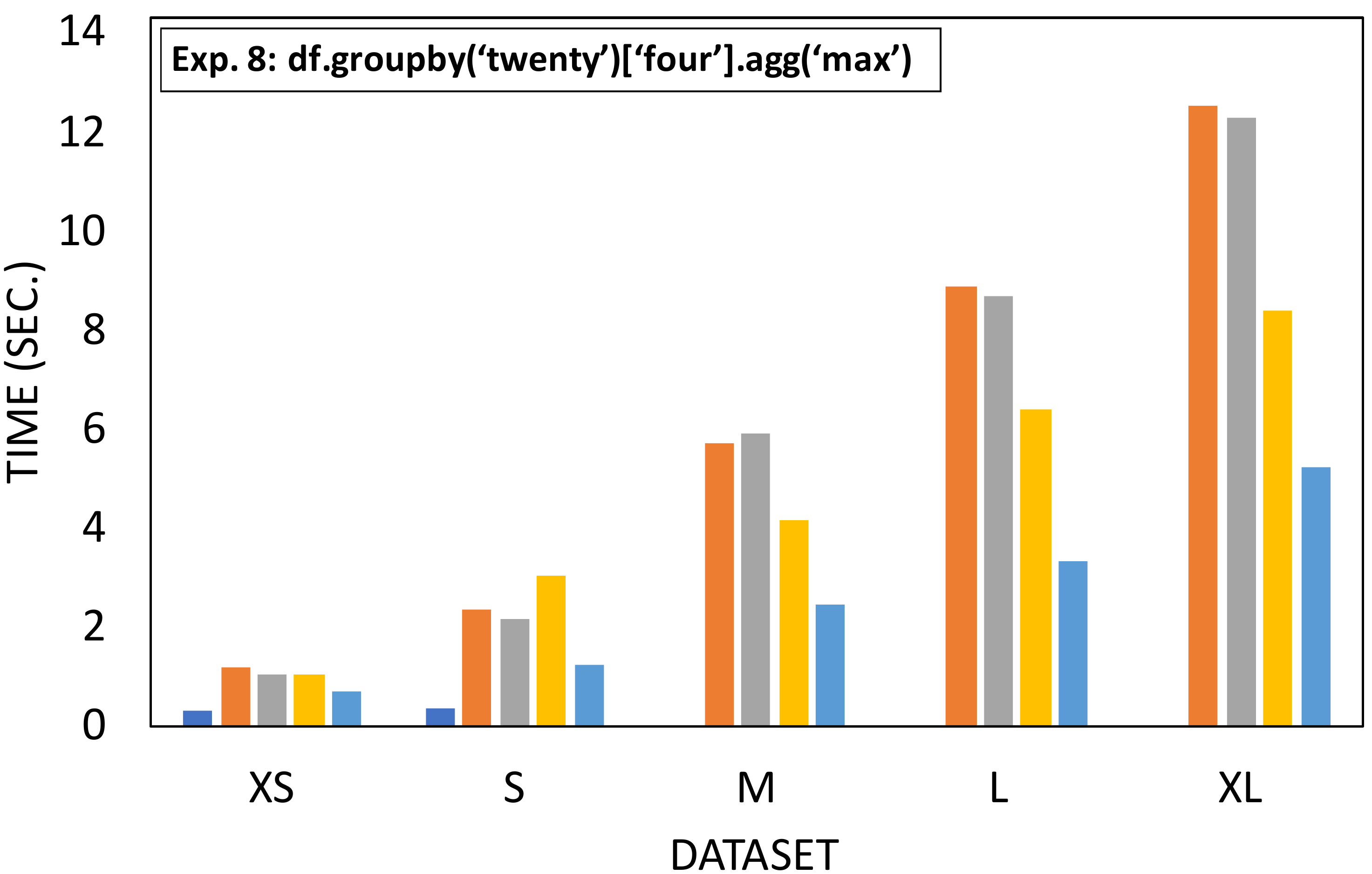}
        \caption{Expression 8: expression-only times}
        \label{fig:q8_single_wo}
    \end{subfigure}

    \begin{subfigure}[t]{0.41\textwidth}
        \includegraphics[trim=1.5 1.5 0cm 1.5,width=\textwidth,height=3.9cm]{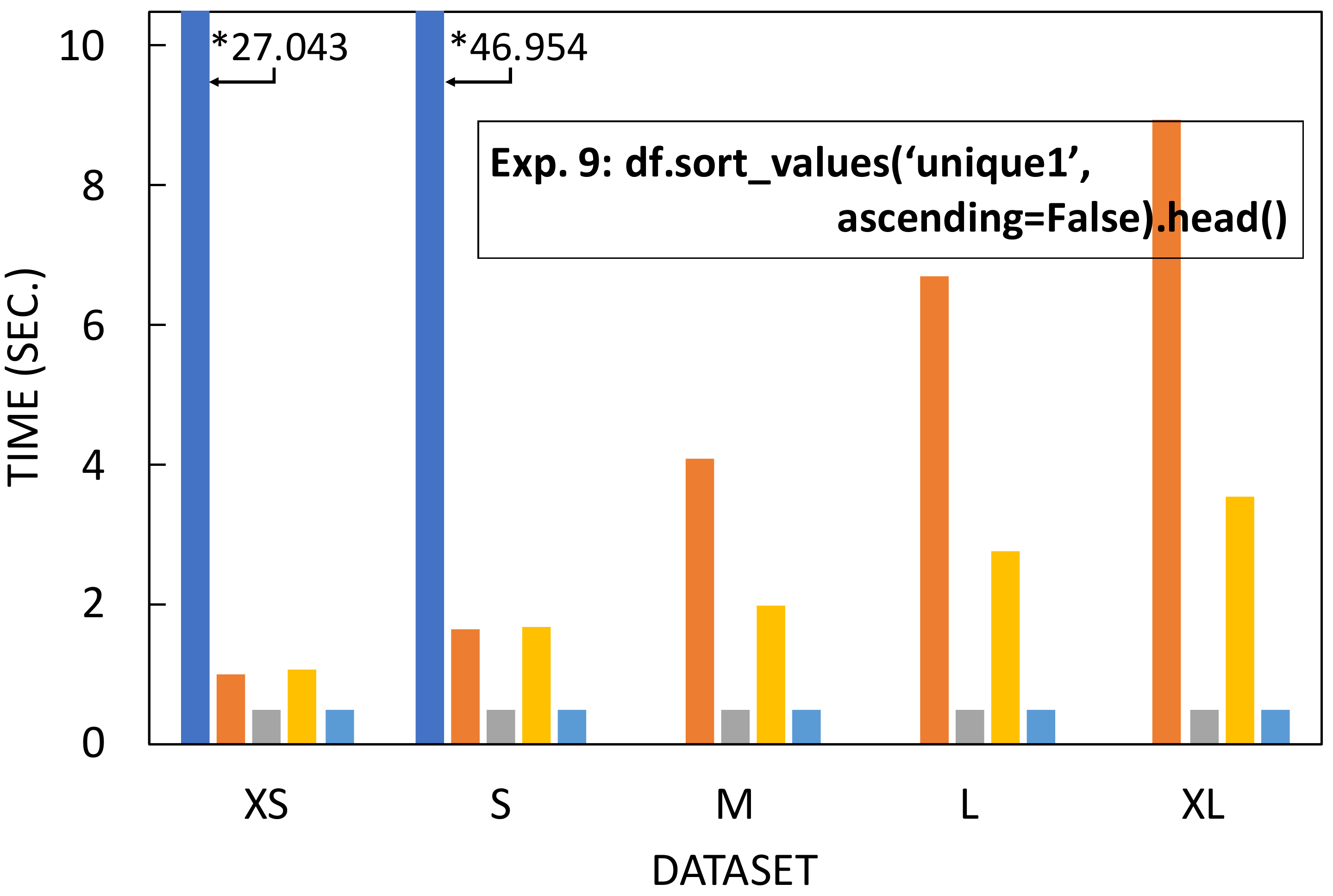}%
        \caption{Expression 9: total times}
        \label{fig:q9_single}
    \end{subfigure}
    \begin{subfigure}[t]{0.41\textwidth}
        \includegraphics[trim=1.5 1.5 0cm 1.5,width=\textwidth,height=3.9cm]{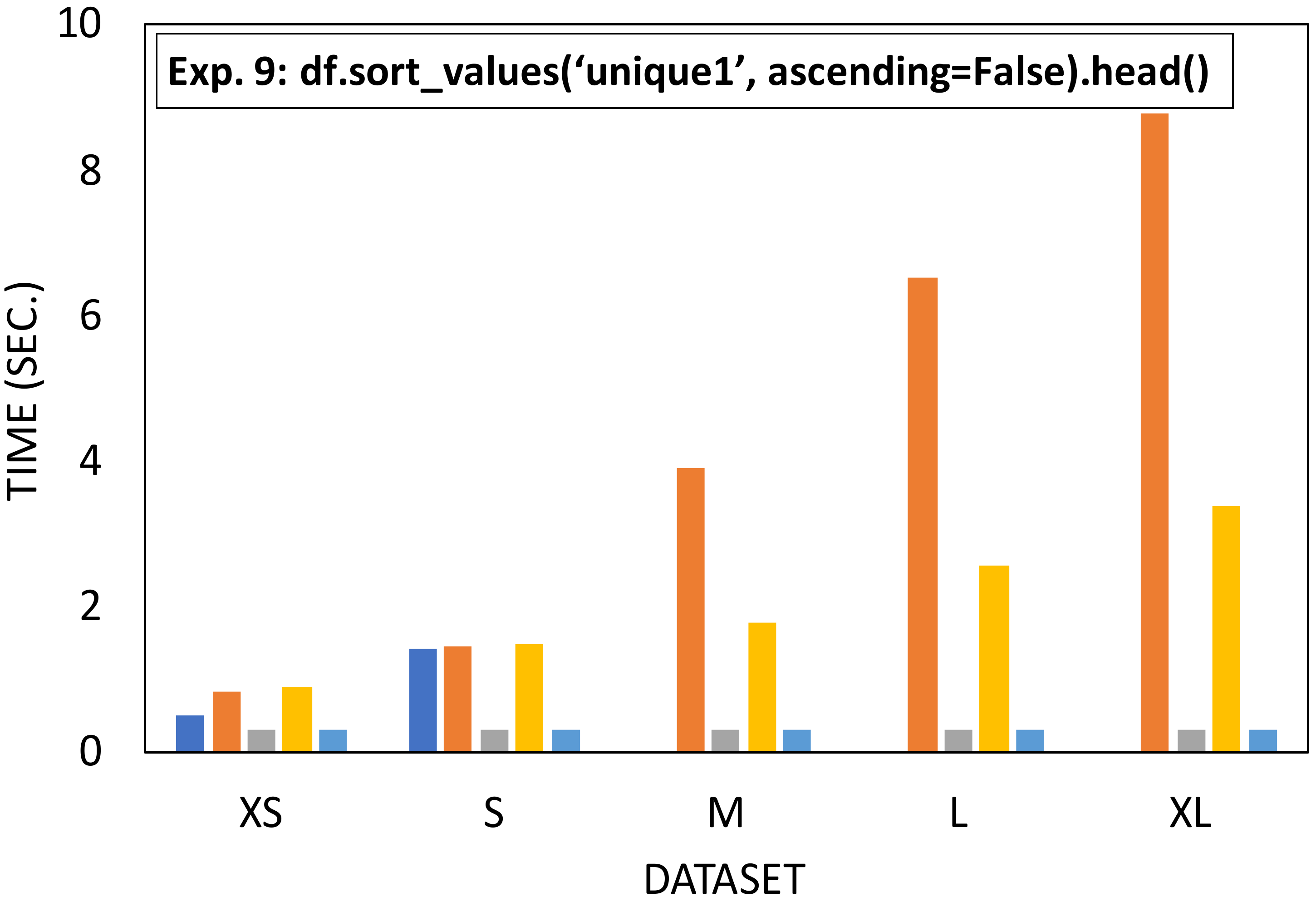}%
        \caption{Expression 9: expression-only times}
        \label{fig:q9_single_wo}
    \end{subfigure}

    \begin{subfigure}[t]{0.41\textwidth}
        \includegraphics[trim=1.5 1.5 0cm 1.5,width=\textwidth,height=3.9cm]{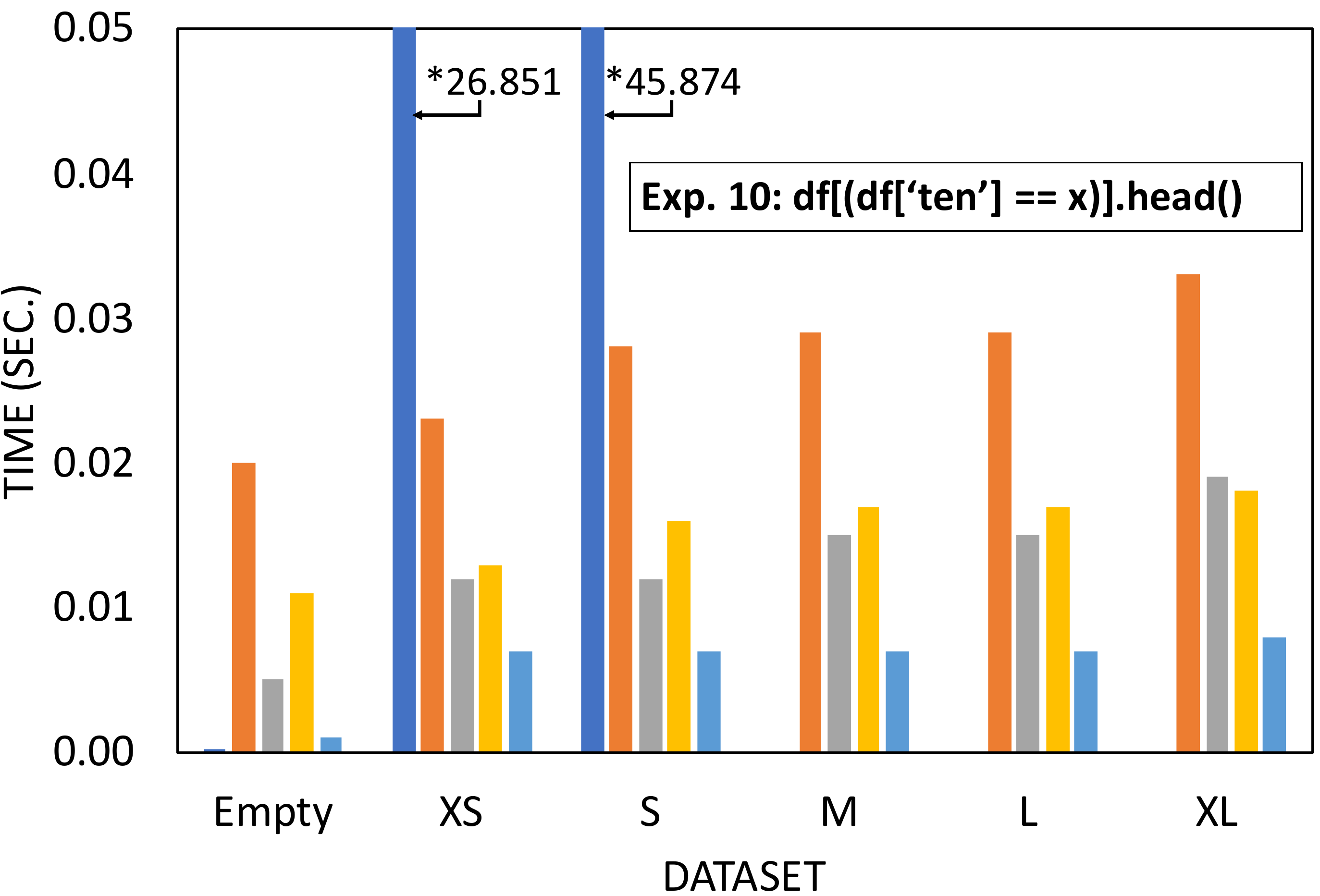}
        \caption{Expression 10: total times}
        \label{fig:q10_single}
    \end{subfigure}
    \begin{subfigure}[t]{0.41\textwidth}
        \includegraphics[trim=1.5 1.5 0cm 1.5,width=\textwidth,height=3.9cm]{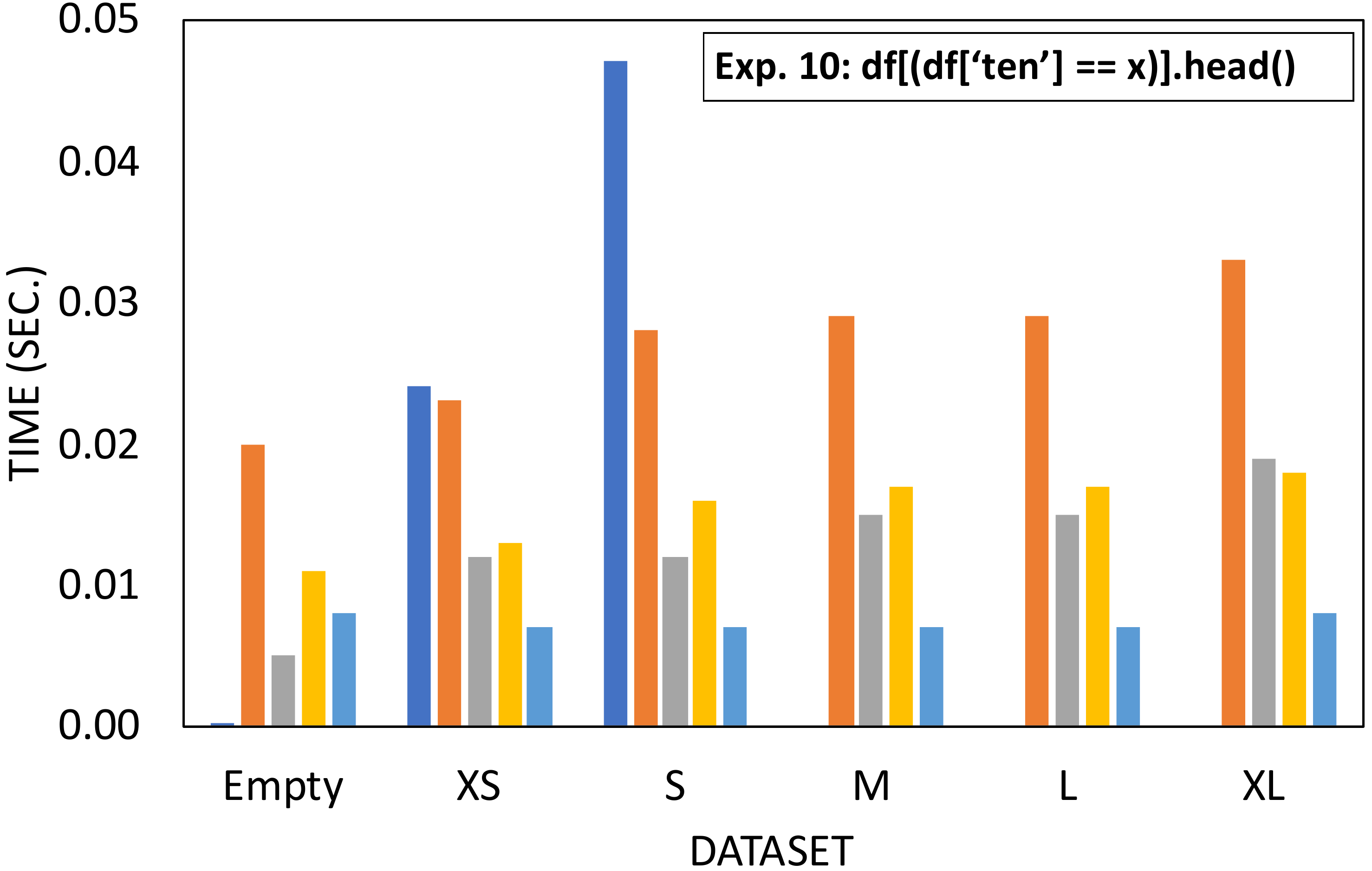}
        \caption{Expression 10: expression-only times}
        \label{fig:q10_single_wo}
    \end{subfigure}

    \caption{Exp.6-10 Single Node Evaluation Results (*=value where the bar ends)}

    \label{fig:6-10_single_results}
    % \vspace{-1.5em}
\end{figure*}

\begin{figure*}[h!]
     \centering
    \begin{subfigure}[t]{0.75\textwidth}
        \includegraphics[trim=1.5 1.5 0cm 1.5,width=\textwidth,height=0.5cm]{figures/legend.pdf}
    \end{subfigure}
    % \hspace{15cm}
    \begin{subfigure}[t]{0.41\textwidth}
        \includegraphics[trim=1.5 1.5 0cm 1.5,width=\textwidth,height=3.9cm]{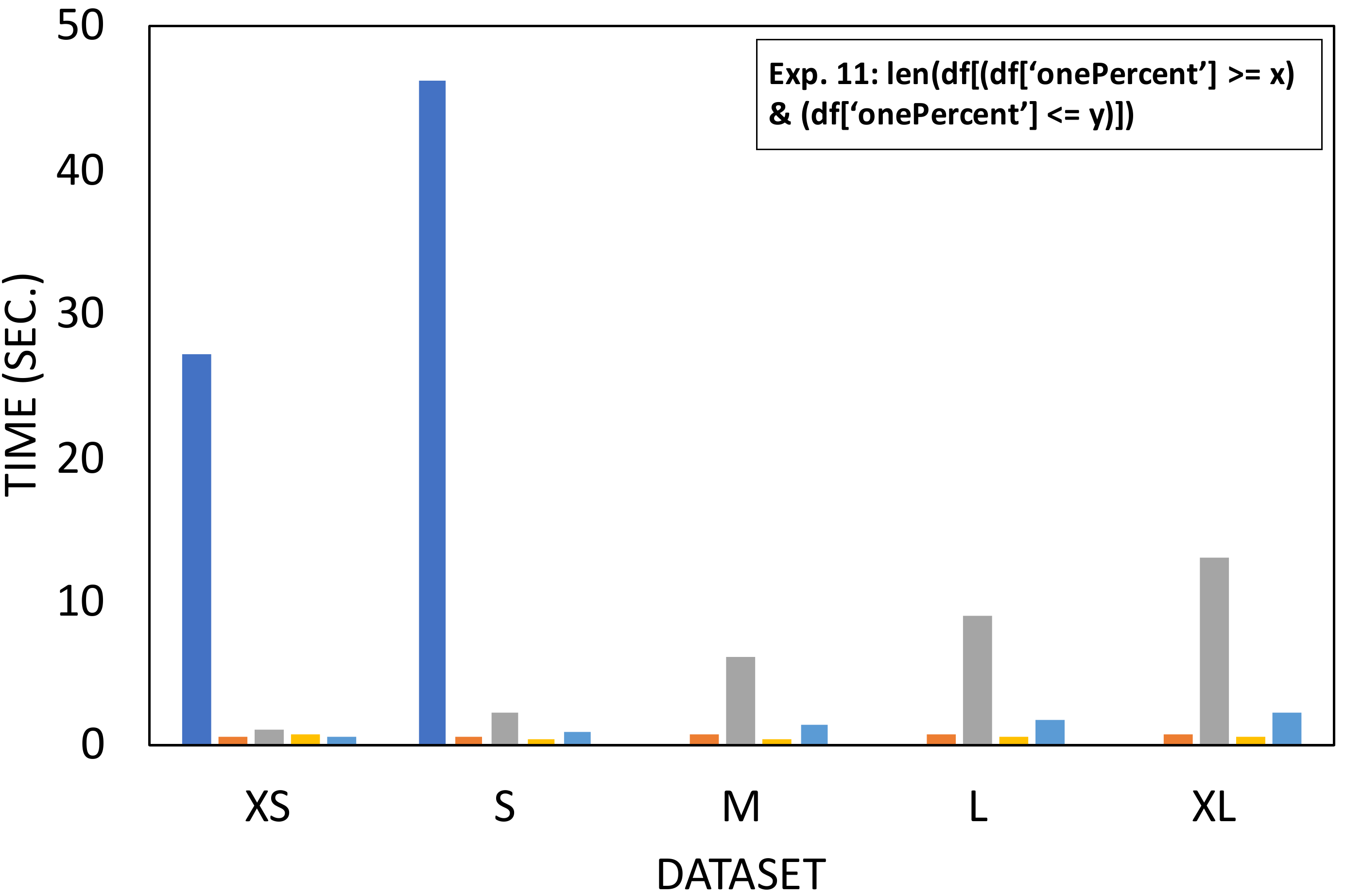}
        \caption{Expression 11: total times}
        \label{fig:q11_single}
    \end{subfigure}
    \begin{subfigure}[t]{0.41\textwidth}
        \includegraphics[trim=1.5 1.5 0cm 1.5,width=\textwidth,height=3.9cm]{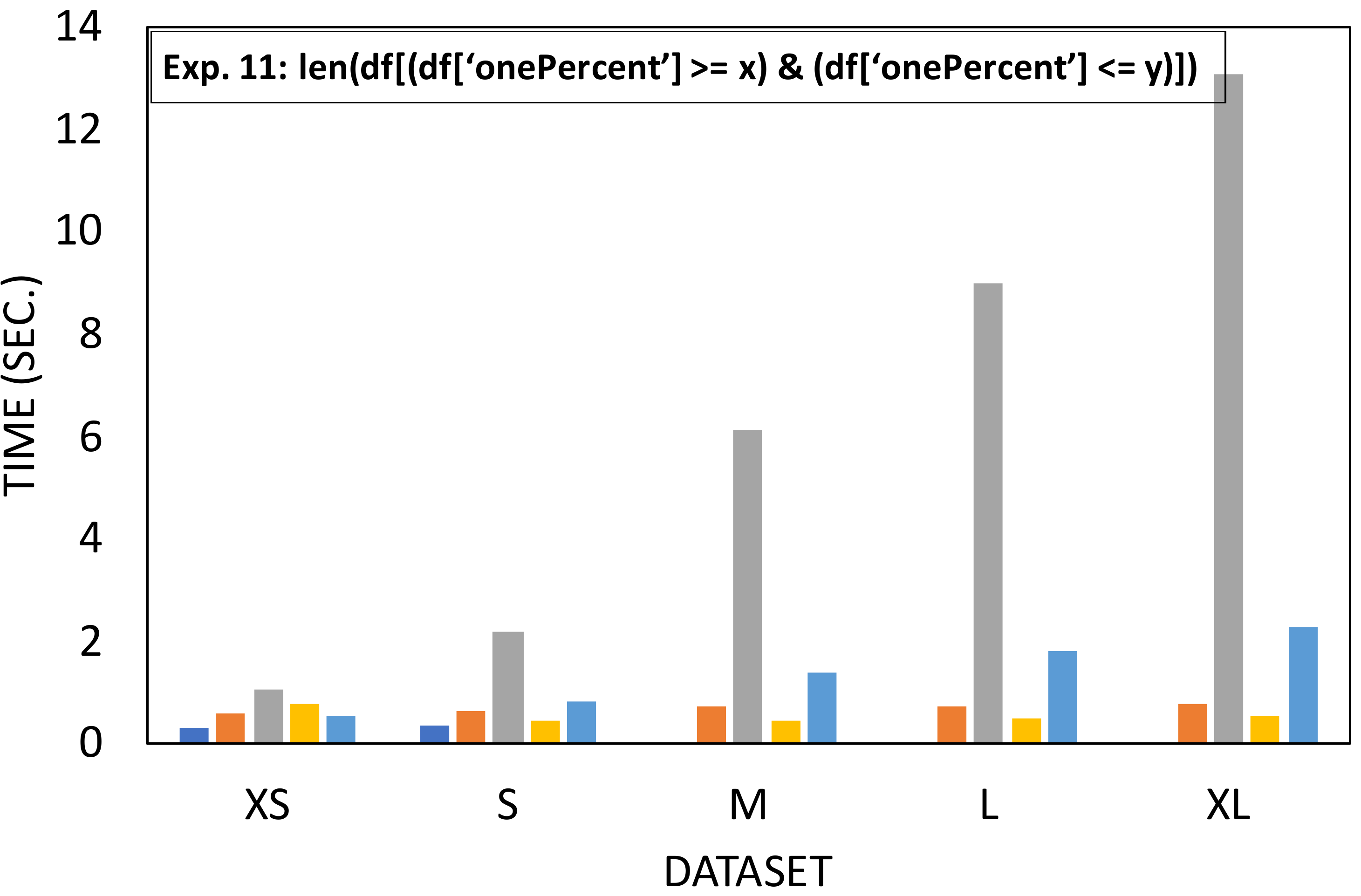}
        \caption{Expression 11: expression-only times}
        \label{fig:q11_single_wo}
    \end{subfigure}
    
    \begin{subfigure}[t]{0.41\textwidth}
        \includegraphics[trim=1.5 1.5 0cm 1.5,width=\textwidth,height=3.9cm]{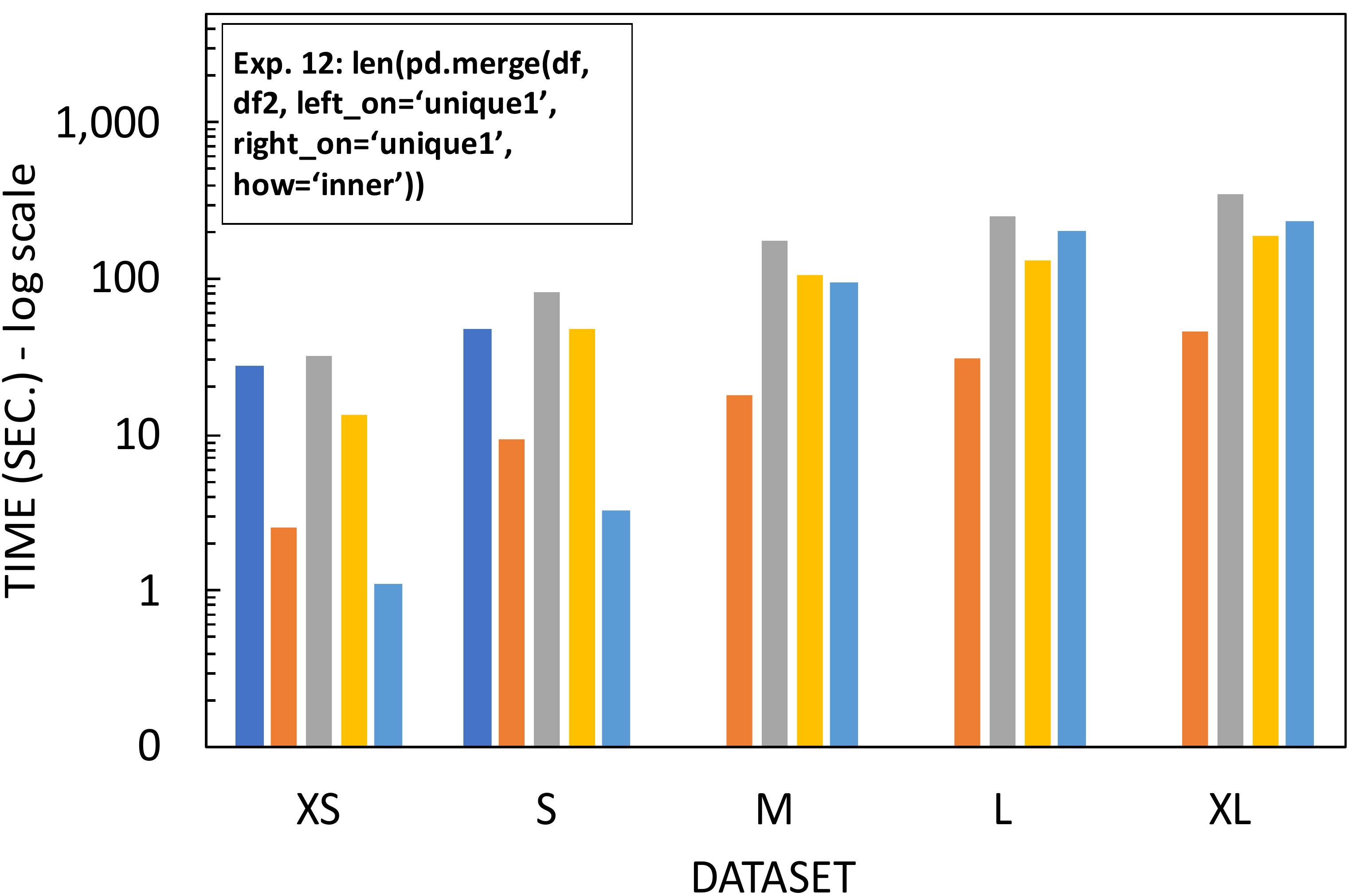}
        \caption{Expression 12: total times}
        \label{fig:q12_single}
    \end{subfigure}
    \begin{subfigure}[t]{0.41\textwidth}
        \includegraphics[trim=1.5 1.5 0cm 1.5,width=\textwidth,height=3.9cm]{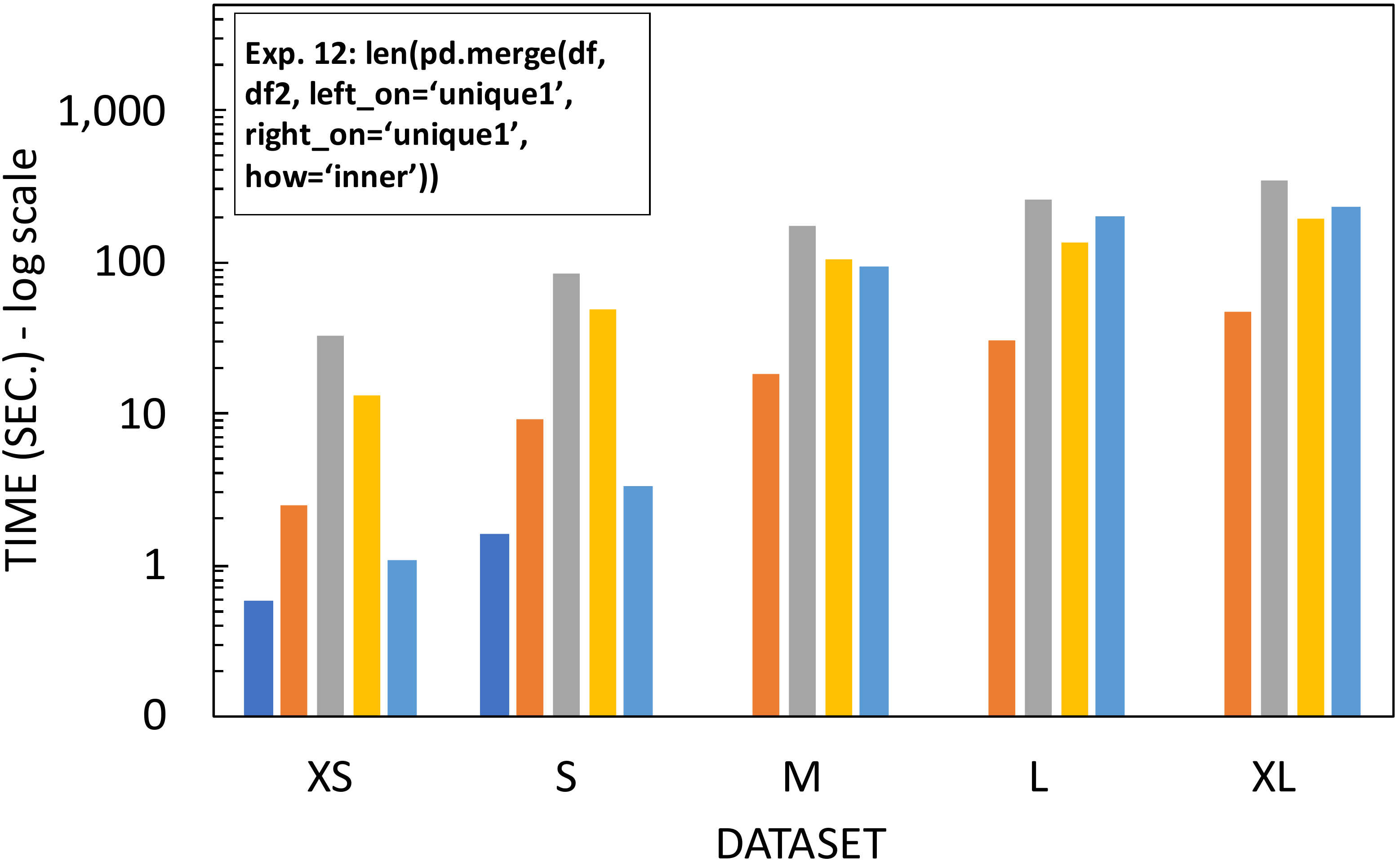}
        \caption{Expression 12: expression-only times}
        \label{fig:q12_single_wo}
    \end{subfigure}
    
    \begin{subfigure}[t]{0.41\textwidth}
        \includegraphics[trim=1.5 1.5 0cm 1.5,width=\textwidth,height=3.9cm]{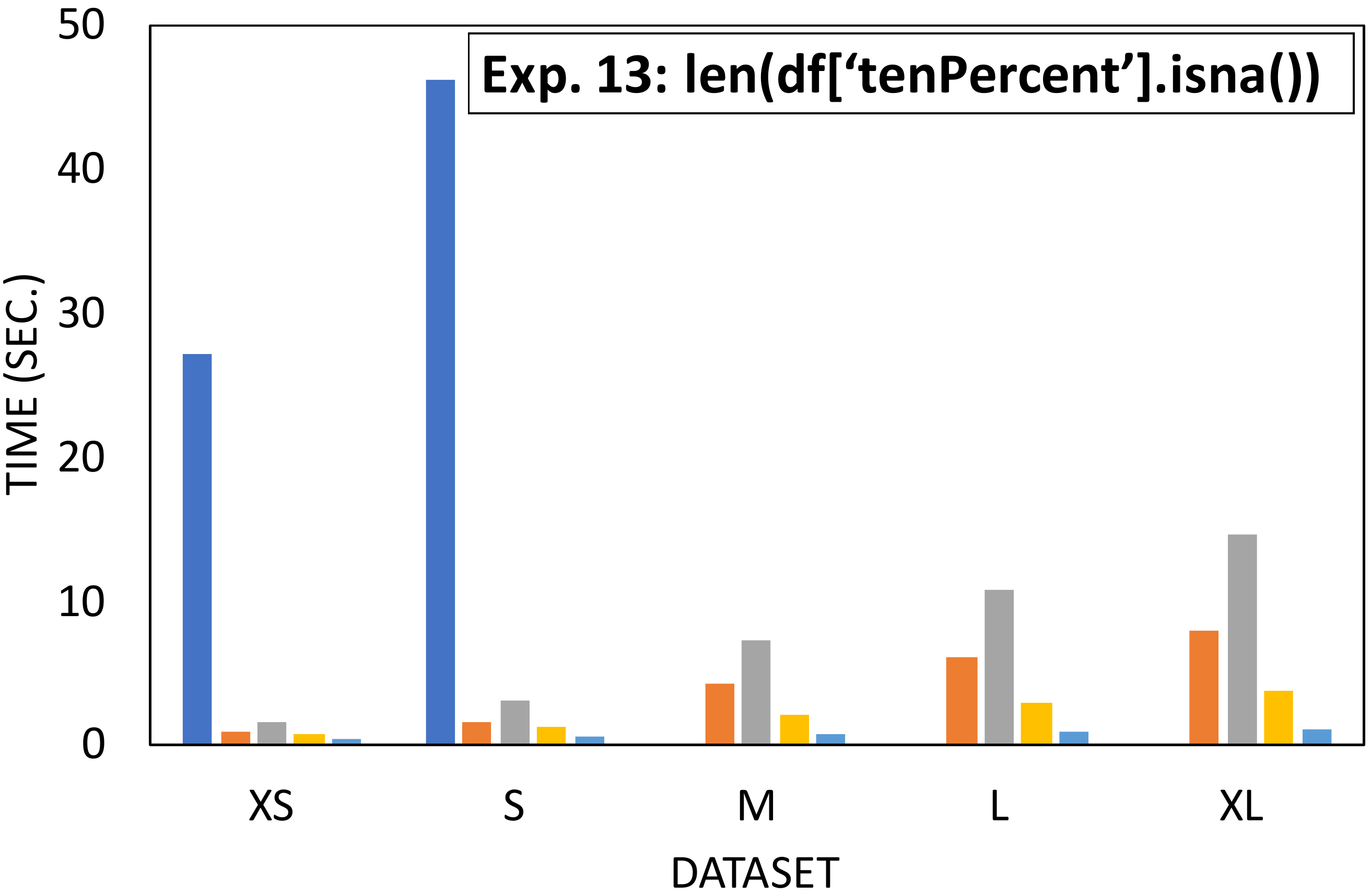}
        \caption{Expression 13: total times}
        \label{fig:q13_single}
    \end{subfigure}
    \begin{subfigure}[t]{0.41\textwidth}
        \includegraphics[trim=1.5 1.5 0cm 1.5,width=\textwidth,height=3.9cm]{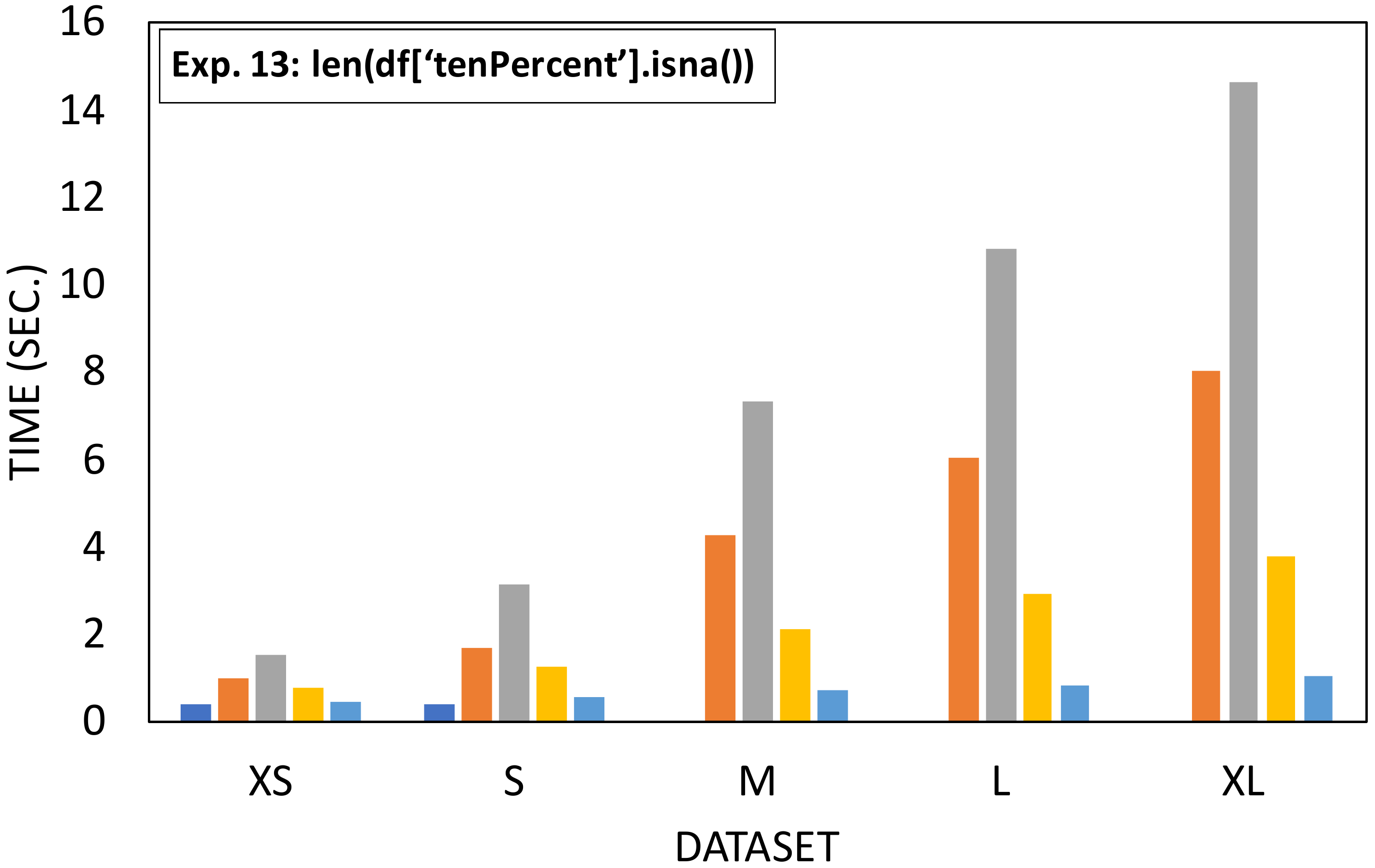}
        \caption{Expression 13: expression-only times}
        \label{fig:q13_single_wo}
    \end{subfigure}

    \caption{Exp.11-13 Single Node Evaluation Results}

    \label{fig:11-13_single_results}
    % \vspace{-1.5em}
\end{figure*}

After the first XS round, we ran the benchmark on the four other dataset sizes, S, M, L and XL, to evaluate the data scalability of Pandas and of PolyFrame on a single node. The single node results are presented in \Cref{fig:1-5_single_results,fig:6-10_single_results,fig:11-13_single_results}. The first thing to note here is that Pandas threw an out-of-memory error on dataset sizes M, L, and XL, while all variants of PolyFrame were able to complete all operations on all of the tested dataset sizes. A programmatic workaround for Pandas could be to partition the data and then programmatically compute and combine intermediate results. However, we did not consider this solution; a partition size would need to be specified and that would directly affect the total runtime. Figure~\ref{fig:1-5_single_results} displays the total runtimes and expression-only runtimes for expressions 1-5. Figure~\ref{fig:6-10_single_results} displays the total and expression-only runtimes for expressions 6-10. Figure~\ref{fig:11-13_single_results} displays the total and expression-only runtimes for expressions 11-13. We discuss the results below; our discussion is based on having inspected the query plans for each operation on each system.

For Expression 1, Figure~\ref{fig:q1_single_wo} shows that PolyFrame operating on Neo4j was the fastest across all data sizes. This is because Neo4j keeps separate data-stores specifically for its nodes and its relationships metadata, so retrieving the count of records is an instant metadata lookup. PolyFrame operating on AsterixDB was competitive and was able to take advantage of a primary key index for this particular expression, while MongoDB and PostgreSQL resorted to table scans. 

Since we are targeting data analysis at scale, for relatively `small’ queries in terms of computation such as expressions 2 and 10, whose timings are in the tens of millisecond range, we also show results for the `Empty’ dataset as a baseline for consideration. Expression 2 is asking for a projection of two attributes from a small subset of data (five records). As shown by the `Empty' results in Figure~\ref{fig:q2_single_wo}, all of the evaluated database systems do have some query preparation overhead, especially AsterixDB (which is designed to operate efficiently on big data rather than being fast on `small’ queries). 

Expressions 5 (Figure~\ref{fig:q5_single_wo}) and 10 (Figure~\ref{fig:q10_single_wo}) are good demonstrations of the advantage of lazy evaluation. These two expressions involve the application of repetitive operations over a small subset of data. For both of these expressions, even in the expression-only runtime case, Pandas was slower than all variants of PolyFrame. This is because PolyFrame's lazy evaluation allows all of the evaluated database systems to take advantage of their indexes and query optimizations to limit the amount of data needed for computation. However, Pandas suffered from eagerly evaluating these expressions and needed to compute intermediate results.

For expressions 6, 7, and 13, PolyFrame running on PostgreSQL was as competitive as Pandas in the case of its expression-only runtime as shown in~\Cref{fig:q6_single_wo,fig:q7_single_wo,fig:q13_single_wo}, respectively. This is because PostgreSQL evaluated expressions 6 and 7 using index-only query plans. For expression 13, PostgreSQL was uniquely able to use an index on the attribute. Even though this particular dataframe expression asks for null or missing data, null and missing values are only recorded in the attribute's index in PostgreSQL. On the other hand, AsterixDB, Cypher and MongoDB do support data with missing attributes, but missing values are not present in their indexes.

Figure~\ref{fig:q9_single_wo} displays the expression-only runtime of Expression 9, which is asking for a sample of records sorted in descending order on an attribute. Even when excluding the DataFrame creation time, PolyFrame operating on MongoDB and PostgreSQL were both faster than Pandas on the smallest dataset, and all variants of PolyFrame were competitive with Pandas for dataset size S. For this expression, MongoDB and PostgreSQL were both able to take advantage of backward index scans to efficiently retrieve the requested records.

The expression-only runtimes for expression 12 are displayed in Figure~\ref{fig:q12_single_wo}. This expression asks for the count of records resulting from a join of two identical datasets. PolyFrame operating on AsterixDB was able to evaluate this expression using an index-only query, while PostgreSQL, Neo4j, and MongoDB each used index nested loop joins followed by data scans.

\begin{figure*}[h!]
     \centering
    \begin{subfigure}[t]{0.44\textwidth}
        \includegraphics[width=\textwidth,height=0.4cm]{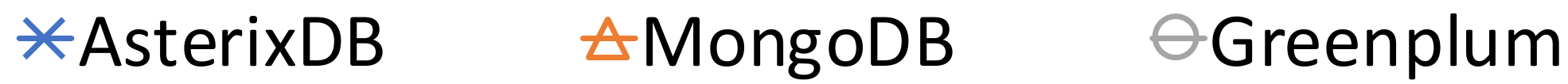}
    \end{subfigure}
    \hspace{15cm}
    \begin{subfigure}[t]{0.24\textwidth}
        \includegraphics[trim=1.5 1.5 0cm 1.5,width=\textwidth,height=3.3cm]{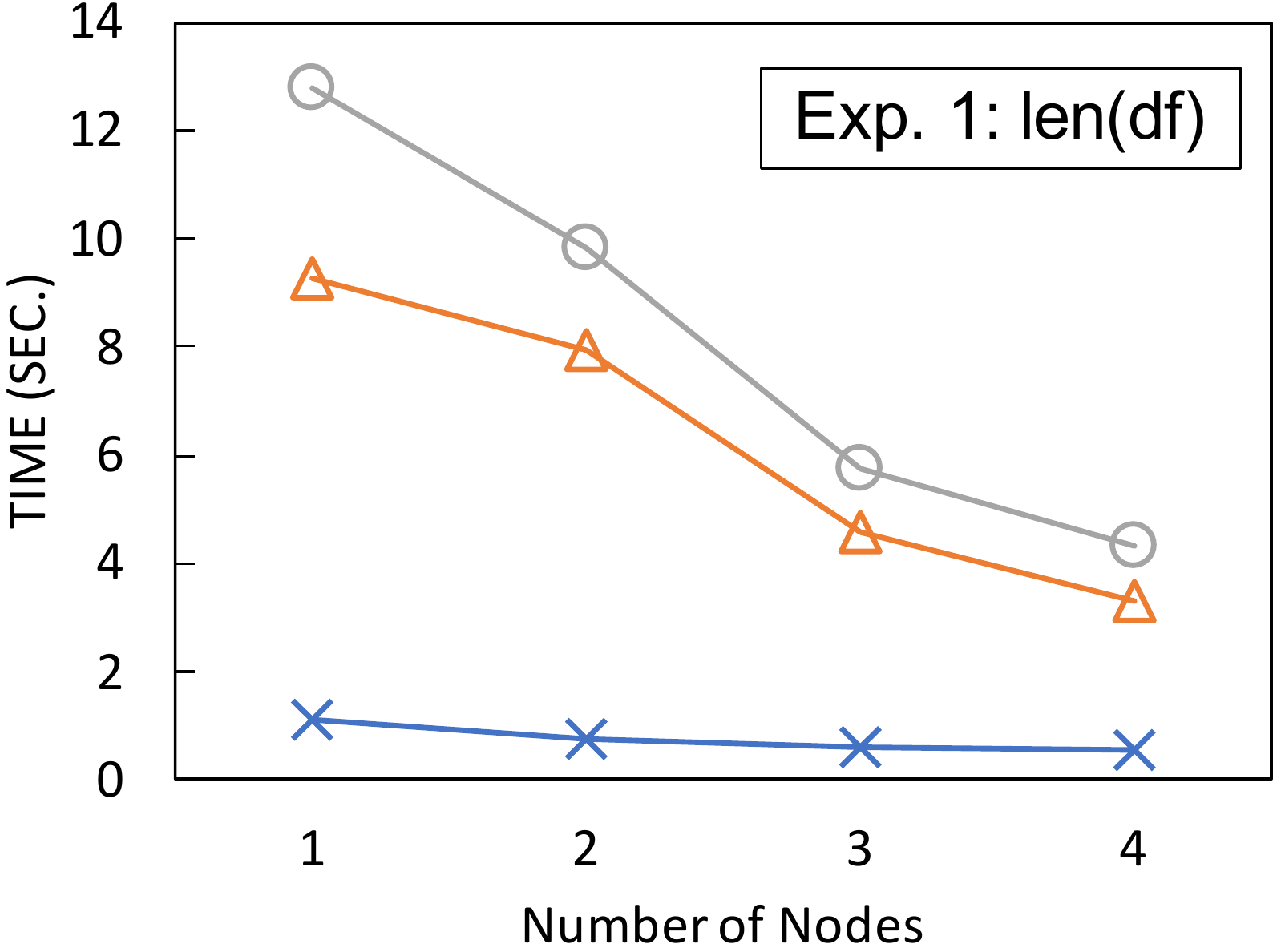}
        \caption{Expression 1}
        \label{fig:q1_speedup}
    \end{subfigure}
    \hfill
    \begin{subfigure}[t]{0.23\textwidth}
        \includegraphics[trim=0.5cm 1.5 0.5cm 1.5,width=\textwidth,height=3.3cm]{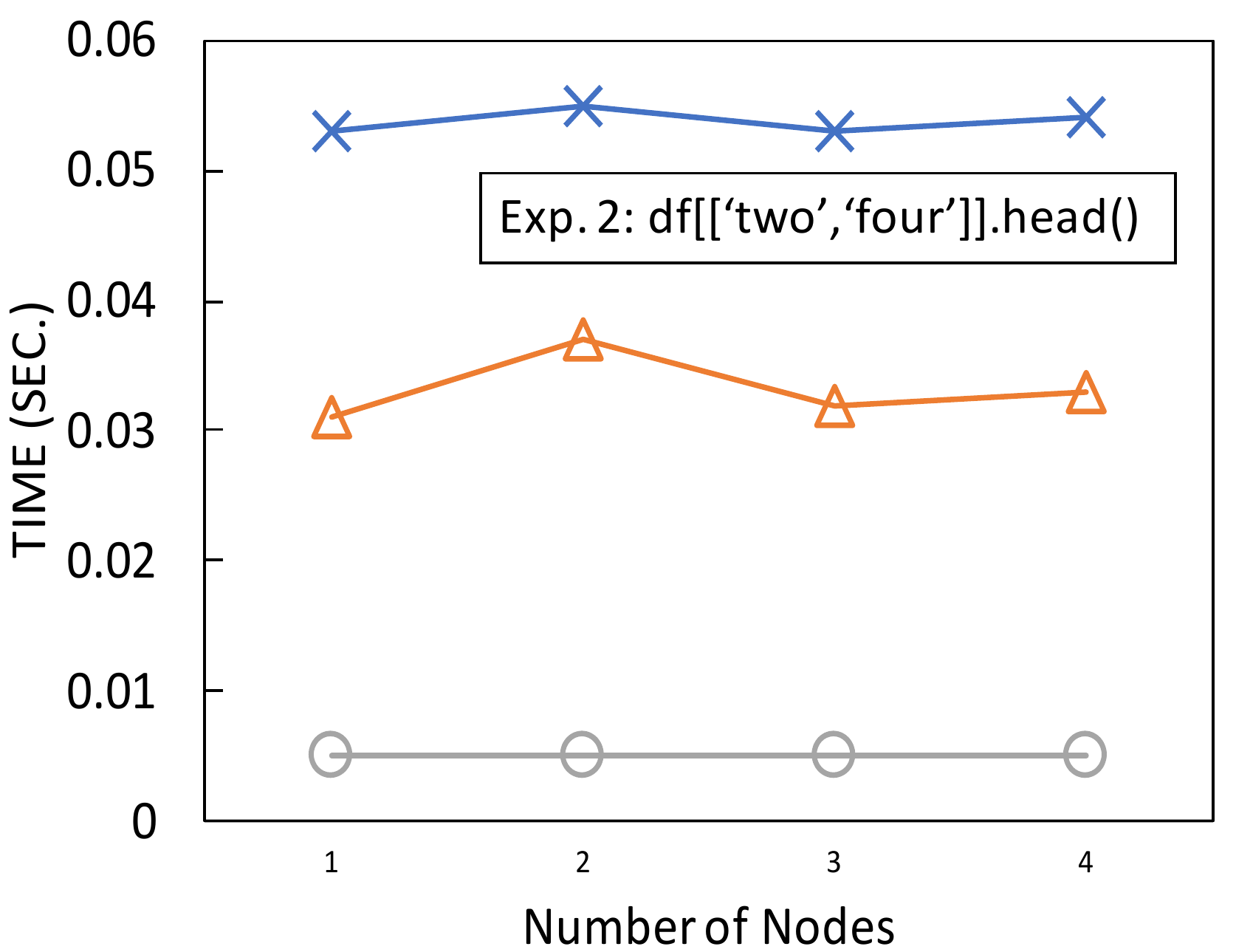}%
        \caption{Expression 2}
        \label{fig:q2_speedup}
    \end{subfigure}
    \hfill
    \begin{subfigure}[t]{0.24\textwidth}
        \includegraphics[trim=1.5 1.5 0cm 1.5,width=\textwidth,height=3.3cm]{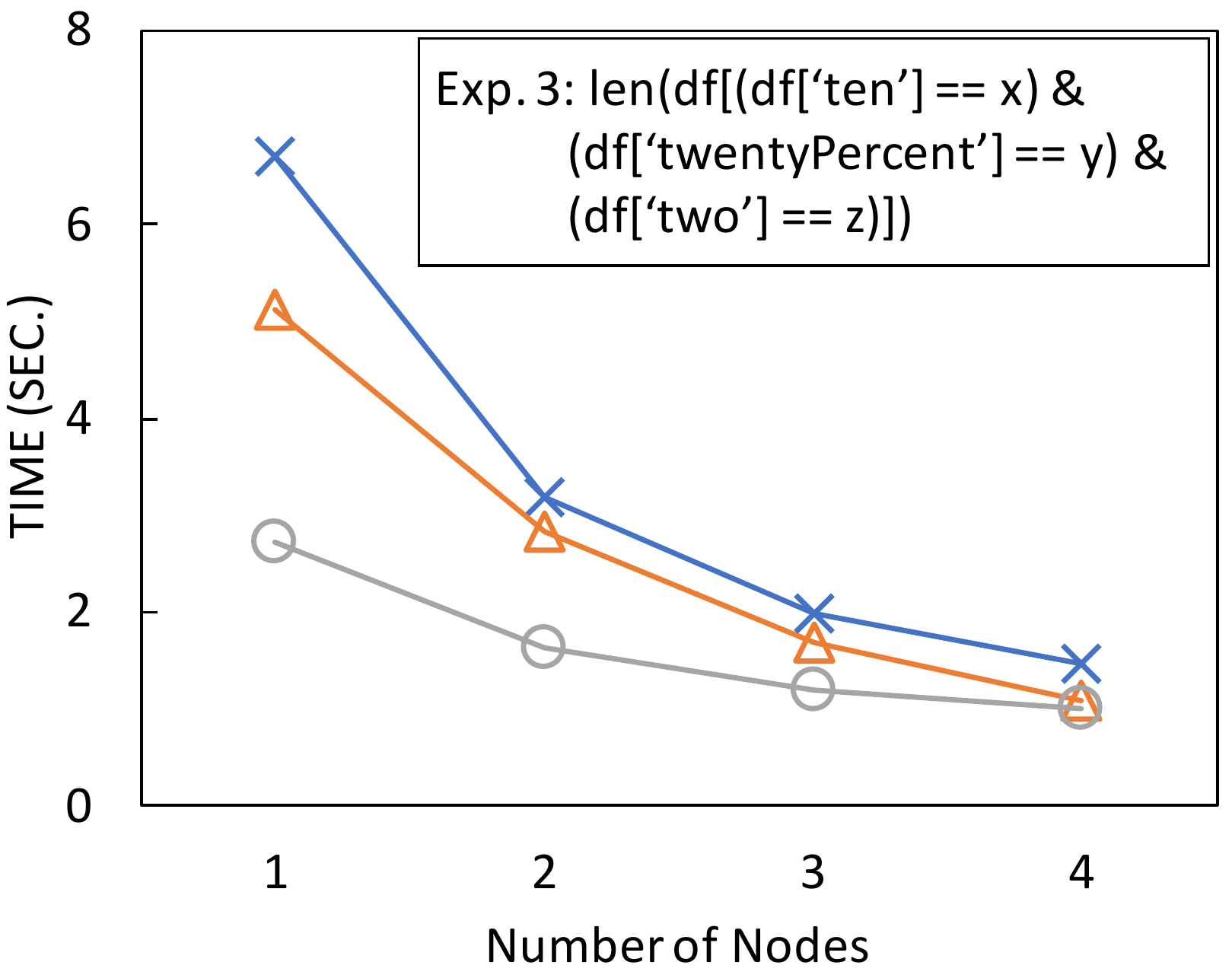}
        \caption{Expression 3}
        \label{fig:q3_speedup}
    \end{subfigure}
    \hfill
    \begin{subfigure}[t]{0.23\textwidth}
        \includegraphics[trim=0.5cm 1.5 0.5cm 1.5,width=\textwidth,height=3.3cm]{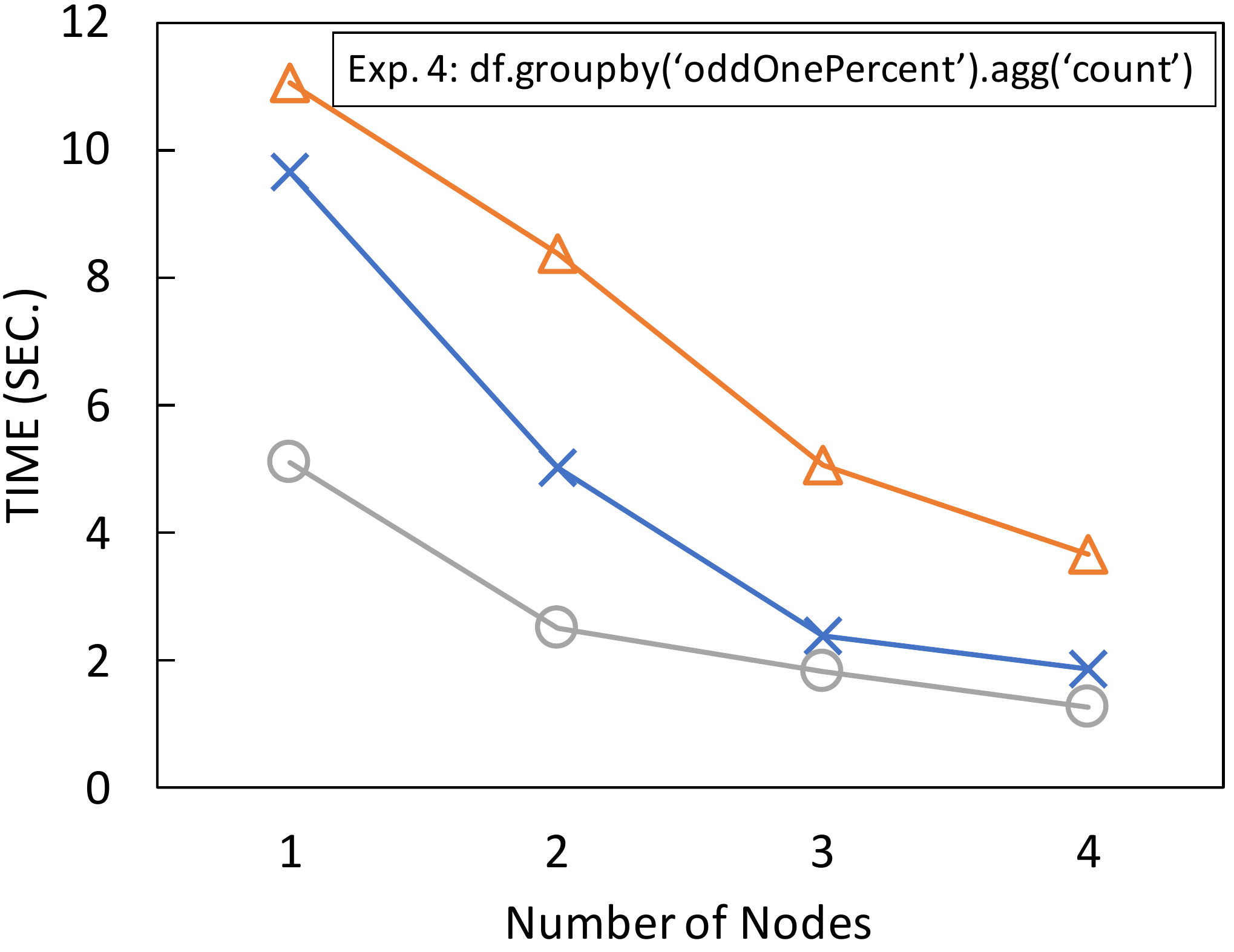}%
        \caption{Expression 4}
        \label{fig:q4_speedup}
    \end{subfigure}
    
    \begin{subfigure}[t]{0.24\textwidth}
        \includegraphics[trim=1.5 1.5 0cm 1.5,width=\textwidth,height=3.3cm]{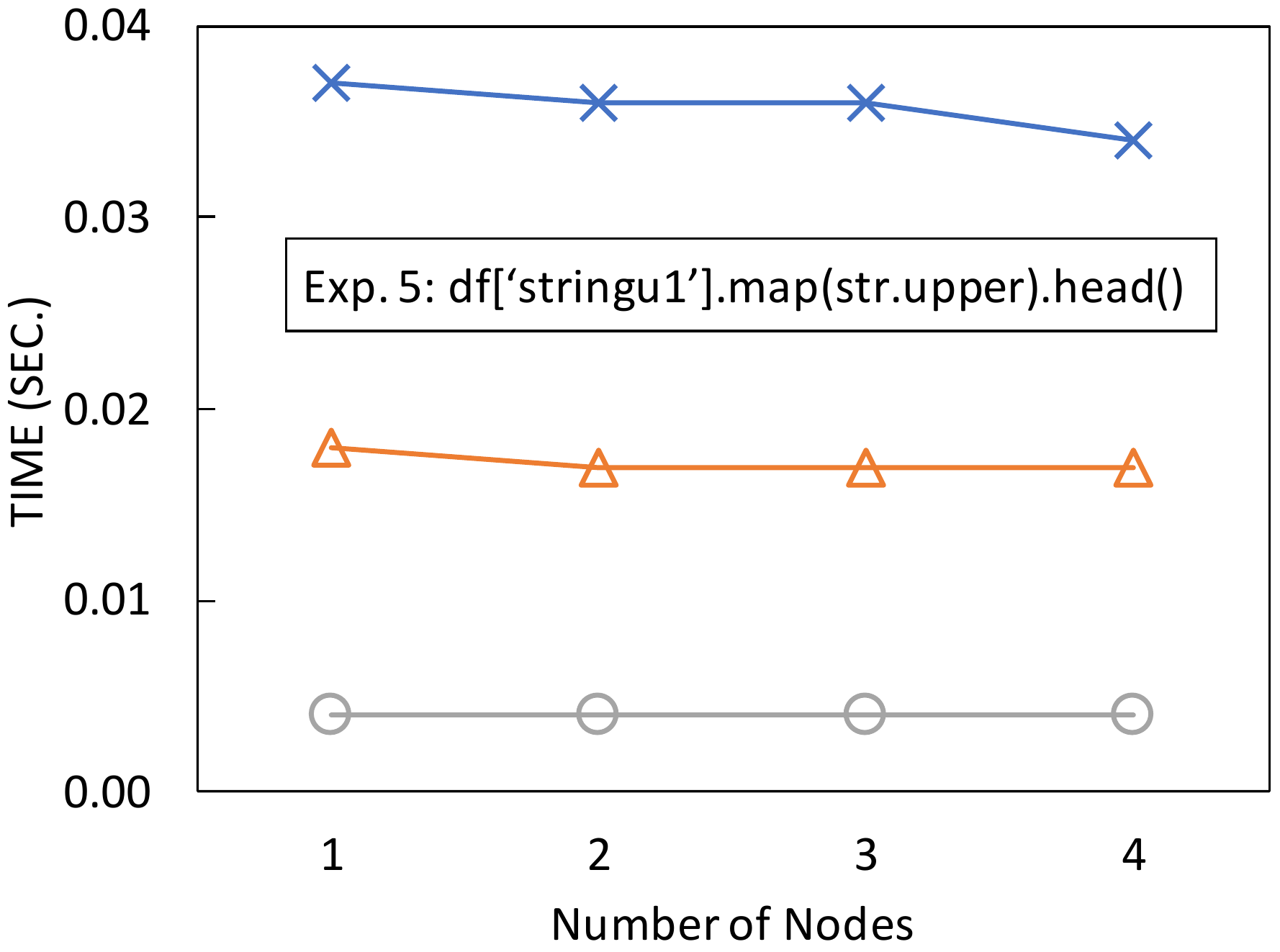}
        \caption{Expression 5}
        \label{fig:q5_speedup}
    \end{subfigure}
    \hfill
    \begin{subfigure}[t]{0.23\textwidth}
        \includegraphics[trim=0.5cm 1.5 0.5cm 1.5,width=\textwidth,height=3.3cm]{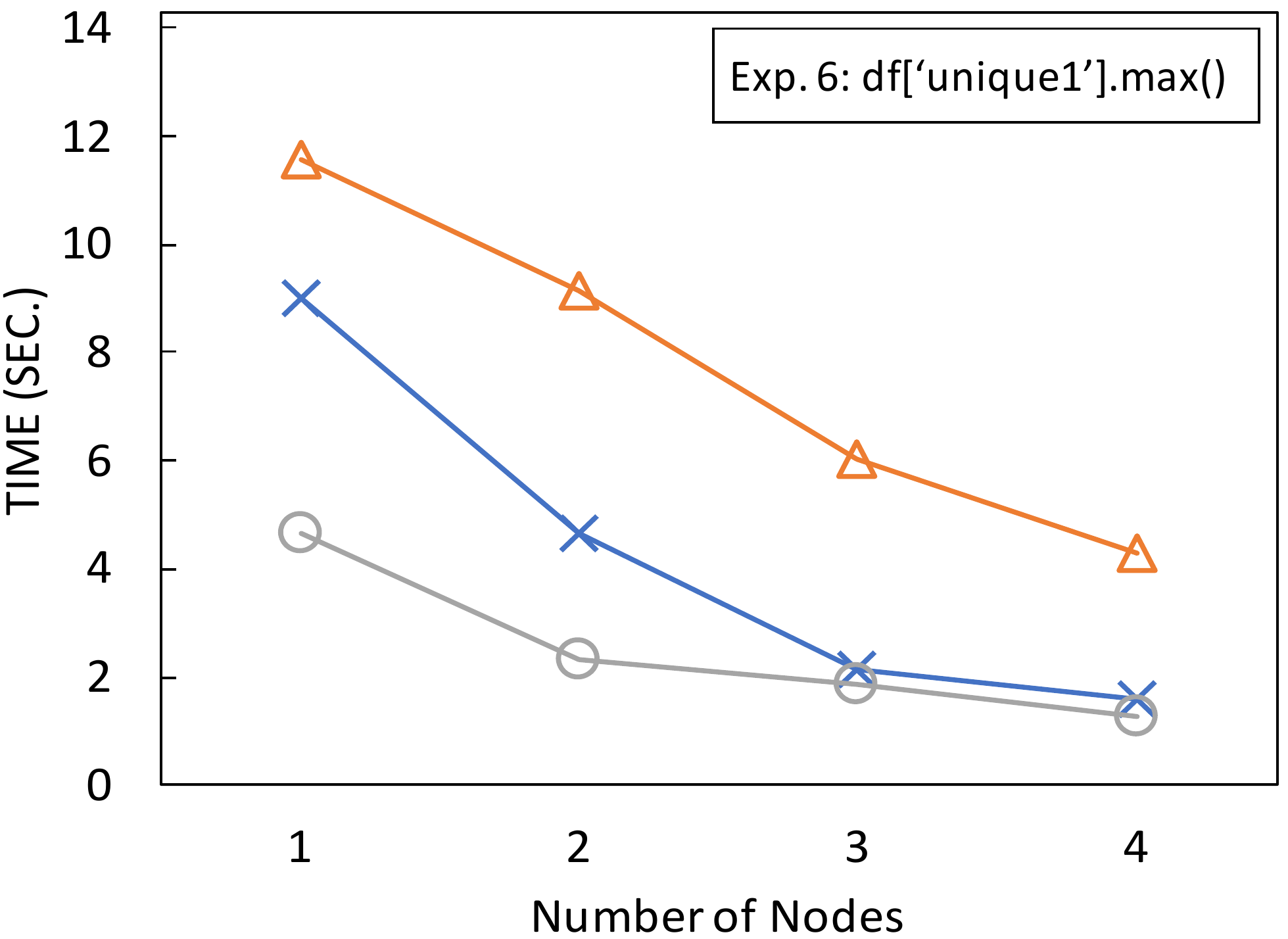}%
        \caption{Expression 6}
        \label{fig:q6_speedup}
    \end{subfigure}
    \hfill
    \begin{subfigure}[t]{0.24\textwidth}
        \includegraphics[trim=1.5 1.5 0cm 1.5,width=\textwidth,height=3.3cm]{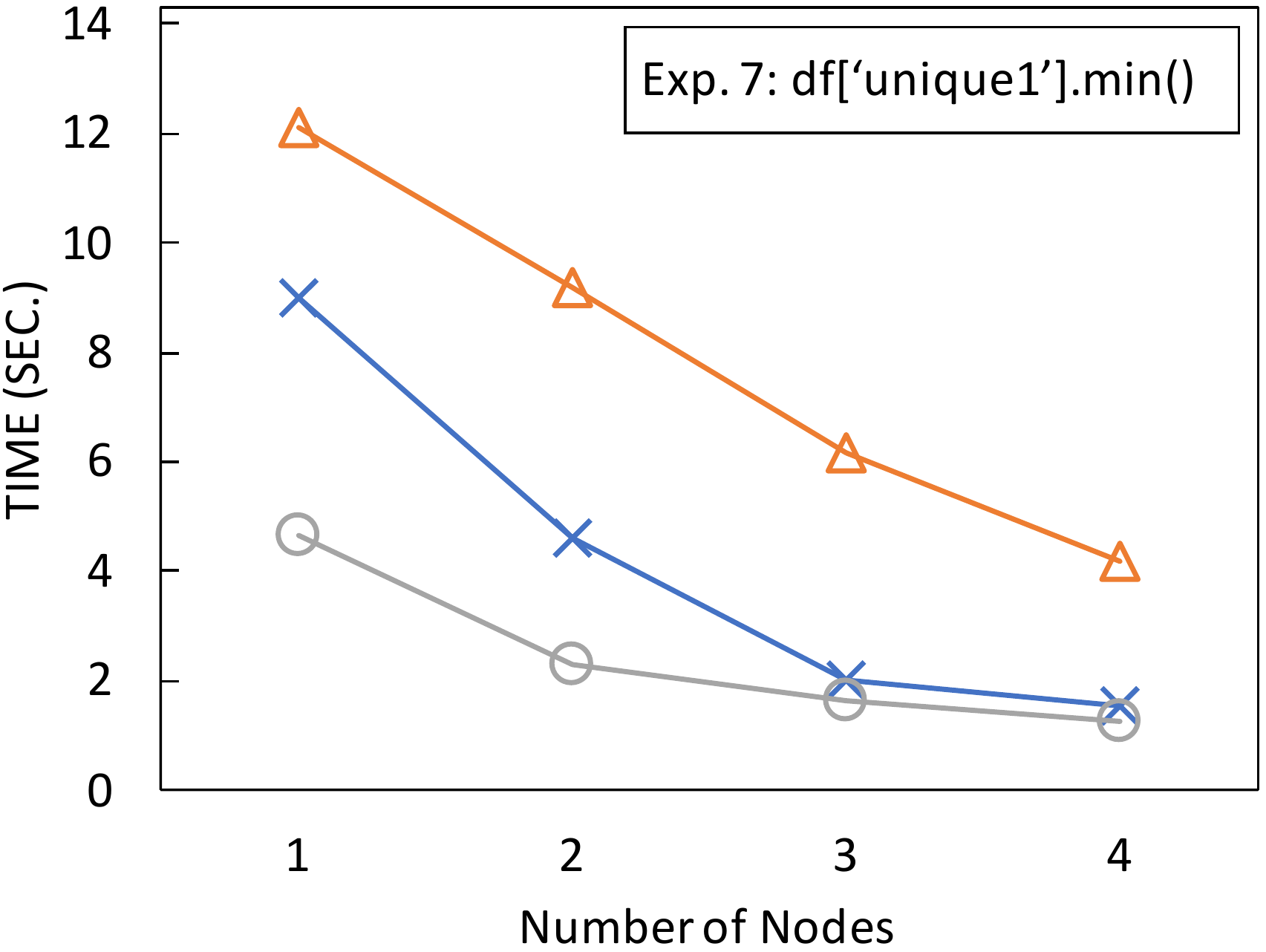}
        \caption{Expression 7}
        \label{fig:q7_speedup}
    \end{subfigure}
    \hfill
    \begin{subfigure}[t]{0.23\textwidth}
        \includegraphics[trim=0.5cm 1.5 0.5cm 1.5,width=\textwidth,height=3.3cm]{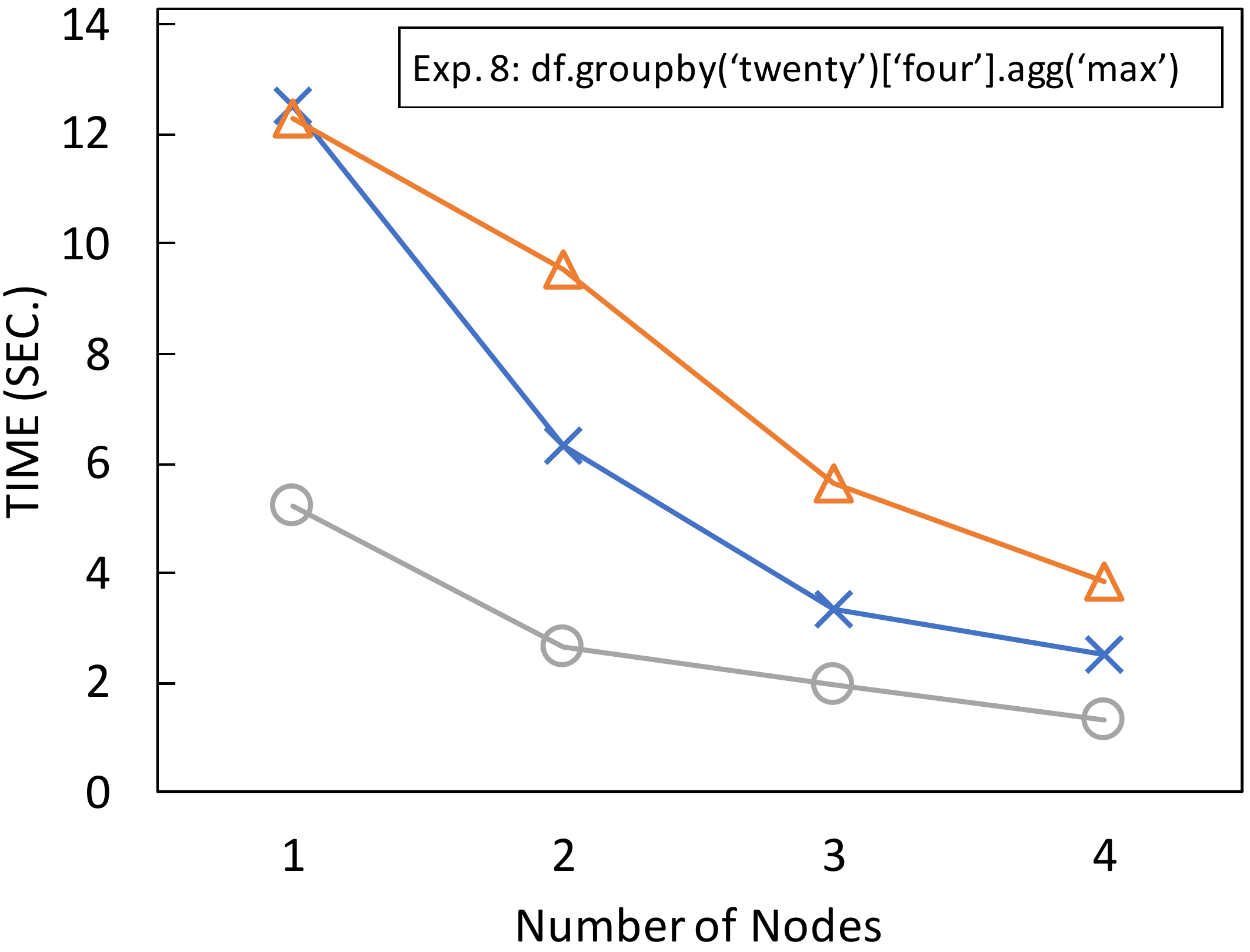}%
        \caption{Expression 8}
        \label{fig:q8_speedup}
    \end{subfigure}

    \begin{subfigure}[t]{0.24\textwidth}
        \includegraphics[trim=1.5 1.5 0cm 1.5,width=\textwidth,height=3.3cm]{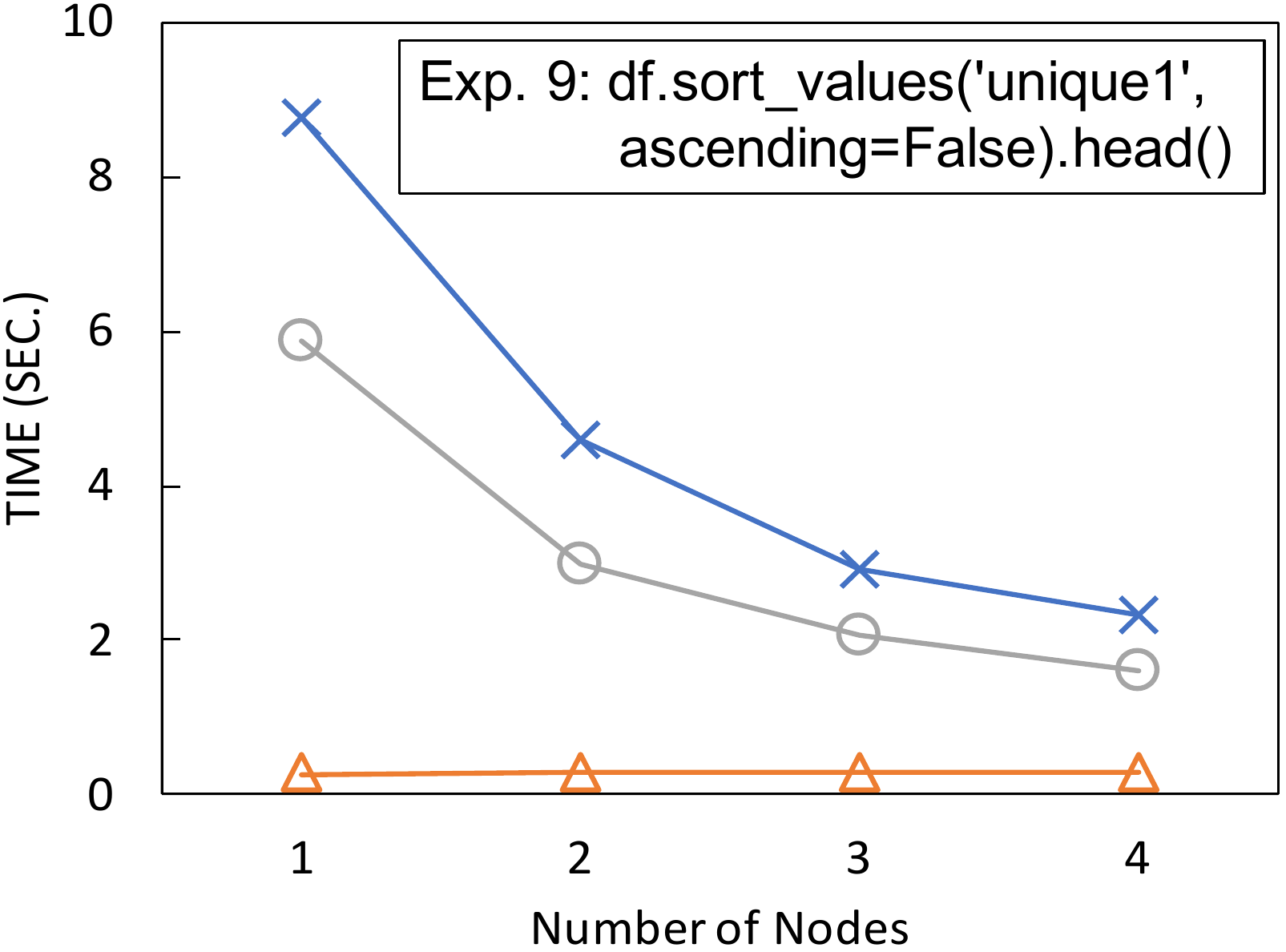}
        \caption{Expression 9}
        \label{fig:q9_speedup}
    \end{subfigure}
    \hfill
    \begin{subfigure}[t]{0.23\textwidth}
        \includegraphics[trim=0.5cm 1.5 0.5cm 1.5,width=\textwidth,height=3.3cm]{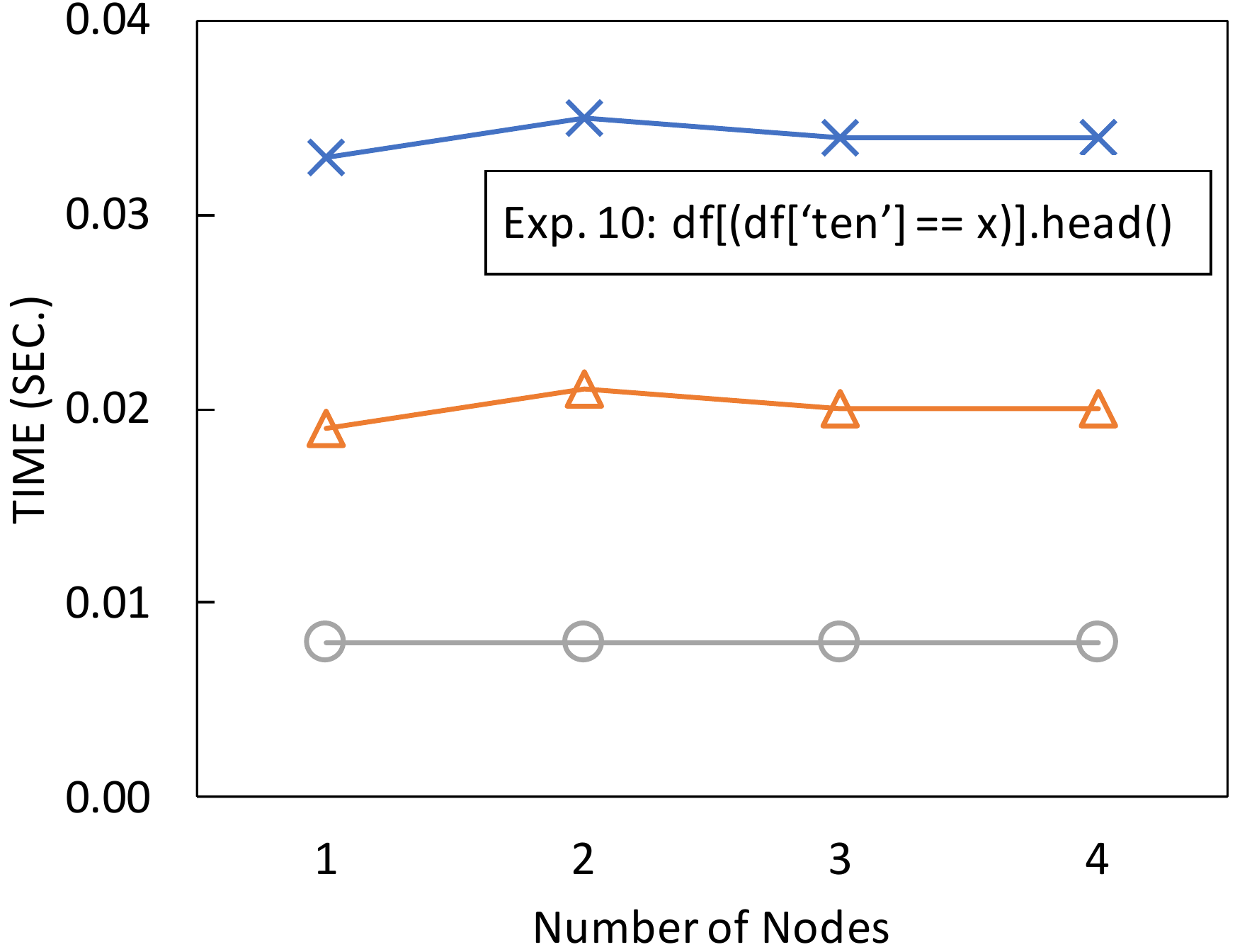}%
        \caption{Expression 10}
        \label{fig:q10_speedup}
    \end{subfigure}
    \hfill
    \begin{subfigure}[t]{0.24\textwidth}
        \includegraphics[trim=1.5 1.5 0cm 1.5,width=\textwidth,height=3.3cm]{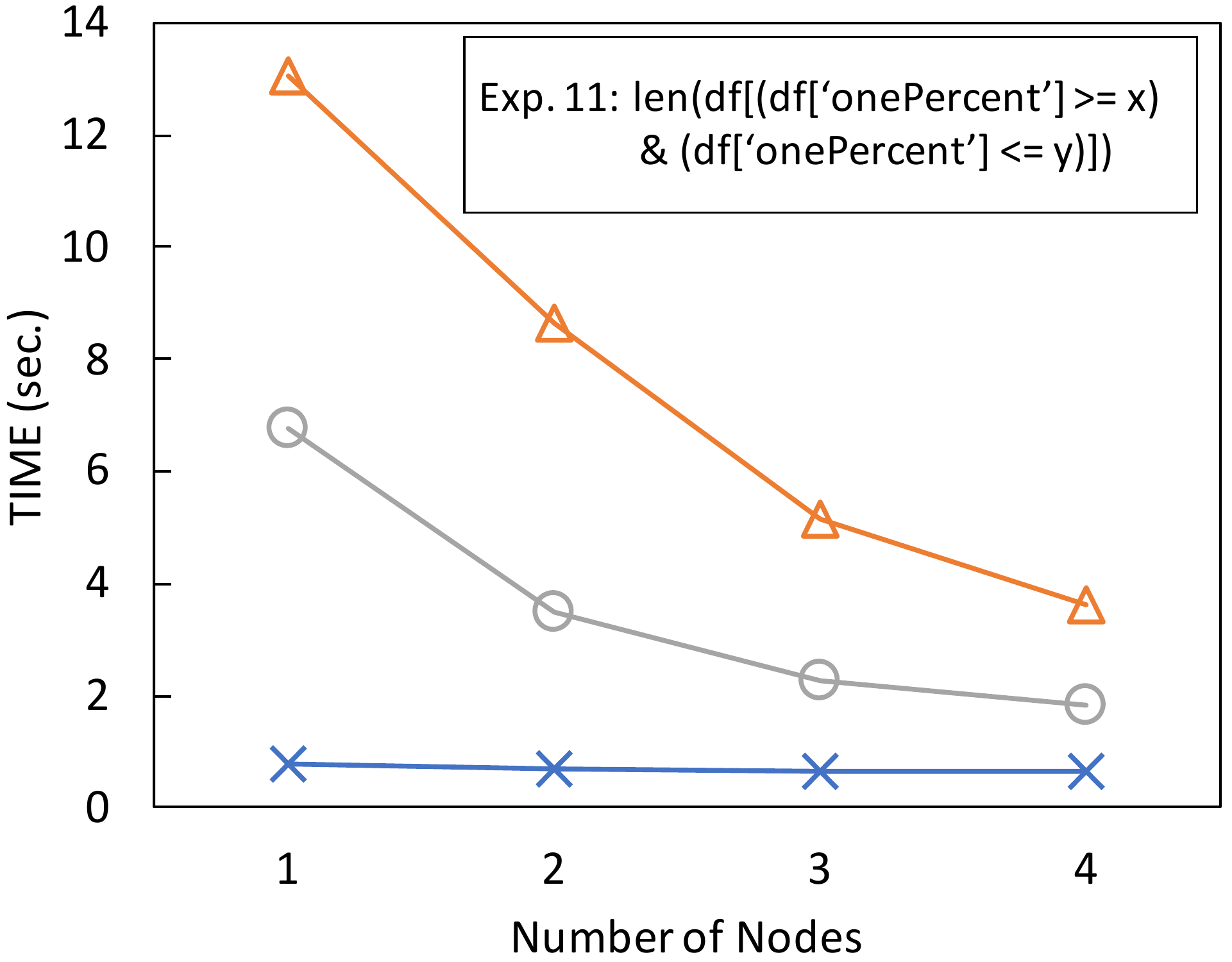}
        \caption{Expression 11}
        \label{fig:q11_speedup}
    \end{subfigure}
    \hfill
    \begin{subfigure}[t]{0.23\textwidth}
        \includegraphics[trim=0.5cm 1.5 0.5cm 1.5,width=\textwidth,height=3.3cm]{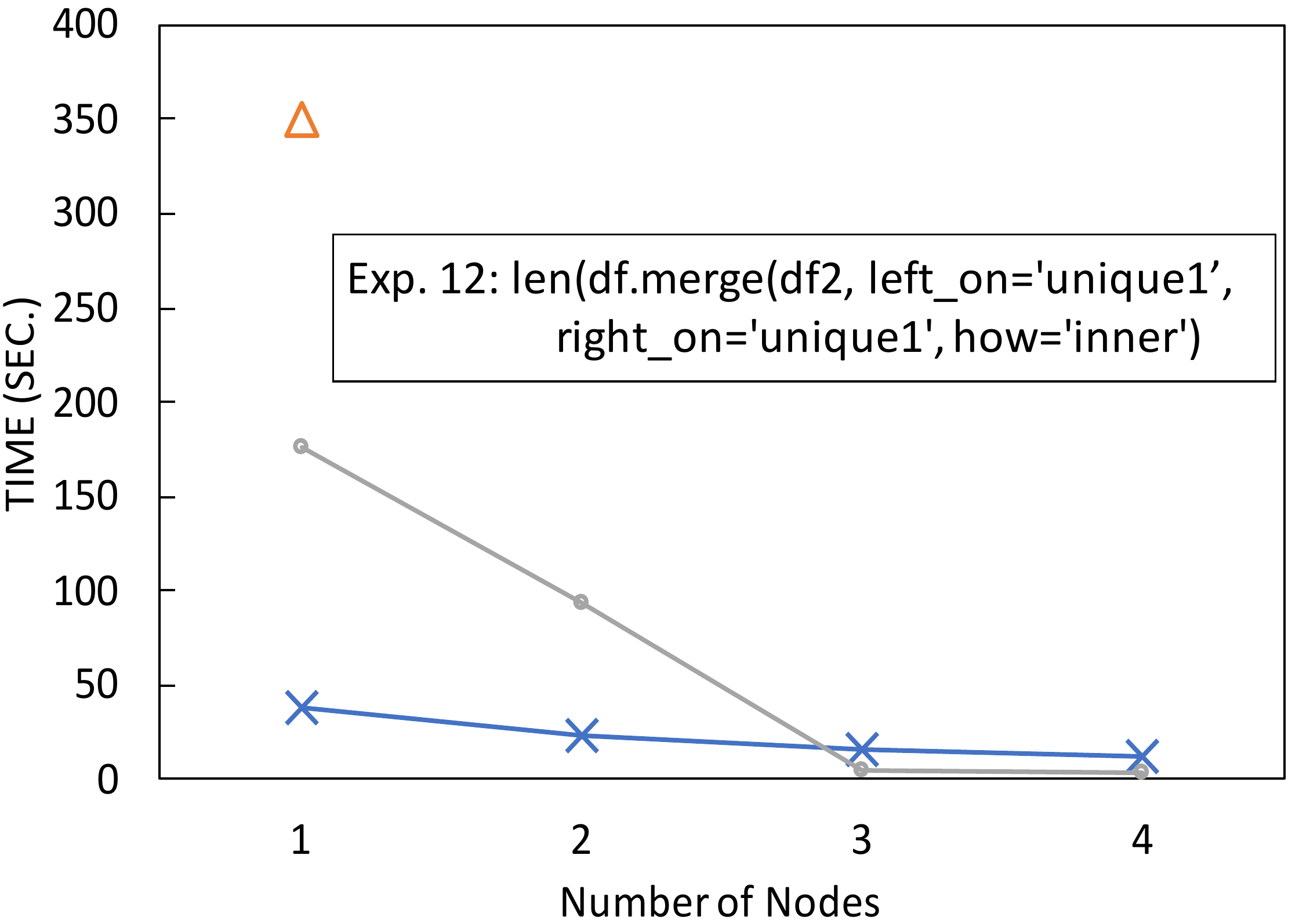}%
        \caption{Expression 12}
        \label{fig:q12_speedup}
    \end{subfigure}
    
    % \hfill
    \begin{subfigure}[t]{0.23\textwidth}
        \includegraphics[trim=0.5cm 1.5 0.5cm 1.5,width=\textwidth,height=3.3cm]{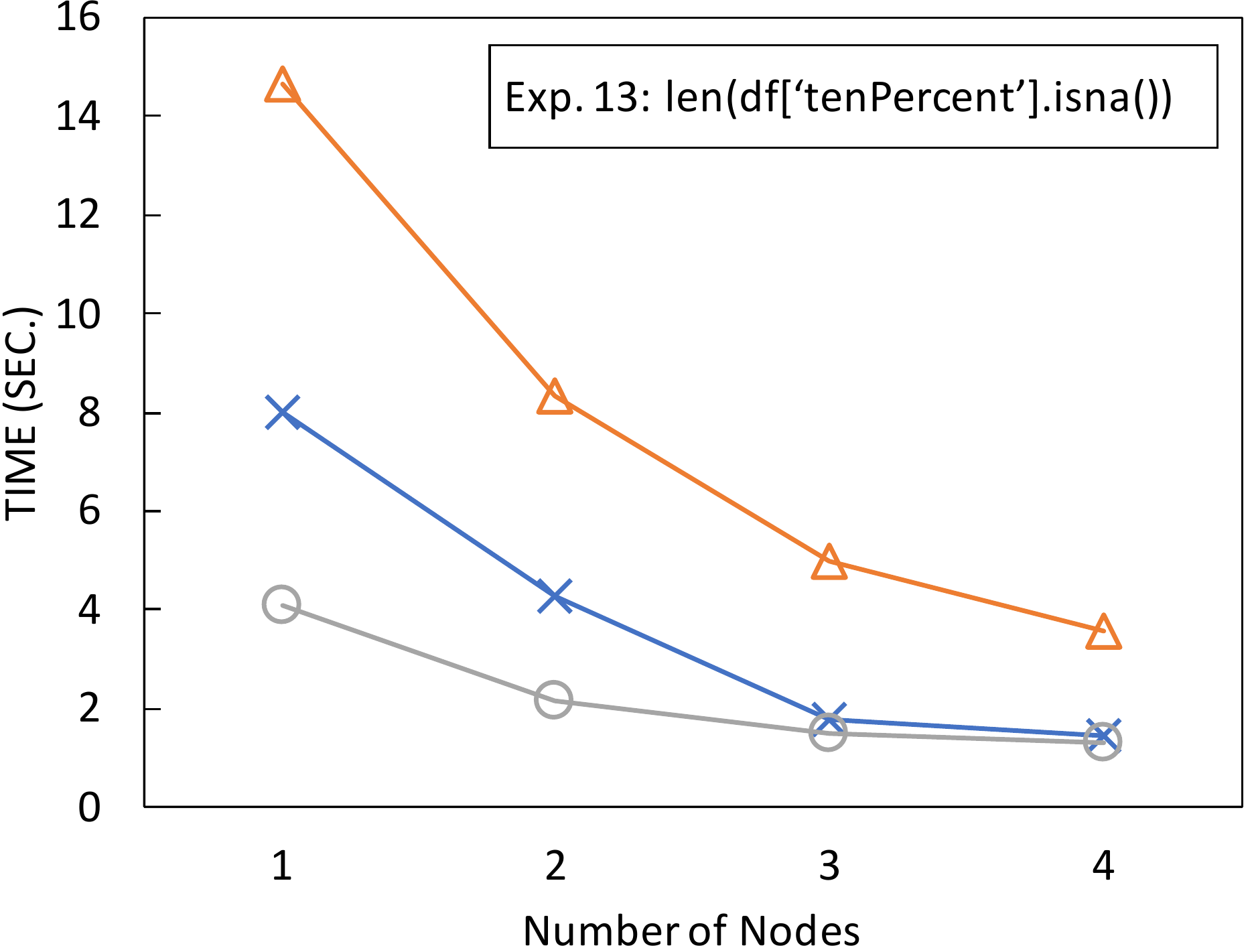}%
        \caption{Expression 13}
        \label{fig:q13_speedup}
    \end{subfigure}
    \caption{Speedup Evaluation Results}
    \label{fig:speedup_results}
\end{figure*}

\begin{figure*}[h!]
     \centering
     \begin{subfigure}[t]{0.44\textwidth}
        \includegraphics[width=\textwidth,height=0.4cm]{figures/cluster_legend.pdf}
    \end{subfigure}
    \hspace{15cm}
    \begin{subfigure}[t]{0.24\textwidth}
        \includegraphics[trim=1.5 1.5 0cm 1.5,width=\textwidth,height=3.3cm]{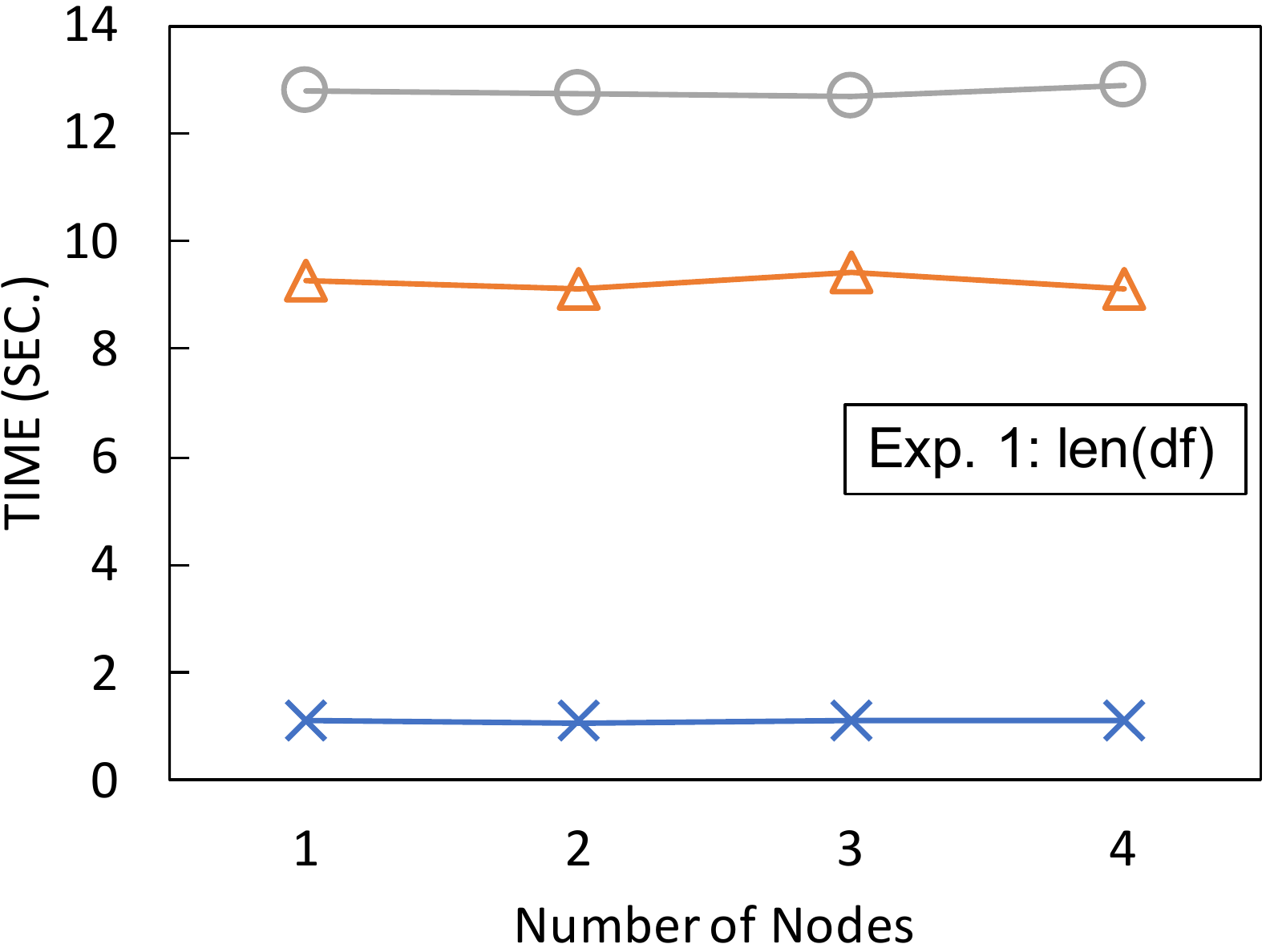}
        \caption{Expression 1}
        \label{fig:q1_scaleup}
    \end{subfigure}
    \hfill
    \begin{subfigure}[t]{0.23\textwidth}
        \includegraphics[trim=0.5cm 1.5 0.5cm 1.5,width=\textwidth,height=3.3cm]{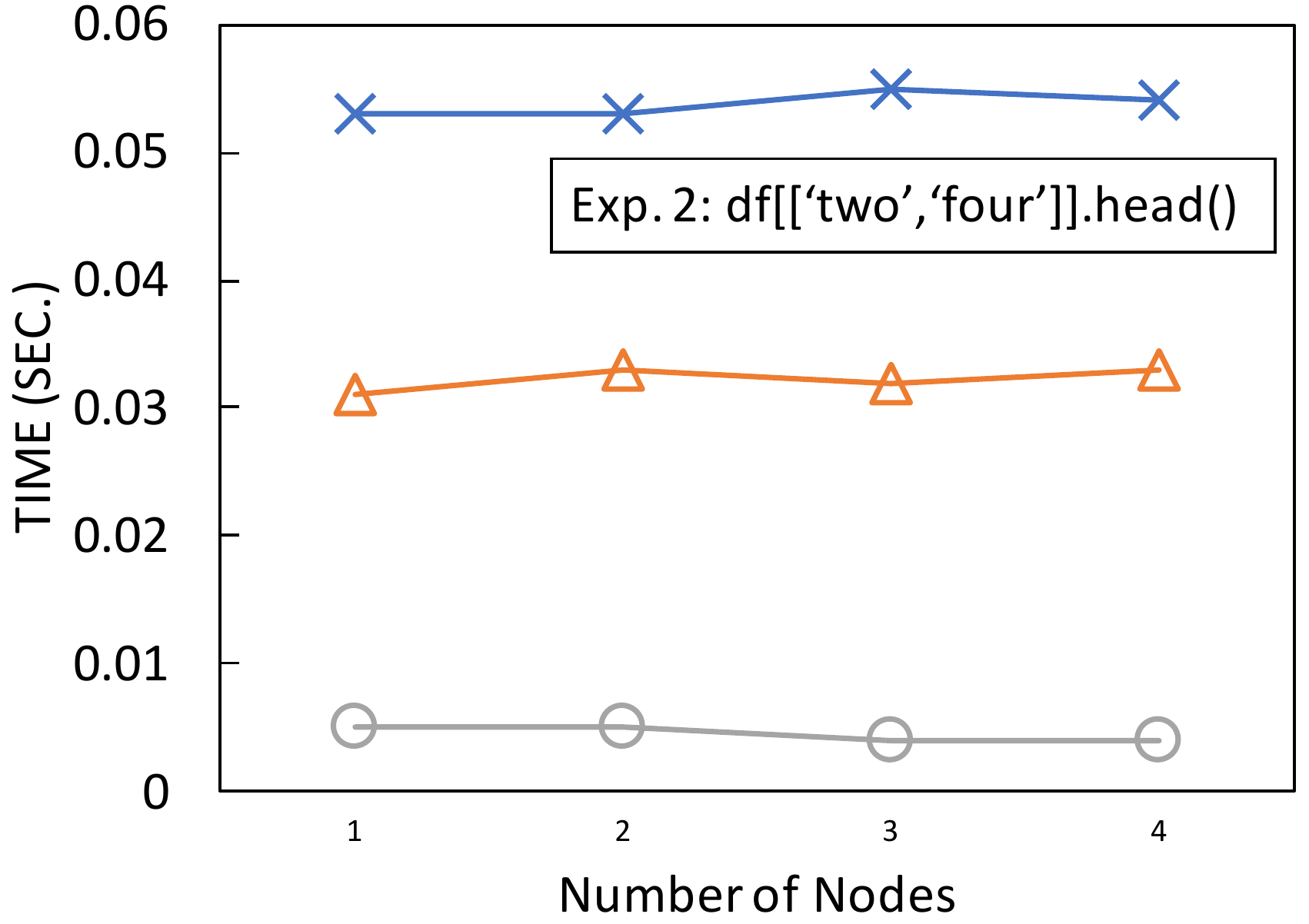}%
        \caption{Expression 2}
        \label{fig:q2_scaleup}
    \end{subfigure}
    \hfill
    \begin{subfigure}[t]{0.24\textwidth}
        \includegraphics[trim=1.5 1.5 0cm 1.5,width=\textwidth,height=3.3cm]{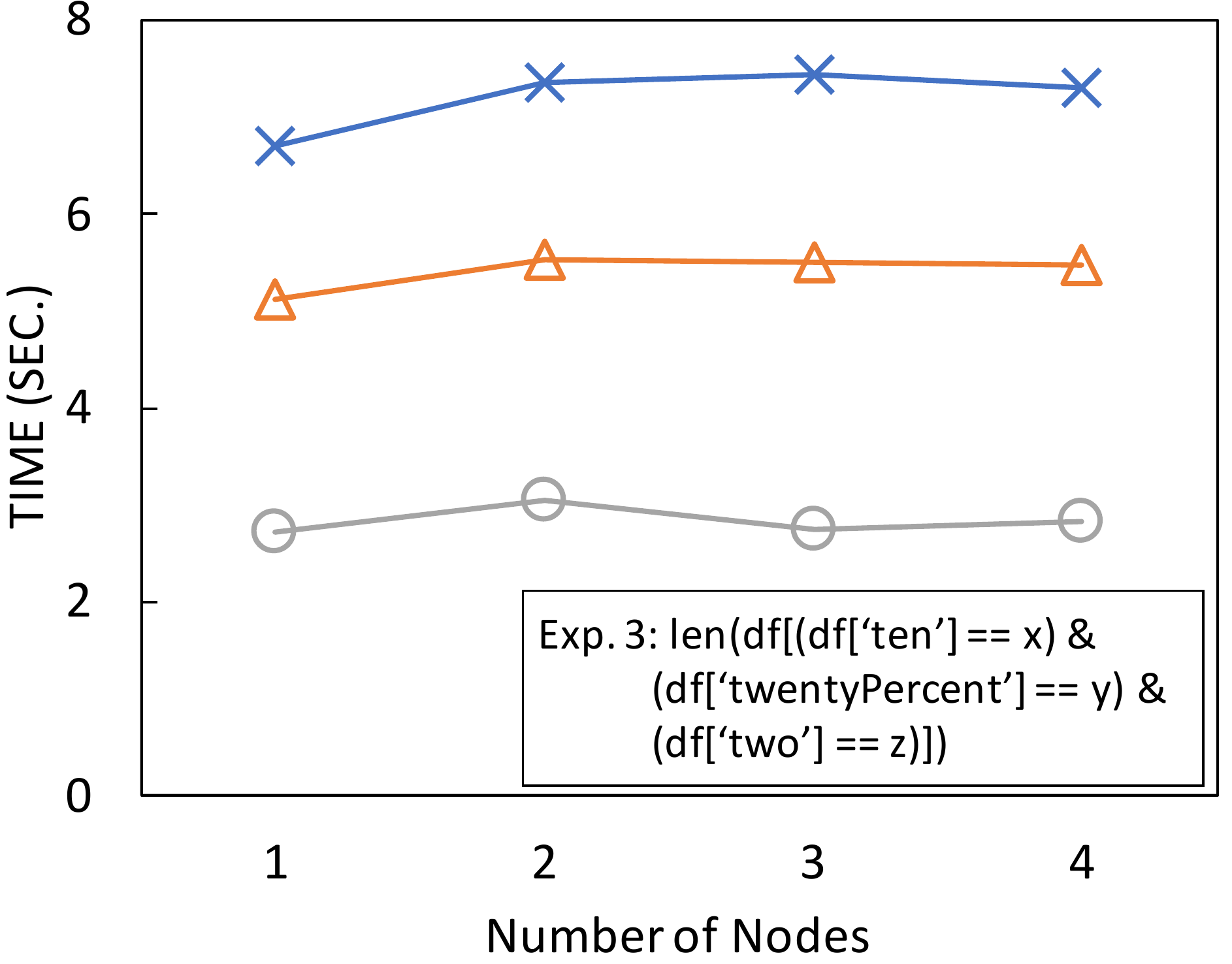}
        \caption{Expression 3}
        \label{fig:q3_scaleup}
    \end{subfigure}
    \hfill
    \begin{subfigure}[t]{0.23\textwidth}
        \includegraphics[trim=0.5cm 1.5 0.5cm 1.5,width=\textwidth,height=3.3cm]{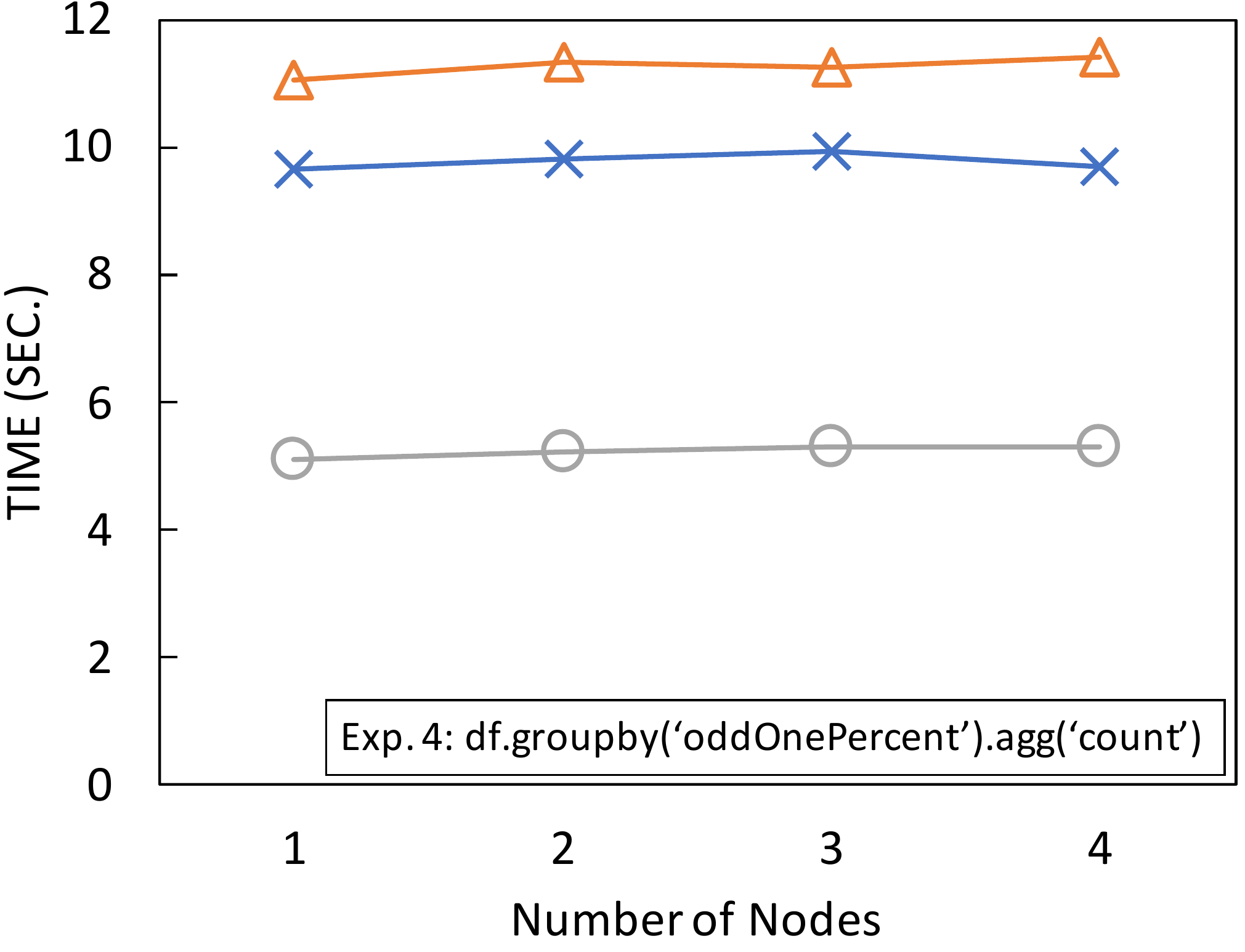}%
        \caption{Expression 4}
        \label{fig:q4_scaleup}
    \end{subfigure}
    \hfill

    \begin{subfigure}[t]{0.24\textwidth}
        \includegraphics[trim=1.5 1.5 0cm 1.5,width=\textwidth,height=3.3cm]{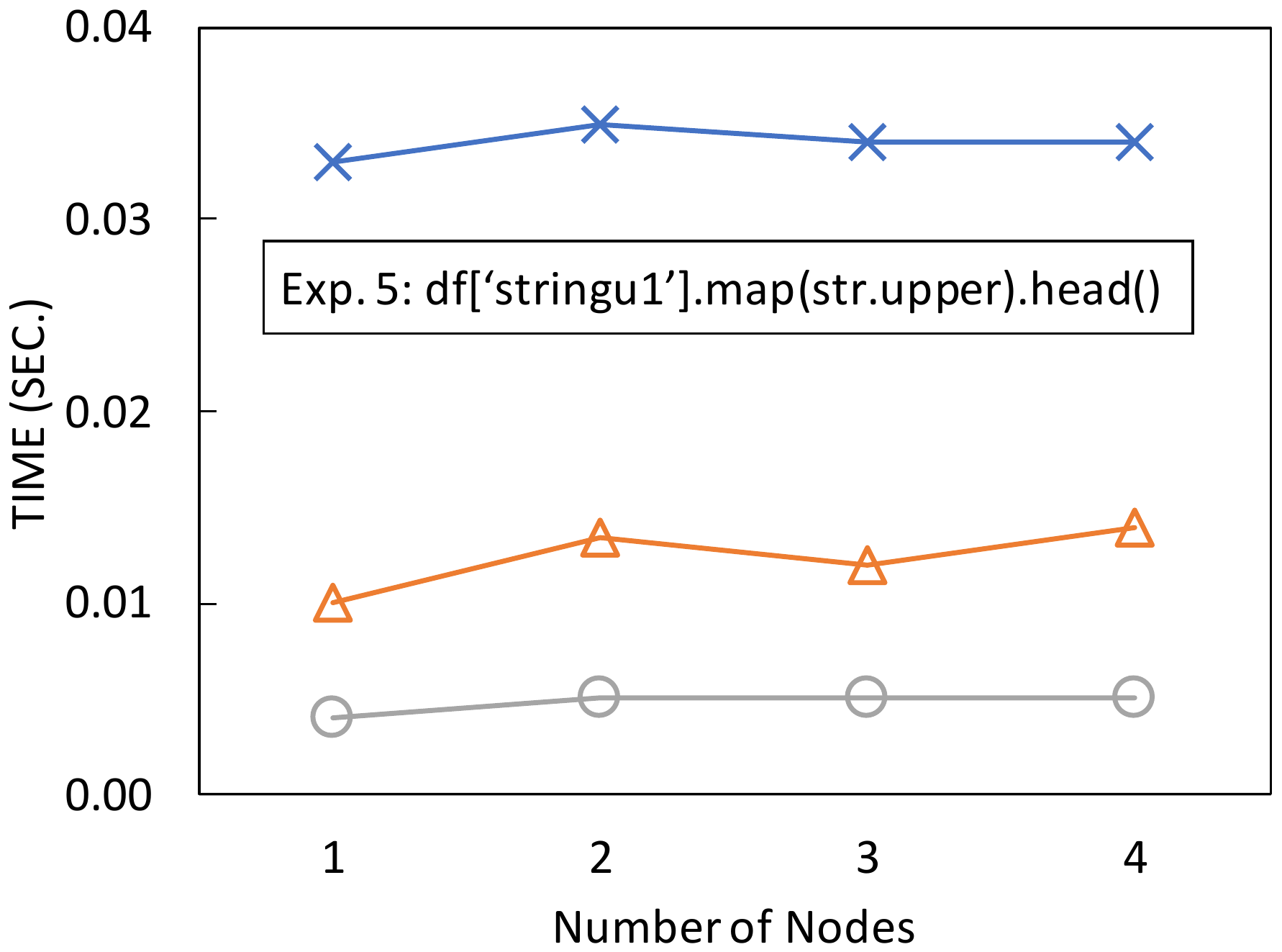}
        \caption{Expression 5}
        \label{fig:q5_scaleup}
    \end{subfigure}
    \hfill
    \begin{subfigure}[t]{0.23\textwidth}
        \includegraphics[trim=0.5cm 1.5 0.5cm 1.5,width=\textwidth,height=3.3cm]{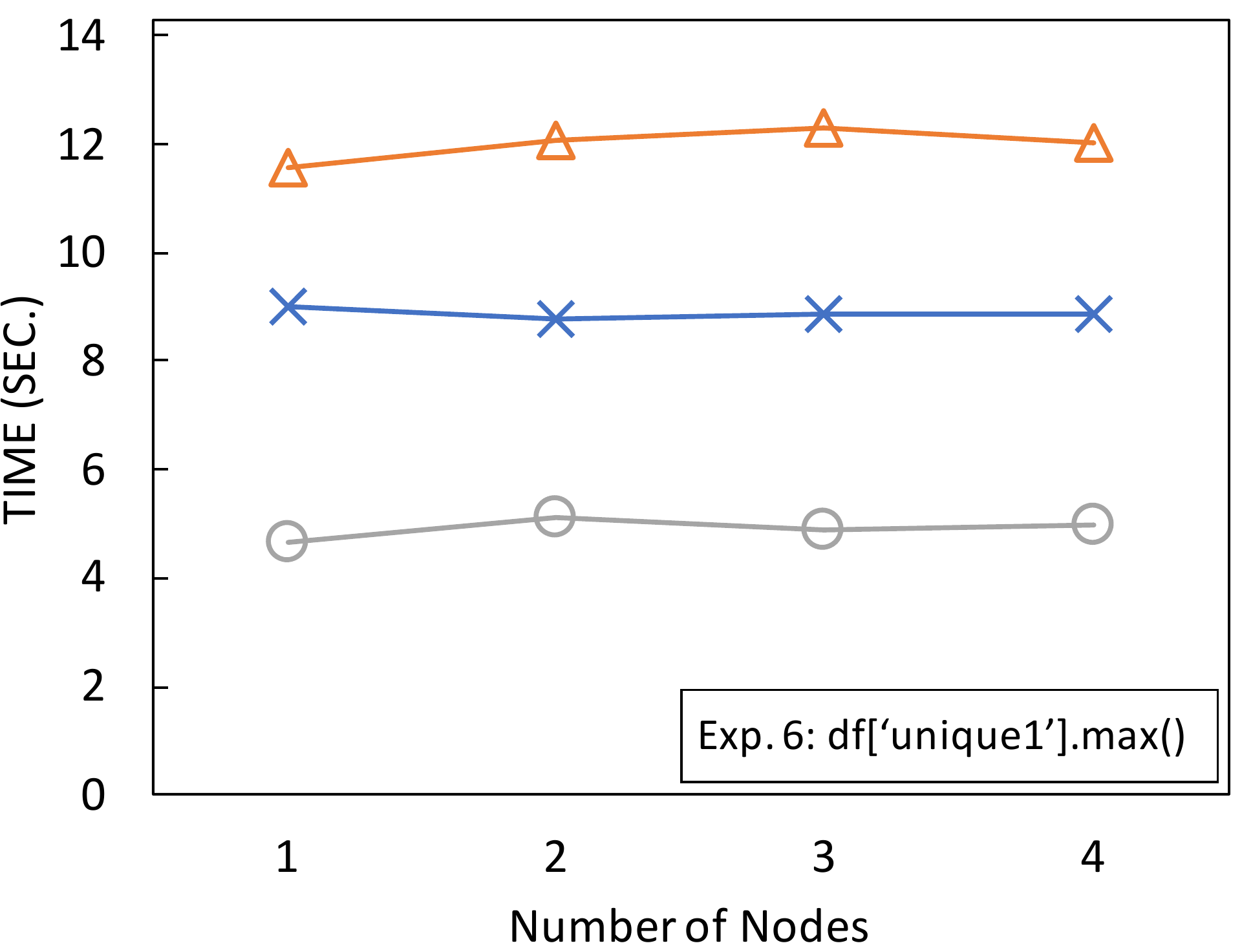}%
        \caption{Expression 6}
        \label{fig:q6_scaleup}
    \end{subfigure}
    \hfill
    \begin{subfigure}[t]{0.24\textwidth}
        \includegraphics[trim=1.5 1.5 0cm 1.5,width=\textwidth,height=3.3cm]{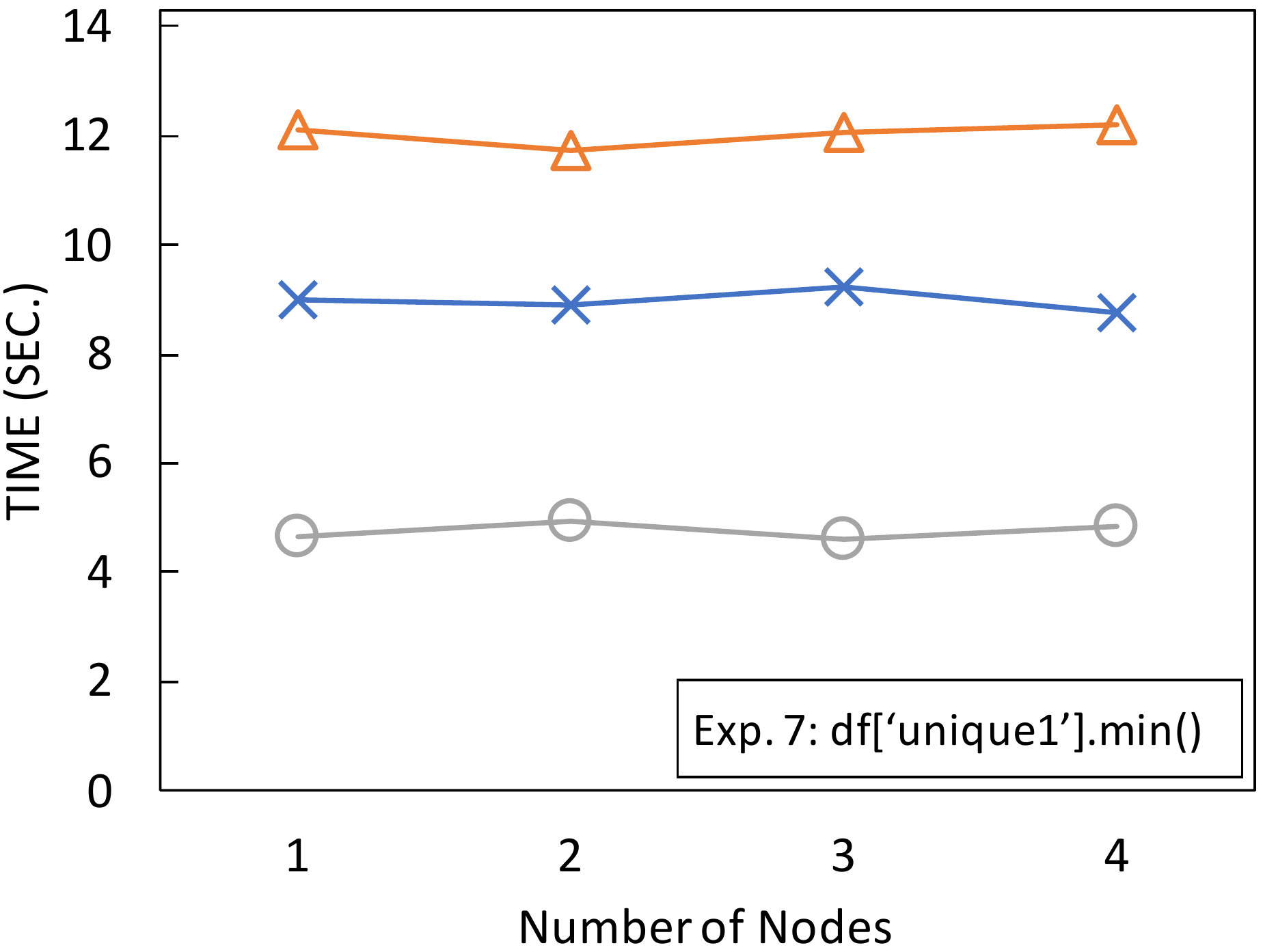}
        \caption{Expression 7}
        \label{fig:q7_scaleup}
    \end{subfigure}
    \hfill
    \begin{subfigure}[t]{0.23\textwidth}
        \includegraphics[trim=0.5cm 1.5 0.5cm 1.5,width=\textwidth,height=3.3cm]{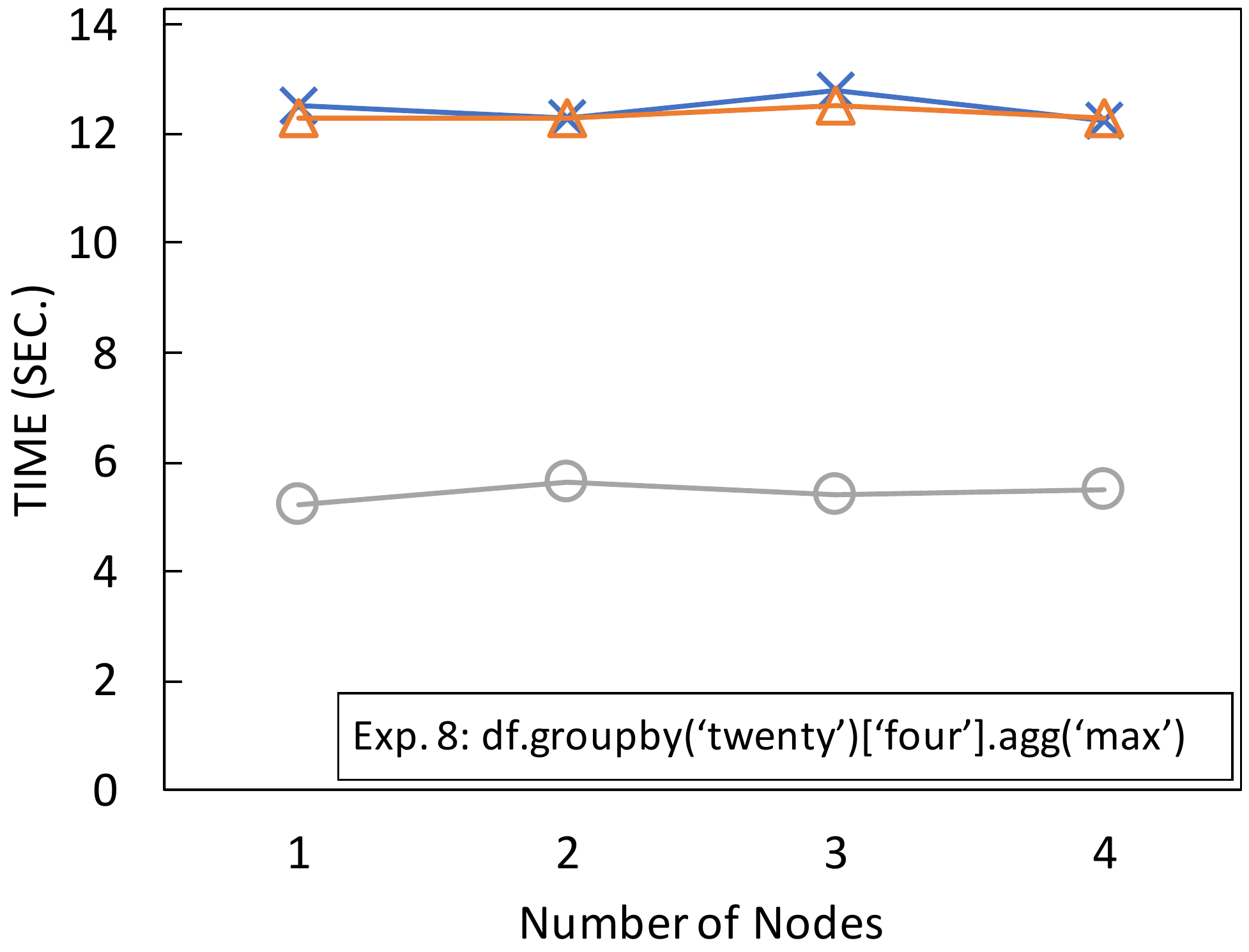}%
        \caption{Expression 8}
        \label{fig:q8_scaleup}
    \end{subfigure}

    \begin{subfigure}[t]{0.24\textwidth}
        \includegraphics[trim=1.5 1.5 0cm 1.5,width=\textwidth,height=3.3cm]{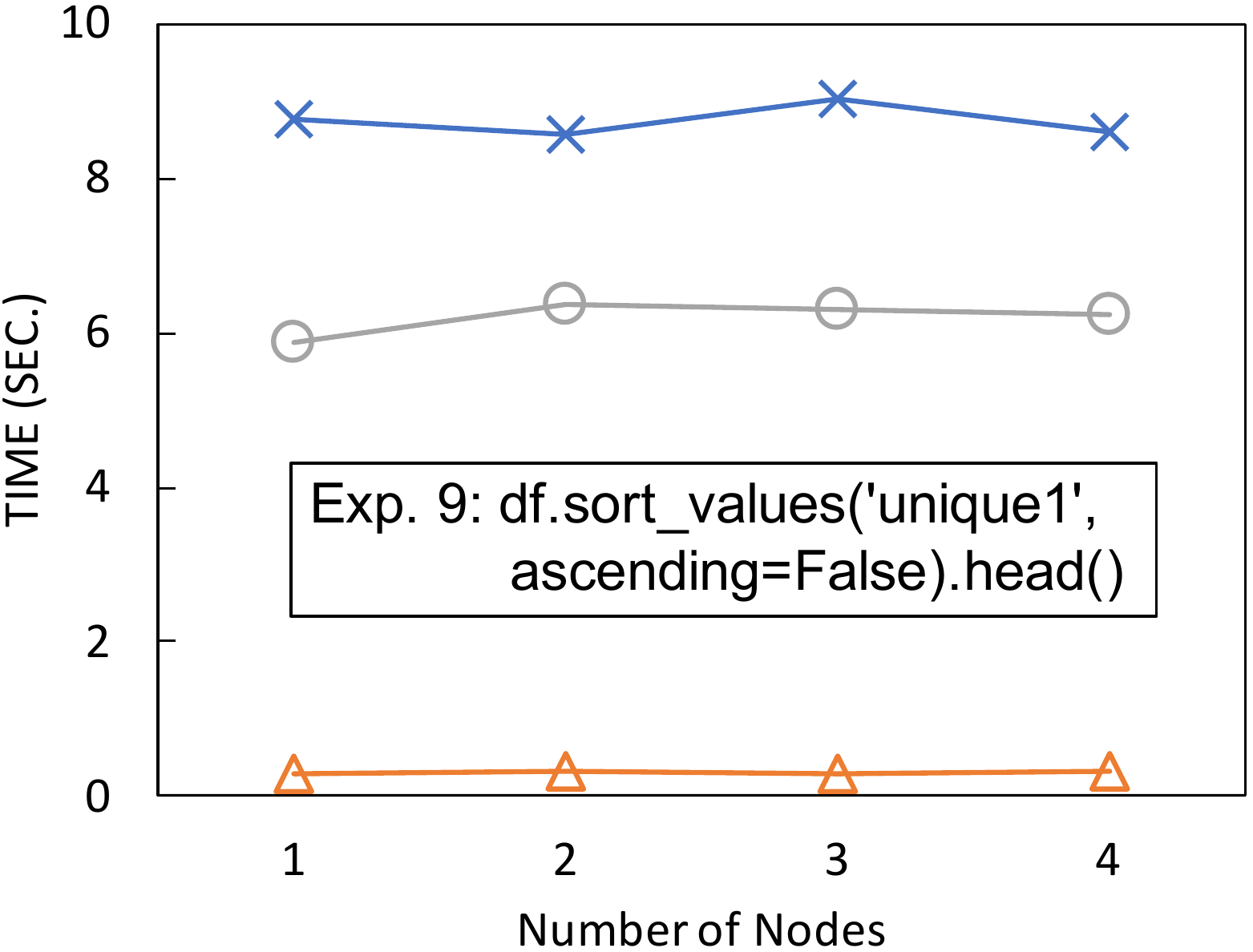}
        \caption{Expression 9}
        \label{fig:q9_scaleup}
    \end{subfigure}
    \hfill
    \begin{subfigure}[t]{0.23\textwidth}
        \includegraphics[trim=0.5cm 1.5 0.5cm 1.5,width=\textwidth,height=3.3cm]{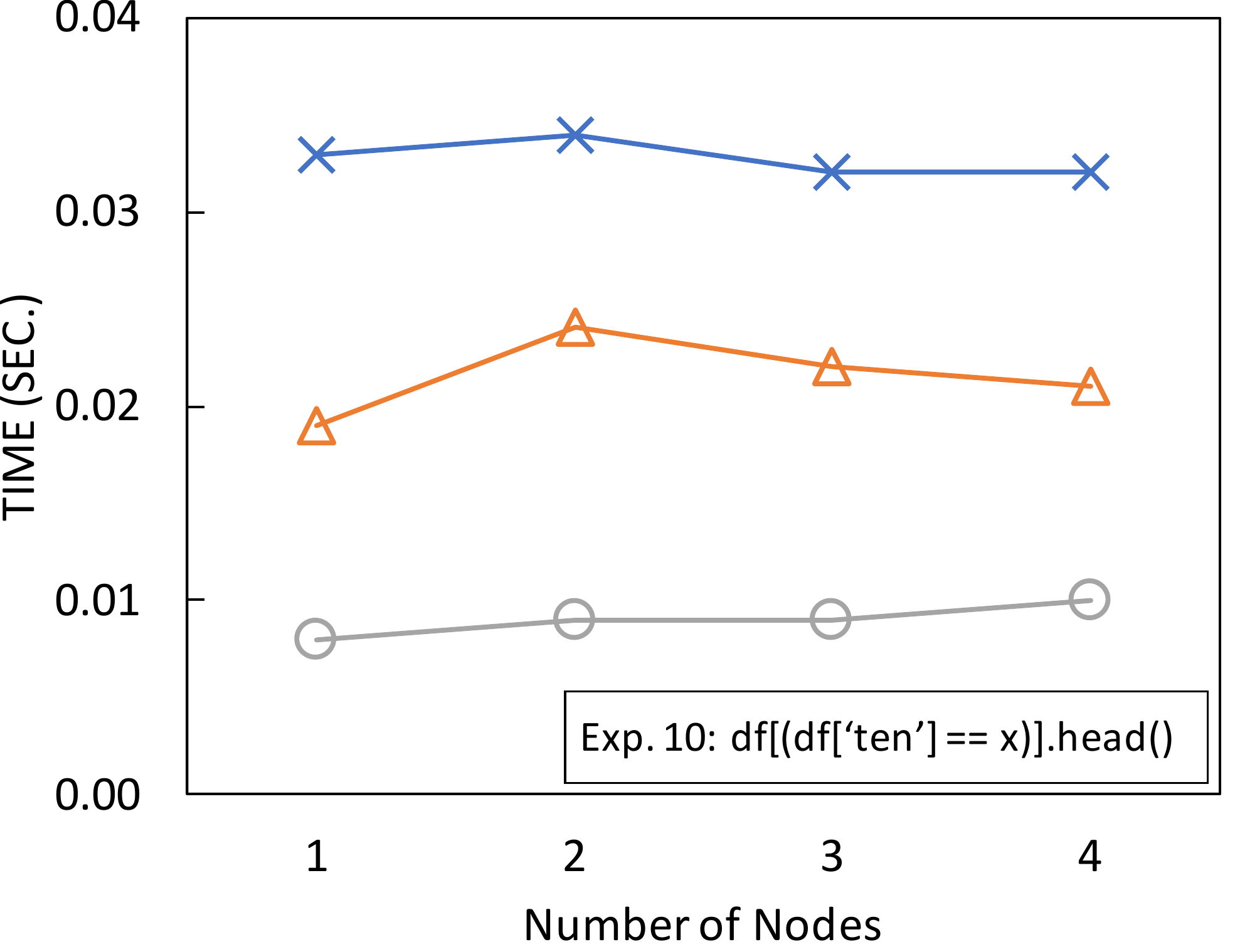}%
        \caption{Expression 10}
        \label{fig:q10_scaleup}
    \end{subfigure}
    \hfill
    \begin{subfigure}[t]{0.24\textwidth}
        \includegraphics[trim=1.5 1.5 0cm 1.5,width=\textwidth,height=3.3cm]{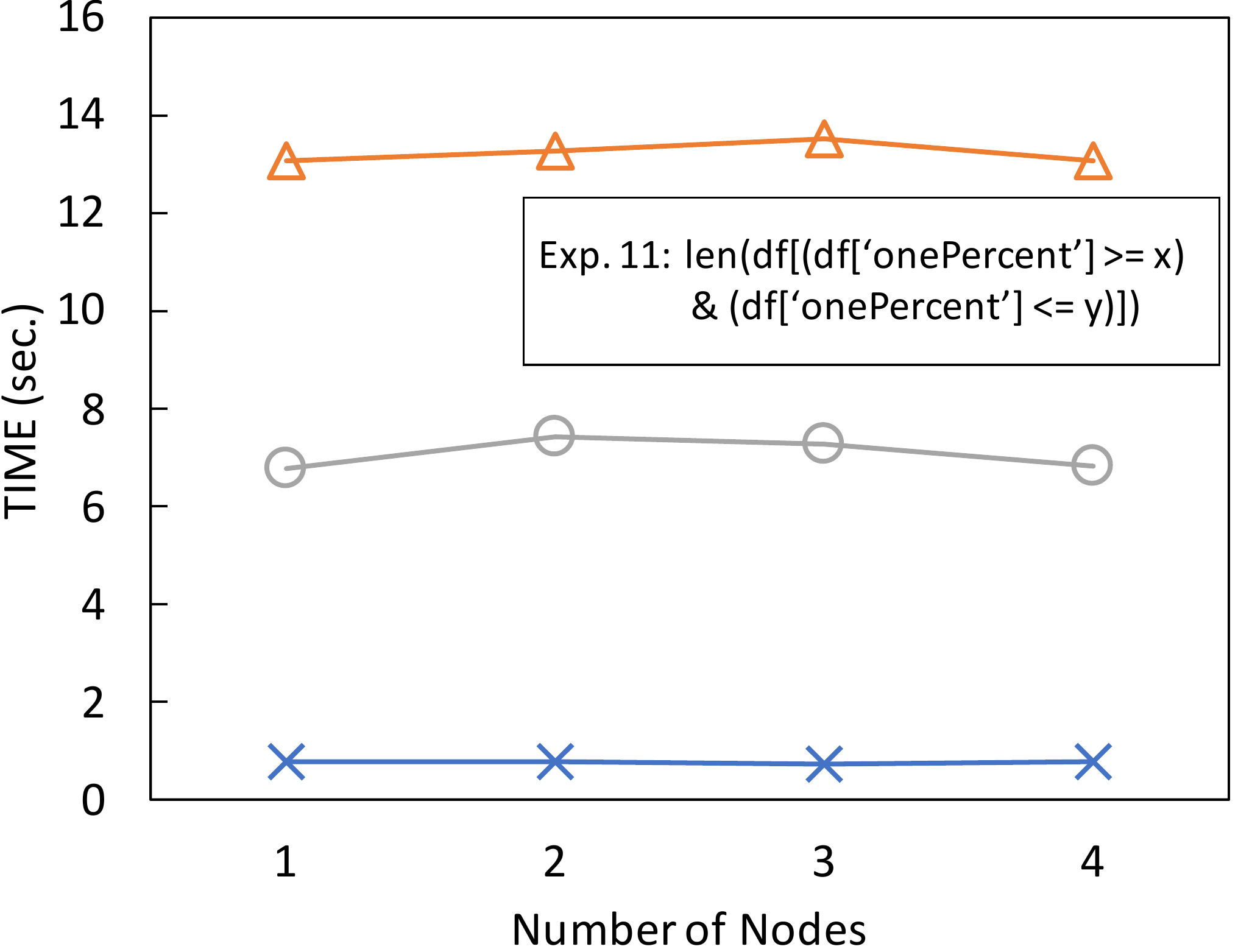}
        \caption{Expression 11}
        \label{fig:q11_scaleup}
    \end{subfigure}
    \hfill
    \begin{subfigure}[t]{0.23\textwidth}
        \includegraphics[trim=0.5cm 1.5 0.5cm 1.5,width=\textwidth,height=3.3cm]{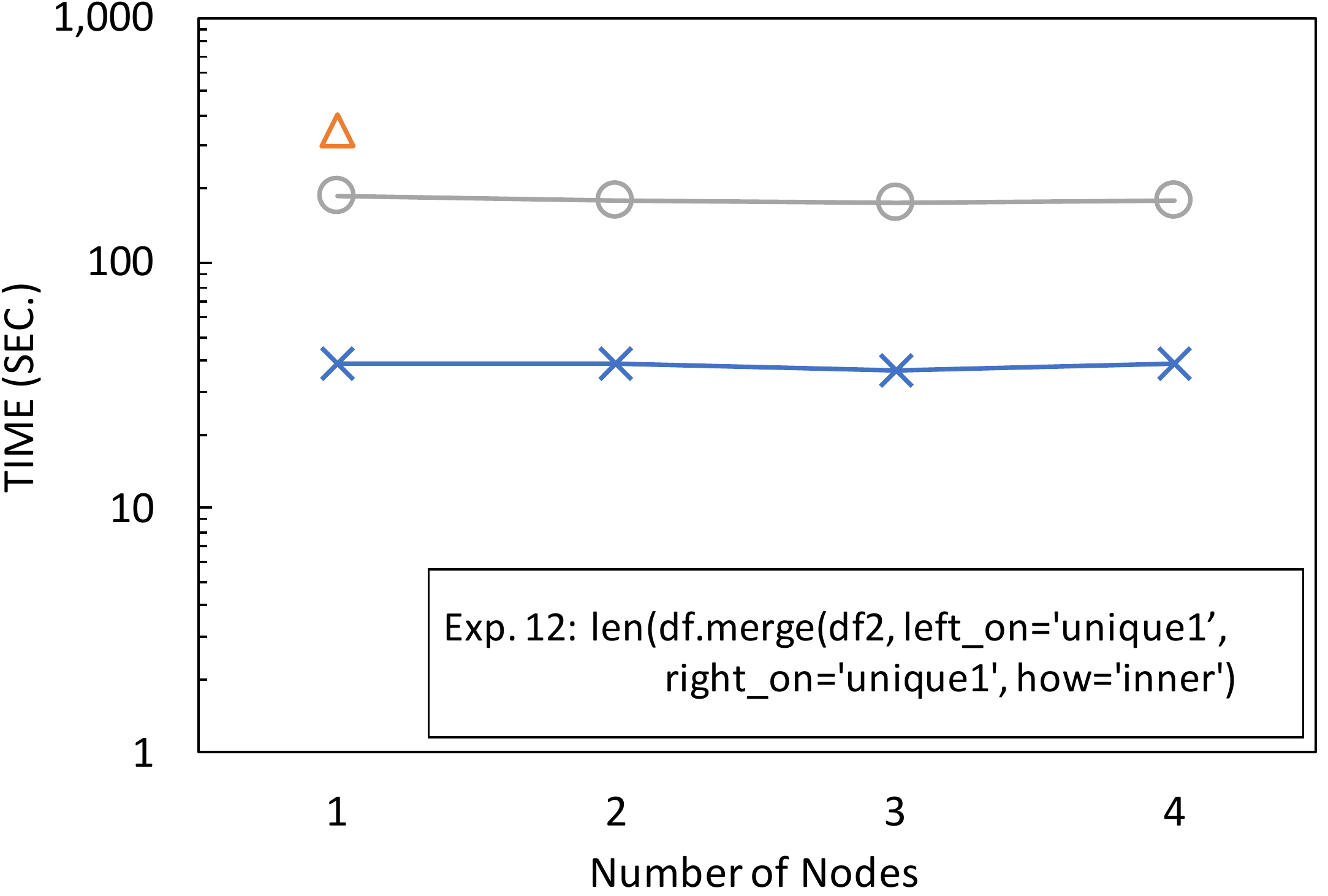}%
        \caption{Expression 12}
        \label{fig:q12_scaleup}
    \end{subfigure}
    
    \begin{subfigure}[t]{0.23\textwidth}
        \includegraphics[trim=0.5cm 1.5 0.5cm 1.5,width=\textwidth,height=3.3cm]{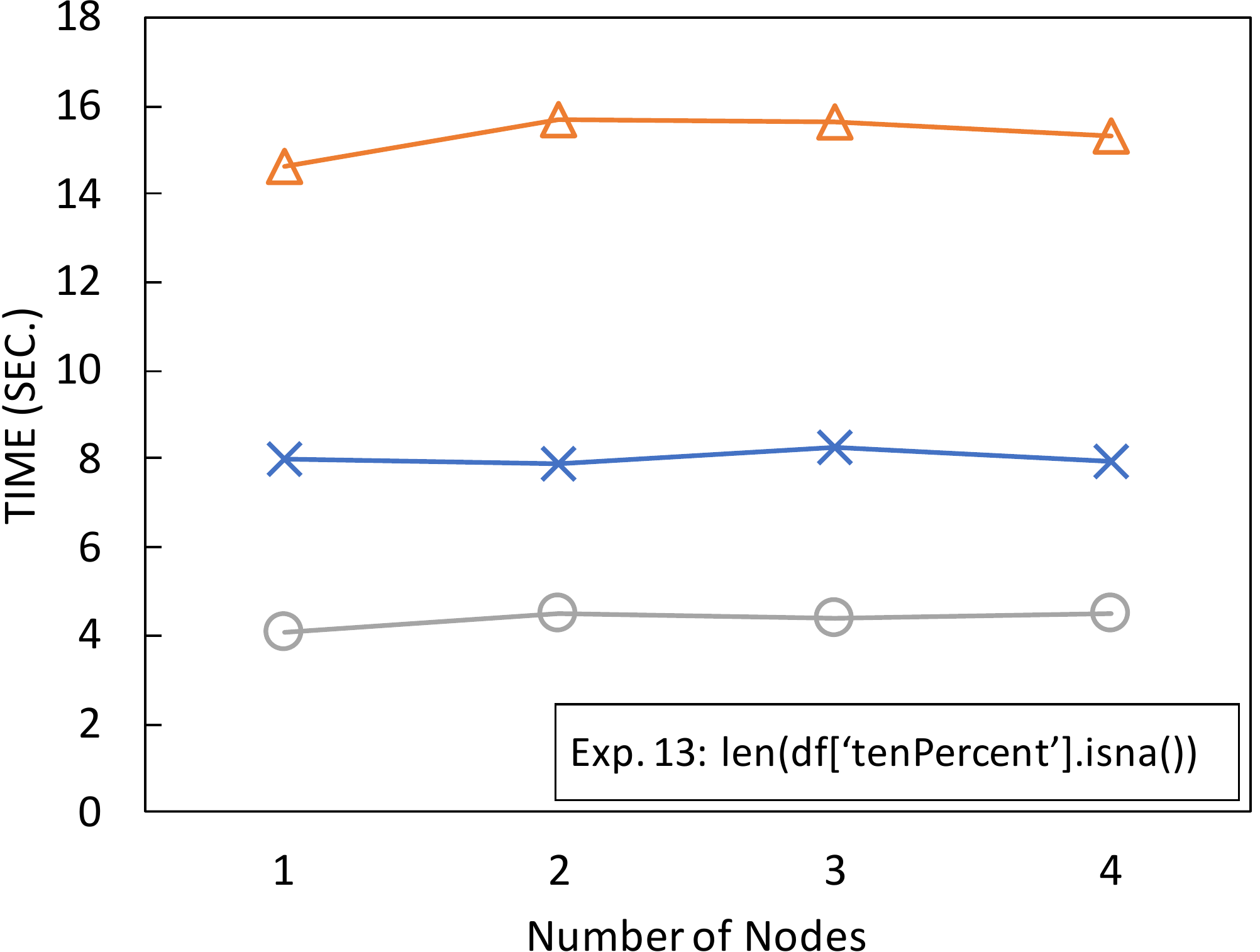}%
        \caption{Expression 13}
        \label{fig:q13_scaleup}
    \end{subfigure}
    
    \caption{Scale-up Evaluation Results}
    \label{fig:scaleup_results}
\end{figure*}

\subsubsection{\textbf{PolyFrame  Heterogeneity (multi-node results)}}
As mentioned in the experiment set up section, we only performed cluster experiments on PolyFrame. We conducted a multi-node evaluation to demonstrate PolyFrame's horizontal scalability on different database systems. We ran the benchmark on PolyFrame operating on AsterixDB, MongoDB, and Greenplum (distributed PostgreSQL) because the Neo4j community edition does not provide support for sharded multi-node clusters. In order to observe the effect of clusters processing data that is larger than the available aggregate memory, we chose to start our multi-node evaluation with the XL (10GB) dataset. Here we evaluated PolyFrame according to both the speedup and scaleup metrics. 

The multi-node evaluation was performed on EC2 machines with the same specifications as the single node evaluation. Figures~\ref{fig:speedup_results} and~\ref{fig:scaleup_results} display the multi-node speedup and scaleup evaluation results respectively. We only display the total time results here since PolyFrame operating on a database system does not require first loading the data into memory.

\begin{itemize}

\item \textbf{Speedup Results}
Figure~\ref{fig:speedup_results} displays the speedup results for expressions 1-13 running on cluster sizes ranging from 1-4 machines. The speedup evaluation results for PolyFrame are mostly consistent with the single node results on the XL dataset except for some of Greenplum's performance. This is due to database optimizations in the latest PostgreSQL version that were not present in the version of PostgreSQL that Greenplum uses. As shown in Figure~\ref{fig:q9_speedup}'s times, Greenplum was not able to use the backward-index scan that the latest PostgreSQL (version 12) used in the single node evaluation; instead it did a table scan. Expressions 6 (Figure \ref{fig:q6_speedup}) and 7 (Figure~\ref{fig:q7_speedup}), which ask for the maximum and minimum values of an attribute, were evaluated as index-only queries in PostgreSQL version 12; however, for the version used in Greenplum, this was not the case.

\item \textbf{Scaleup Results}
Figure~\ref{fig:q1_scaleup} displays the scaleup results for expressions 1-13. No single system performed the best across all tasks, but all systems were able to operate at scale when we increased the workload in proportion to the number of processing machines. The scale-up evaluation results for PolyFrame running on AsterixDB, MongoDB, and Greenplum are also consistent with the single node evaluation on the XL dataset with the same exceptions discussed in the speedup results section. Again, the exceptions are due to the older PostgreSQL version used by Greenplum. 

\end{itemize}

\subsection{Discussion}
We conducted the Spark experiments to show important differences between utilizing database optimizations versus using an optimized compute engine to read and then process the data. It is important to note that passing queries down to a database can significantly lower the amount of transfer data. However, in Spark, doing so requires data scientists to be familiar with the database's query language in order to fully optimize Spark performance. Intuitively, if users can generate the needed database queries and then execute them directly on a database system, that yields the most optimal execution results, as shown in our experiments. However, it significantly reduces the benefits offered by the Dataframe abstraction.

Pandas performed competitively on all tasks when data fits in memory. However, due to its eager evaluation approach, it needs to accommodate intermediate computation results, which leads to higher memory consumption. In addition, Pandas suffered from under-resourced utilization and scalability as it only utilizes a single processing core and only operates on a single machine. On repetitive tasks that return a small subset of larger data, Pandas did not perform as well as PolyFrame operating on database systems that employ parallel processing and that utilize indexes to retrieve only a small subset of the data.

PolyFrame utilizes lazy evaluation by only sending queries to an underlying database system when an action is invoked. This allows PolyFrame to take advantage of database systems' optimizations. As mentioned before, we conducted our single node evaluation to compare PolyFrame's lazy evaluation approach and Pandas' eager in-memory evaluation approach rather than comparing the performance of the different database systems. We demonstrated that operating on top of the database systems allows PolyFrame to take advantage not only of optimizations such as indexes and query optimization, but also of the data management capabilities that go beyond memory limits. PolyFrame does not require loading data into memory prior to computing expression results; this results in lower total runtimes across all benchmark expressions. In terms of the expression-only runtime, lazy evaluation utilizing database indexes and query optimization has better performance than eagerly evaluating certain repetitive tasks. For the multi-node evaluation, we have demonstrated the horizontal scalability of PolyFrame operating on three parallel database systems for both speedup and scale up metrics.

By configuring PolyFrame to work against a sampling of significantly different database systems and query languages, we have also demonstrated the generality and feasibility of its language rewrite rules. The flexibility of our rewrite rules allows PolyFrame to take advantage of each database system’s optimizations while maintaining efficiency. As a side effect of our experimental study, we were also able to identify certain potentially effective optimizations across the evaluated database systems, as discussed next.

PolyFrame operating on top of PostgreSQL was able to take advantage of index-only query plans, backward index scans to retrieve a subset of records sorted in descending order, and null value statistics using its indexes.

For Neo4j, apart from the usual database optimizations (e.g., index search, fast metadata lookup), we found that its storage layout and record structure also contribute to its performance, especially for this particular benchmark dataset. Neo4j stores attributes (node properties) as a linked list of fixed-size records. It stores string attributes in a separate record store, storing only pointers to these string attributes in the attribute linked list. Because of this feature, Neo4j's record structure is particularly suitable for the Wisconsin benchmark dataset since it contains multiple long string attributes. Since only one of our benchmark expressions required access to the string attributes, Neo4j had the advantage of scanning shorter records than the other (row store based) database systems. 

As mentioned before in the case of MongoDB, PolyFrame utilizes its aggregation pipeline language in order to preserve AFrame’s incremental query formation and to support the language rewriting process. As a result, certain query optimizations were not considered by MongoDB. For example, MongoDB also supports a fast lookup of certain metadata (similar to Neo4j) that includes the total count of records in a collection. However, this particular optimization is not enabled as part of a MongoDB aggregation pipeline, so PolyFrame was not able to take advantage of this optimization. Another limitation for MongoDB is related to joins. MongoDB only supports the joining of unsharded data, which requires at least one of the joined datasets to be stored on a single machine. As such, we could not run expression 12 on MongoDB in the distributed environment.

\section{Conclusions and Future Work}
In this work, we have shown the practicality of retargeting AFrame’s incremental query formation approach onto a range of query-based database systems in order to scale dataframe operations without requiring users to have distributed or database systems expertise. The flexibility of our language rewrite rules enables database-specific optimizations and makes extending the Pandas DataFrame API to custom languages and systems possible. We evaluated PolyFrame versus Pandas DataFrames through a set of analytical benchmark operations. As a result, we have also shown that lazy evaluation, which takes advantage of database optimizations, is an efficient (and important) solution to data analysis at scale.

In its current stage, PolyFrame has already shown promising results for enabling a scale-independent data analysis experience. Moving forward, we would like to extend the PolyFrame framework's support to cover the extensive list of Pandas operations. A recent paper~\cite{modin_algebra} has given a formal definition to dataframe operators. It may be worthwhile to incorporate their dataframe algebra with our generic rewrite rules to provide an intermediate abstraction for query language mapping. Another research problem being explored in other similar dataframe libraries is how to support Pandas’ notion of row label efficiently in a distributed environment. There is not yet an efficient solution to enable row-indexing on unordered data (which is the data model used in most existing database systems); currently, an order is required in the form of either system-generated internal identifiers or sorted data to enable such a capability in a distributed environment. This results in a performance trade-off that we would like to eliminate if possible.

% \balance
% \bibliographystyle{IEEEtran}
\bibliographystyle{abbrv}
\bibliography{references}
% \vspace{12pt}

\clearpage

\appendix

\lstdefinestyle{pythonstyle}
{
    language = python,
    showstringspaces=false,
    basicstyle=\ttfamily,
    keywordstyle=\color{blue},
    commentstyle=\color{gray}\ttfamily,
    otherkeywords = {min, groupby,head,max,sort_values,map,agg,len, merge,count, read_json},
    morekeywords = [2]{min},
    morekeywords = [3]{groupby},
    morekeywords = [4]{head},
    morekeywords = [5]{AFrame},
    frame=single
}

\lstdefinestyle{sqlstyle}
{
    language = sql,
    showstringspaces=false,
    basicstyle=\ttfamily,
    keywordstyle=\color{blue},
    otherkeywords = {value, count, group by, START, FEED, TO, WITH, TYPE, CLOSED, DATASET},
    morekeywords = [2]{value},
    morekeywords = [3]{group by},
    morekeywords = [4]{from},
    frame=single
}

\lstdefinestyle{cypherstyle}
{
    % language = sql,
    showstringspaces=false,
    basicstyle=\ttfamily,
    keywordstyle=\color{blue},
    otherkeywords = {MATCH, WITH, RETURN, LIMIT,ORDER,BY,WHERE,DESC,AS,COUNT,AND,IS,NULL},
    morekeywords = [2]{return},
    morekeywords = [3]{match},
    morekeywords = [4]{with},
    morekeywords = [5]{limit},
    frame=single
}
\lstdefinestyle{mongostyle}
{
    % language = sql,
    showstringspaces=false,
    basicstyle=\ttfamily,
    keywordstyle=\color{blue},
    otherkeywords = {match, project, limit, expr,eq,\$,gte,lte,unwind,count,lookup,group,addFields,sum,and,sort},
    morekeywords = [2]{toUpper},
    morekeywords = [3]{match},
    morekeywords = [4]{project},
    morekeywords = [5]{limit},
    morekeywords = [6]{expr},
    morekeywords = [7]{eq},
    morekeywords = [8]{gte},
    morekeywords = [9]{lt},
    morekeywords = [10]{unwind},
    frame=single
}

\lstdefinelanguage{ini}
{
basicstyle=\ttfamily,
columns=fullflexible,
morecomment=[s][\color{black}\bfseries]{[}{]},
morecomment=[l]{\#},
morecomment=[l]{;},
commentstyle=\color{gray}\ttfamily,
morekeywords={},
otherkeywords={=,:},
keywordstyle={\color{blue}\bfseries}
}

\subsection{PolyFrame Translated Queries}
\label{sec:language_translate}
\footnotesize
\textbf{Pandas Dataframe Expression}
\begin{lstlisting}[style=pythonstyle]
df[df['lang'] == 'en'][['name', 'address']].head(10)
\end{lstlisting}

\textbf{SQL++ Translation}
\begin{lstlisting}[style=sqlstyle]
SELECT t.name, t.address 
FROM (SELECT VALUE t 
      FROM (SELECT VALUE t 
	    FROM Test.Users t) t 
            WHERE t.lang = 'en') t 
LIMIT 10;
\end{lstlisting}

\textbf{SQL Translation}
\begin{lstlisting}[style=sqlstyle]

SELECT t.name, t.address 
FROM (SELECT * 
      FROM (SELECT * 
	    FROM Test.Users t) t 
            WHERE t.lang = 'en') t 
LIMIT 10;

\end{lstlisting}

\textbf{MongoDB Query Language Translation}
\begin{lstlisting}[style=mongostyle]
Test.Users.aggregate([
    { "$match": {} },
    { "$match": {"$expr": {"$eq":["$lang", "en"]}}},
    { "$project": { "name": 1, "address": 1 } },
    { "$project": { "_id": 0 } },
    { "$limit" : 10 }
])
\end{lstlisting}

\textbf{Cypher Translation}
\begin{lstlisting}[style=cypherstyle]
MATCH(t: Users)
WITH t WHERE t.lang = "en"
WITH t{`name`:t.name, `address`:t.address}
RETURN t
LIMIT 10

\end{lstlisting}

\subsection{Sample Language-specific rewrite rules for Cypher}
\label{sec:cypher_rewrites}
\footnotesize
\begin{lstlisting}[language={ini}]
;Below are query explanations
;q1: select all records from a collection
;q2: project an attribute
;q3: return total count of records
;q4: sort based on an attribute in descending order
;q5: sort based on an attribute in ascending order
; .
; .
; .
[QUERIES]
q1 = MATCH(t: $collection)
q2 = $subquery
    WITH t{$attribute_alias}
q3 = $subquery
    RETURN COUNT(*) AS t
q4 = $subquery
    WITH t ORDER BY $sort_desc_attr DESC

q5 = $subquery
    WITH t ORDER BY $sort_asc_attr
...
[ATTRIBUTE ALIAS]
single_attribute = t.$attribute
attribute_alias = `$alias`: $attribute
sort_asc_attr = t.$attribute
sort_desc_attr = t.$attribute
attribute_separator = $left, $right

[ARITHMETIC STATEMENTS]
add = $left + $right
sub = $left - $right
mul = $left * $right
div = $left / $right
mod = $left %% $right

[LOGICAL STATEMENTS]
and = $left AND $right
or = $left OR $right
not = NOT $left

[COMPARISON STATEMENTS]
eq = $left = $right
ne = $left != $right
gt = $left > $right
lt = $left < $right
ge = $left >= $right

[TYPE CONVERSION]
to_str = apoc.convert.toInteger($statement)
to_int = apoc.convert.toInteger($statement)

[LIMIT]
limit = $subquery
        RETURN t
        LIMIT $num
return_all = $subquery
              RETURN t
            
[FUNCTIONS]
min = min(t.$attribute)
max = max(t.$attribute)
avg = avg(t.$attribute)
std = stDevP(t.$attribute)
count = count(t.$attribute)
\end{lstlisting}

\subsection{Sample Language-specific rewrite rules for MongoDB}
\label{sec:mongo_rewrites}
\footnotesize
\begin{lstlisting}[language={ini}]
[QUERIES]
q1 = { "$match": {} }
q2 = $subquery,
    { "$project": { $attribute_alias } }
q3 = $subquery,
    { "$count": "count" }
q4 = $subquery,
    { "$sort": { $sort_desc_attr } }
q5 = $subquery,
    { "$sort": { $sort_asc_attr } }
...
[ATTRIBUTES]
single_attribute = $attribute
attribute_alias = "$alias": { $attribute }
sort_asc_attr = "$attribute": 1
sort_desc_attr = "$attribute": -1
attribute_separator = $left, $right

[ARITHMETIC STATEMENTS]
add = "$add": [ "$$left", $right ]
sub = "$subtract": [ "$$left", $right ]
mul = "$multiply": [ "$$left", $right ]
div = "$divide": [ "$$left", $right ]
mod = "$mod": [ "$$left", $right ]

[LOGICAL STATEMENTS]
and = "$and": [ { $left }, { $right } ]
or = "$or": [ { $left }, { $right } ]
not = "$not": [ { $left } ]

[COMPARISON STATEMENTS]
eq = "$eq": ["$$left", $right]
ne = "$ne": ["$$left", $right]
gt = "$gt": ["$$left", $right]
lt = "$lt": ["$$left", $right]
ge = "$gte": ["$$left", $right]

[TYPE CONVERSION]
to_int = "$toInt": { $statement }
to_str = "$toString": { $statement }

[LIMIT]
limit = $subquery,
        { "$project": { "_id": 0 } },
        { "$limit" : $num }
return_all = $subquery

[FUNCTIONS]
min = "$min": "$$attribute"
max = "$max": "$$attribute"
avg = "$avg": "$$attribute"
std = "$stdDevPop": "$$attribute"
abs = "abs": "$$attribute"

[SAVE RESULTS]
to_collection = $subquery,
    { "$out": "$collection" }
to_view = CREATE FUNCTION $namespace.$collection()
          {$subquery};
\end{lstlisting}

\subsection{Benchmark Timing Points}
\label{sec:timingAppendix}
\begin{itemize}

\item\textbf{Pandas Timing}
\vspace{-0.5em}
\begin{lstlisting}[style=pythonstyle]
# DataFrame creation time
df = pd.read_json(file_path)
# Expression-only time
df.head()
\end{lstlisting}

\item\textbf{Spark Timing}
\begin{lstlisting}[style=pythonstyle]
# DataFrame creation time
df = spark.read.format("mongo")
     .option("uri", "mongodb://x.x.x.x")
     ...
     .load()
# Expression-only time
df.head(5) 
\end{lstlisting}

\item\textbf{PolyFrame Timing}
\begin{lstlisting}[style=pythonstyle]
# DataFrame creation time
df = AFrame(namespace, collection, DBConnector) 
# Expression-only time
df.head() 
\end{lstlisting}

\end{itemize}

\subsection{Benchmark Translated SQL++ Queries}
\label{sec:sqlpp_queries}
\footnotesize
\begin{lstlisting}[style=sqlstyle]
1.  SELECT VALUE COUNT(*) FROM data;
2.  SELECT two, four 
    FROM (SELECT VALUE t FROM data t) t 
    LIMIT 5;
3.  SELECT VALUE COUNT(*) 
    FROM (SELECT VALUE t 
          FROM (SELECT VALUE t FROM data t) t 
          WHERE ten = x 
            AND twentyPercent = y 
            AND two = z) t;
4.  SELECT oddOnePercent, 
           COUNT(oddOnePercent) AS cnt 
    FROM (SELECT VALUE t FROM data) t 
    GROUP BY oddOnePercent; 
5.  SELECT VALUE UPPER(stringu1) 
    FROM (SELECT VALUE t FROM data) t 
    LIMIT 5;
6.  SELECT MAX(unique1) 
    FROM (SELECT unique1 
          FROM (SELECT VALUE t FROM data) t) t;
7.  SELECT MIN(unique1) 
    FROM (SELECT unique1 
          FROM (SELECT VALUE t FROM data) t) t;
8.  SELECT twenty, MAX(four) AS max_four 
    FROM (SELECT VALUE t FROM data) t 
    GROUP BY twenty;
9.  SELECT VALUE t 
    FROM (SELECT VALUE t FROM data) t 
    ORDER BY unique1 DESC 
    LIMIT 5;
10. SELECT VALUE t 
    FROM (SELECT VALUE t FROM data) t 
    WHERE ten = x 
    LIMIT 5;
11. SELECT VALUE COUNT(*)
    FROM (SELECT VALUE t 
        FROM (SELECT VALUE t FROM data) t 
        WHERE onePercent >= x 
    	    AND onePercent <= y) t;
12. SELECT VALUE COUNT(*) 
    FROM (SELECT l,r 
              FROM leftData l JOIN rightData r 
              ON l.unique1 = r.unique1) t;
13. SELECT VALUE COUNT(*)
    FROM (SELECT VALUE t 
        FROM (SELECT VALUE t FROM data) t 
        WHERE tenPercent IS UNKNOWN) t;
x,y,z = random values within range
\end{lstlisting}

\subsection{Benchmark Translated SQL Queries}
\label{sec:sql_queries}
\footnotesize
\begin{lstlisting}[style=sqlstyle]
1.  SELECT COUNT(*) FROM (SELECT * FROM data) t;
2.  SELECT "two", "four" 
    FROM (SELECT * FROM data) t LIMIT 5;
3.  SELECT COUNT(*) 
    FROM (SELECT * 
          FROM (SELECT * FROM data) t
          WHERE "ten" = x 
            AND "twentyPercent" = y 
            AND "two" = z) t;
4.  SELECT "oddOnePercent", 
            COUNT("oddOnePercent") AS cnt 
    FROM (SELECT * FROM data) t 
    GROUP BY "oddOnePercent"; 
5.  SELECT upper("stringu1") 
    FROM (SELECT stringu1 
          FROM (SELECT * FROM data) t) t 
    LIMIT 5;
6.  SELECT MAX("unique1") 
    FROM (SELECT unique1 
          FROM (SELECT * FROM data) t) t;
7.  SELECT MIN("unique1") 
    FROM (SELECT unique1 
          FROM (SELECT * FROM data) t) t;
8.  SELECT "twenty", MAX("four") 
        FROM (SELECT * FROM data) t 
        GROUP BY "twenty";
9.  SELECT * 
    FROM (SELECT * FROM data) t 
    ORDER BY unique1 DESC 
    LIMIT 5;
10. SELECT * 
    FROM (SELECT * FROM data) t 
    WHERE "ten" = x 
    LIMIT 5;
11. SELECT COUNT(*)
    FROM (SELECT * 
         FROM (SELECT * FROM data t) t 
         WHERE "onePercent" >= x 
            AND "onePercent" <= y) t;
12. SELECT COUNT(*) 
    FROM (SELECT l.*,r.* 
          FROM (SELECT * FROM left) l 
          INNER JOIN (SELECT * FROM right) r 
          ON l.unique1 = r.unique1) t;
          
13. SELECT COUNT(*)
    FROM (SELECT * 
        FROM (SELECT * FROM data) t 
        WHERE "tenPercent" IS NULL) t;
        
x,y,z = variables representing random values within 
        range
\end{lstlisting}

\subsection{Benchmark Translated Cypher Queries}
\label{sec:cypher_queries}
\footnotesize
\begin{lstlisting}[style=cypherstyle]
1.  MATCH(t: data)
    RETURN COUNT(*) AS t
2.  MATCH(t: data)
    WITH t{`two`:t.two, `four`:t.four}
    RETURN t
3.  MATCH(t: data)
    WITH t WHERE t.ten = x 
            AND t.twentyPercent = y 
            AND t.two = z
    RETURN COUNT(*) AS t
4.  MATCH(t: data)
    WITH {`oddOnePercent`: t.oddOnePercent, 
        `count`: count(t.oddOnePercent)} AS t
    RETURN t
5.  MATCH(t: data)
    WITH t{`stringu1`:t.stringu1}
    WITH t{`upper(t.stringu1)`:upper(t.stringu1)}
    RETURN t
    LIMIT 5
6.  MATCH(t: data)
    WITH t{`unique1`:t.unique1}
    WITH {`max_unique1`: max(t.unique1)} AS t
    RETURN t
7.  MATCH(t: data)
    WITH t{`unique1`:t.unique1}
    WITH {`min_unique1`: min(t.unique1)} AS t
8.  MATCH(t: data)
    WITH {`twenty`: t.twenty, 
          `max_four`: max(t.four)} AS t
    RETURN t
9.  MATCH(t: data)
    WITH t ORDER BY t.unique1 DESC
    RETURN t
    LIMIT 5
10. MATCH(t: data)
    WITH t WHERE t.ten = x
    RETURN t
    LIMIT 5
11. MATCH(t: data)
    WITH t WHERE t.onePercent >= x 
            AND t.onePercent <= y
    RETURN COUNT(*) AS t
12. MATCH(t: data)
    MATCH (t),(r:wisconsin2)
    WHERE t.unique1 = r.unique1
    WITH t{.*, r}
    RETURN COUNT(*) AS t
13. MATCH(t: data)
    WITH t WHERE t.tenPercent IS NULL
    RETURN COUNT(*) AS t
x,y,z = random values within range
\end{lstlisting}

\subsection{Benchmark Translated MongoDB Queries}
\label{sec:mongo_queries}
\footnotesize
\begin{lstlisting}[style=mongostyle]
1.  namespace.collection.aggregate([
    {"$match":{}} ,
    {"$count":"count"}
    ])
2.  namespace.collection.aggregate([
    {"$match":{}},
    {"$project":{"two":1,"four":1}},
    {"$project":{ "_id":0}},
    {"$limit":5}
    ])
3.  namespace.collection.aggregate([
    {"$match":{}},
    {"$match":{"$expr":{"$and":[{"$and":[
                    {"$eq":["$ten",x]}, 
                    {"$eq":["$twentyPercent", y]}]}, 
                    {"$eq":["$two",z]}]}}},
    {"$count":"count"}])
4.  namespace.collection.aggregate([
    {"$match":{}},
    {"$group":{
        "_id":{ "oddOnePercent":"$oddOnePercent" }, 
        "count_oddOnePercent":{"$sum":1}}},
    {"$addFields": { "oddOnePercent": 
                    "$_id.oddOnePercent"} },
    {"$project": {"_id": 0 } }
    ]) 
5.  namespace.collection.aggregate([
    {"$match":{}},
    {"$project":{"stringu1":1}},
    {"$project":{"stringu1:{"$toUpper":"$stringu1"}}},
    {"$project":{"_id":0}},
    {"$limit":5}
    ])
6.  namespace.collection.aggregate([
    {"$match":{}},
    {"$project":{"unique1":1}},
    {"$group":{"_id":{},"max":{"$max":"$unique1"}}},
    {"$project":{"_id":0}}
    ])
7.  namespace.collection.aggregate([
    {"$match":{}},
    {"$project":{"unique1":1}},
    {"$group":{"_id":{},"min":{"$min":"$unique1"}}},
    {"$project":{"_id":0}}
    ])
8.  namespace.collection.aggregate([
    {"$match":{}},
    {"$group":{"_id":{"twenty":"$twenty"},
                "max":{"$max":"$four"}}},
    {"$addFields":{"twenty":"$_id.twenty"}},
    {"$project":{"_id":0}}
    ])
9.  namespace.collection.aggregate([
    {"$match":{}},
    {"$sort":{"unique1":-1}},
    {"$project":{"_id":0}},
    {"$limit":5}
    ])
10. namespace.collection.aggregate([
    {"$match":{}},
    {"$match":{"$expr":{"$eq":["$ten",x]}}},
    {"$project":{"_id":0}},
    {"$limit":5}   
    ])
11. namespace.collection.aggregate([
    {"$match":{}},
    {"$match":{"$expr":{"$and":[
                    {"$gte":["$onePercent",x]},
                    {"$lte":["$onePercent",y]}]}}},
    {"$count":"count"}
    ])
12. namespace.collection.aggregate([
    {"$lookup":{"from":"collection2",
            "as":"collection2",
            "let":{"left":"$unique1"},
            "pipeline":[{"$match":{}},
                {"$match":{"$expr":
                  {"$eq":["$unique1","$$left"]}}}]}},
    {"$unwind":{"path":"$collection2",
            "preserveNullAndEmptyArrays":false}},
    {"$count":"count"}
    ])
13. namespace.collection.aggregate([
   {"$match":{}},
   {"$match":{"$expr":{"$lt":["$tenPercent",null]}}},
   {"$count":"count"}
   ])
   
x,y,z = variables representing random values within 
        range
        
        
        
\end{lstlisting}

\end{document}